\documentclass[structabstract]{aa}  

\usepackage{graphicx}
\usepackage[varg]{txfonts}
\usepackage{amsmath}
\usepackage{longtable}
\usepackage{longtable,lscape}
\usepackage{lscape}
\usepackage{epsfig}
\usepackage{color}

\newcommand\be{\begin{equation}}
\newcommand\en{\end{equation}}

\usepackage{txfonts}
\usepackage{hyperref}
\begin{document} 

\title{IRAM and Gaia views of multi-episodic star formation in IC\,1396A:}
 
\subtitle{The origin and dynamics of the Class 0 protostar at the edge of an HII region}

\author{Aurora Sicilia-Aguilar\inst{1,2}, Nimesh Patel\inst{3}, Min Fang\inst{4}, Veronica Roccatagliata\inst{5,6},
Konstantin Getman\inst{7}, Paul Goldsmith\inst{8}}

\institute{\inst{1} SUPA, School of Science and Engineering, University of Dundee, Nethergate, Dundee DD1 4HN, UK \\
	\email{a.siciliaaguilar@dundee.ac.uk}\\
\inst{2} SUPA, School of Physics and Astronomy, University of St Andrews, North Haugh, St Andrews KY16 9SS, UK\\
\inst{3} Harvard-Smithsonian Center for Astrophysics, 60 Garden Street, Cambridge, MA 02138, USA \\
\inst{4} Department of Astronomy, University of Arizona, 933 North Cherry Avenue, Tucson, AZ 85721, USA\\ 
\inst{5} INAF/Osservatorio Astrofisico di Arcetri, Largo E. Fermi 5, 50125, Firenze, Italy \\
\inst{6} Dipartimento di Fisica ``Enrico Fermi", Universit\`{a} di Pisa, Largo Pontecorvo 3, 56127 Pisa , Italy\\
\inst{7} Department of Astronomy \& Astrophysics, 525 Davey Laboratory, Pennsylvania State University, University Park PA 16802\\
\inst{8} Jet Propulsion Laboratory, M/S 180-703, 4800 Oak Grove Drive, Pasadena, CA 91109
	}
	
   \date{Submitted April 11, 2018, accepted December 17, 2018}

\abstract
{IC\,1396A is a cometary globule that contains the Class 0 source IC1396A-PACS-1, discovered with \emph{Herschel}.}
{We use IRAM 30m telescope and Gaia DR2 data to explore the star-formation history of IC\,1396A and investigate the possibilities of triggered star formation.}
{IRAM and \emph{Herschel} continuum data are used to obtain dust temperature and  column density maps. Heterodyne data reveal the velocity structure of the gas. Gaia DR2 proper motions for the stars complete the kinematics of the region.  }
{IC1396A-PACS-1 presents molecular emission similar to a hot corino with warm 
carbon chain chemistry due to the UV irradiation.
The source is embedded in a dense clump surrounded by 
gas at velocities significantly different from the velocities of the Tr~37 cluster.
CN emission reveals photoevaporation, while continuum data and 
high density tracers (C$^{18}$O, HCO$^+$, DCO$^+$, N$_2$D$^+$) reveal distinct
gaseous structures with a range of densities and masses.}
{Combining the velocity, column density, and temperature information and Gaia DR2 kinematics, we
confirm that the globule has suffered various episodes of star formation.
IC1396A-PACS-1 is probably the last intermediate-mass protostar that will form within IC\,1396A, showing evidence of triggering by radiative driven implosion.  Chemical signatures such as CCS
place IC1396A-PACS-1 among the youngest protostars known.
Gaia DR2 data
reveal velocities in the plane of the sky  $\sim$4\,km/s for IC\,1396A with respect to Tr~37.
The total velocity difference (8 km/s) between the Tr~37 cluster and IC\,1396A
 is too small for IC\,1396A to have undergone substantial rocket acceleration,
which imposes constraints on the distance to the ionizing source in time and the possibilities of triggered star formation.
The three stellar populations in the globule 
reveal that objects located within relatively close distances ($<$0.5\,pc)
can be formed in various star-forming episodes
within $\sim$1-2 Myr period. Once the remaining cloud disperses,
we expect substantial differences in evolutionary stage and initial conditions for the 
resulting objects and their protoplanetary disks, which may affect their evolution. 
Finally, evidence for short-range feedback from 
the embedded protostars and, in particular, the A-type star V390 Cep is also observed. }


\authorrunning{Sicilia-Aguilar et al.}

\titlerunning{Origin of IC1396A-PACS-1}

\maketitle


\section{Introduction \label{intro}}

The IC~1396A dark globule is one of the classical examples of 
bright-rimmed clouds (BRC) at the
edge of an H~II region \citep{sharpless59,osterbrock89,patel95}. 
The globule is illuminated by the
O6.5 trapezium-like system HD~206267 in the center of the Tr\,37 cluster 
\citep{kun90,peter12}. HD~206267 is located 
at about 4.5 pc projected distance, considering the 870 pc distance to Tr\,37 \citep{contreras02}. 
The main structure consist of a dark cloud about $\sim$5.4 arcmin ($\sim$1.4 pc) in size.
Behind the tail of this cometary-shaped BRC, dark globules and
ionized rims extend over more than half a degree. 
These structures cover only a small part of the large bubble-shaped nebula around the Tr\,37 cluster,
which has been beautifully imaged in H$\alpha$ \citep{barentsen11} and by IR space
missions, such as AKARI \citep{huang13}.

Early molecular-line observations suggested the presence of highly embedded sources
and the potential of the region to undergo a substantial episode of star formation
\citep{loren75}. The total gas content was estimated to be about 200\,M$_\odot$
\citep{patel95}. Several very young objects
were confirmed with the advent of the mid-IR observatories,
starting with IRAS \citep{sugitani91} and continuing with the \emph{Spitzer Space
Telescope}. \citet{reach04} and \citet{sicilia06a} identified more than 40 embedded
young stars, most of them low-mass Class I protostars and T Tauri stars. \emph{Spitzer} data also
revealed protoplanetary disks around many of the young T Tauri stars around the globule, and of
variable accretion/variable obscuration in some of them \citep{morales09}. 
The interaction between the optical stars and the nebula is also clear.
\emph{Spitzer} revealed a delicate structure of filaments and reddened objects interlaced within
the globule, including heated structures behind the ionization rim 
(and around the \textit{eye}-shaped hole
containing V 390 Cep) that confirm the interaction between stars and the globule, and also knots suggestive of 
heating by jets and outflows from the embedded population. Imaging in [S II] showed
jets and outflows, often associated with the known optical and IR sources \citep{sicilia13}.

\begin{figure*}
\centering
\includegraphics[width=0.9\linewidth]{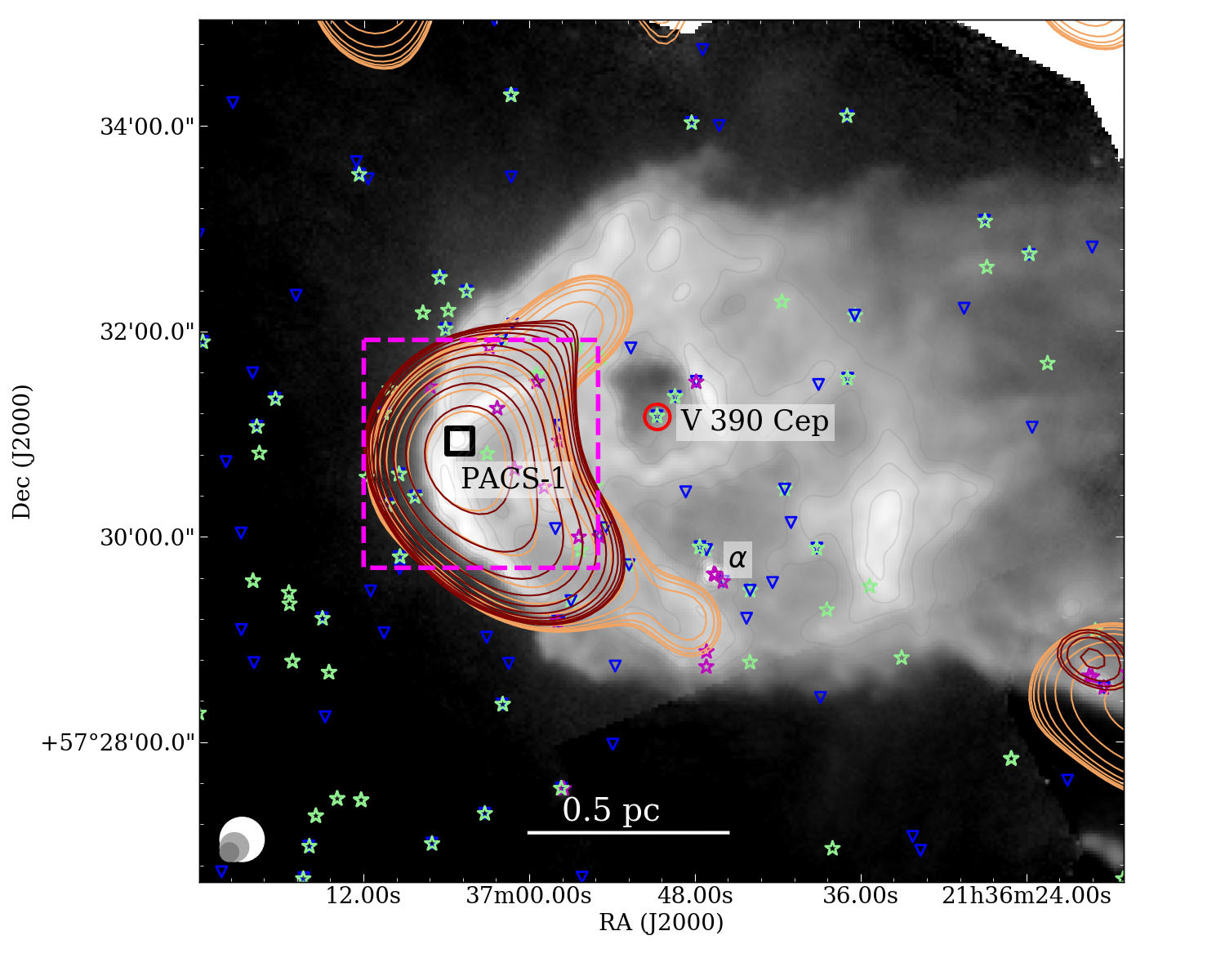}
\caption{Observations of IC1396A. Gray-scale background image: \emph{Herschel}/PACS 70~$\mu$m. Orange contours: 
NIKA 1.3mm data (15 contours, starting at 3$\sigma$=0.007 Jy/beam up to 0.3 Jy/beam in log scale). Brown contours: NIKA 2mm
data (15 contours, starting at 3$\sigma$=0.003 Jy/beam up to 0.14 Jy/beam in log scale). The rms increases towards the edges of the field.
The field observed with EMIR is marked with a pink dashed box. Known young objects are marked as green stars
\citep[detected in the optical;][]{sicilia04,sicilia05,sicilia13,barentsen11}, 
magenta stars \citep[detected in the IR;][]{reach04,sicilia06a,morales09}, and blue inverted triangles \citep[X-ray detections;][]{getman12}.
The Class 0 source, marked with a black square, is located in the coolest, densest
part of the globule. V 390 Cep is marked with a red circle. The protostar $\alpha$ \citep{reach04} is also labeled. The beams for 1.3, 2, and 3mm are shown in the lower left corner.\label{nika-fig}}
\end{figure*}

Deep optical/near-IR photometry 
and spectroscopy \citep{sicilia05,sicilia13,getman12} led to the 
identification and classification of numerous young stars in and around the globule,
and X-ray imaging also confirmed a substantial young population \citep{getman12}.
The sources in IC~1396A, younger and less evolved than the population in the Tr~37
cluster, were also strong candidates for triggered
or sequential star formation. 
Gas dynamics showed expansion of the H\,II region and ionization
front around HD 206267, which could
lead to triggered or sequential star formation in the
dense globules around the massive star \citep{patel95}. The young members
of IC~1396A would be part of
the multiple star-forming episodes within the entire Cep~OB2 region
($\sim$120 pc in diameter), which has suffered triggered or sequential
bursts of star formation starting some 10-12 Myr ago \citep{patel98}.
The ages derived from gas dynamics were also in agreement with the isochronal ages
of young stars in Tr\,37 and IC\,1396A \citep{sicilia05,getman12}
and with the evolutionary status of
the stars and disks as seen with \emph{Spitzer} \citep{reach04,sicilia06a}.
The presence of younger
stars could be equally well explained by either triggered star formation by the action of a previously-formed population
and the expansion of the H\,II region in a radiation-driven implosion scenario 
\citep[RDI;][]{sandford80,bertoldi89}, or by time-sequential formation across the molecular cloud.
Dynamical evidence of triggering is usually elusive, and velocity observations in clouds similar to
IC~1396A are often inconclusive regarding triggering on large scales \citep[e.g.][]{mookerjea12}.

Although early millimeter and molecular-line observations showed an overdensity at 
the tip of IC~1396A \citep{loren75,patel95}, the large beams used did not
allow resolving any point source at this location.
Our \emph{Herschel}/PACS observations of IC~1396A revealed a remarkable object at 
the very tip of the cloud and facing the ionized rim: IC1396A-PACS-1
\citep[][from now on Paper I]{sicilia14}. The object is the brigthest 70~$\mu$m point-source detected, and it has
some extended structure running along the BRC rim at 160~$\mu$m. Its spectral energy distribution (SED) 
agrees with that of a Class 0 source with a very low temperature ($\sim$16-20 K; Paper I).
Its very early evolutionary state is also in agreement with the lack of any other positive identification
at optical and IR wavelengths. None of the outflows detected in [S II] \citep{sicilia13}
is related to IC1396A-PACS-1. The fact that this object is located in the coolest and densest
part of the globule suggested that it is the most embedded and the youngest among all the members of
IC~1396A (Paper I).

The discovery of the
Class 0 source with the \emph{Herschel Space Telescope} is the motivation for the detailed
continuum and molecular line observations with the IRAM telescope presented here.
This paper is the first part of our study to understand the formation
history and structure of the object from a dynamical point of view. A second paper
(Sicilia-Aguilar et al. in prep) will deal with the chemical analysis of the region.
The study of the region will be completed by SMA observations at higher angular resolution
\citep[][Patel et al. in prep]{patel15}. The IRAM observations and ancillary data are
described in Section \ref{observations-sect}. In Section \ref{analysis-sect} we derive the
physical and dynamical parameters of the region. The formation history of IC~1396A is discussed
in Section \ref{discussion-sect}, and our results are summarized in Section \ref{conclu}.

\section{Observations and data reduction \label{observations-sect}}

\subsection{IRAM EMIR and NIKA data \label{observations-iram}}

The observations were obtained using the EMIR heterodyne receiver \citep{carter12} for molecular-line
observations, and the bolometer camera NIKA \citep{monfardini10} for continuum observations. Both instruments
have the advantage that they allow simultaneous observations in two bands, with several 
options for the case of EMIR, and 1.3 and 2.1 mm in the case of NIKA. 
We used NIKA to produce a uniform map of the entire IC~1396A globule,
and EMIR to obtain line observations of the Class 0 source and its surroundings. 
The beam size of IRAM at 1.3mm ($\sim$11.8"; the beam size at 2mm is 17.5". ) 
is comparable to the
FWHM of \emph{Herschel}/PACS at 160~$\mu$m, providing excellent spatial resolution that allows us
to study the same structures detected with \emph{Herschel}. 

The NIKA data were obtained on 2014-02-28  using the limited bolometer array
before the upgrade to NIKA-2. 
Given the large size of the globule, the region was divided in 
four 4'$\times$4' maps, which was the most efficient arrangement in terms of 
total time, following the exposure time calculations for NIKA. 
Each map took a total of
2.5 h of on-the-fly (OTF) mapping (including scan and cross-scan maps), achieving
 a rms $\sim$11.5 mJy at 1.3~mm
and 2.5 mJy at 2.0~mm in clean areas.
The data were reduced by the NIKA team following the procedures for  the 
NIKA Data Products v1\footnote{http://www.iram.es/IRAMES/mainWiki/Continuum/NIKA/DataReduction} which builds on the techniques described in \cite{catalano14}. First, clean time ordered information is created, and bad pixels are flagged. Instrumental effects are flagged and filtered and cosmic rays are removed by flagging peaks at 5$\sigma$ level. The data are then corrected for atmospheric absorption and calibrated.
The IC1396A field
has the complication that it includes both point-like and extended emission. For this project,
we focused in the
detection of the Class 0 source and nearby structure, which means that part of the fainter
cloud may not be properly extracted. A total of 160 scans  with typical integration times 140s 
were combined in a single on-the-fly map. The atmospheric opacity, measured
via skydips, ranged from $\tau$=0.01 to 0.33 at 1mm (average 0.18) and 0.01 to 0.25 at 2mm (average 0.14).  
The atmospheric and electronic noise
decorrelation is done following the iterative procedure masking the point source as described in \citet{catalano14}. The final map is created avoided the flagged data and weighting each detector sample by the inverse variance of the detector timeline. The errors induced by the
filtering have been estimated to be around 5\%, while the nominal calibration errors of the 1mm and 2mm NIKA channels are 15\% and 10\%, respectively \citep{catalano14}.
The final map reveals strong emission at the tip of the globule, with IC1396A-PACS-1 being detected as a
point-like source (see Figure \ref{nika-fig}).

\begin{table*}
\caption{Molecular line observations summary.}              
\label{obs-table}     
\centering                                     
\begin{tabular}{l c c c c c c c}       
\hline\hline                        
Center Line &  Resolution & Coverage & Int. Time & $\tau_{ave}$ & T$_{sys}$ & F$_{eff}$/B$_{eff}$ & $rms_{ave}$ \\ 
 	    &   (km/s)     & (MHz)    & (min) &   &  (K)  &   & (mK)   \\ 
\hline                                  
$^{12}$CO(2-1) &   0.25 &  225060-232840  &  261 & 0.10 & 230 &  1.56  & 20-15-50  \\
$^{12}$CO(2-1) &  0.063  &  226398-228200, 229670-231497  & 105 & 0.18  & 230 &  1.56  & 50    \\
$^{13}$CO(2-1) &   0.20 & 214920-222700 & 209 & 0.10 & 170 &  1.53  & 10  \\
C$^{18}$O(2-1) $^a$ &  0.067  & 215420-217240, 218700-220520 & 160 & 0.16 & 220 &  1.52   & 50  \\
C$^{18}$O(2-1) &  0.20  &  214080-221860 & 313 & 0.09 & 220  & 1.52   & 10-20    \\
HCO$^+$(1-0) &  0.66  & 83710-91490  & 261 & 0.03 & 120 & 1.18   & 6-20   \\
HCO$^+$(1-0) &   0.16 &  85050-86870, 88330-90148 & 110 & 0.03 &  120 & 1.18   & 30    \\
N$_2$H$^+$(1-0) &  0.63  & 87680-97475  &  313 & 0.03 & 110 &  1.18  & 10  \\
N$_2$H$^+$(1-0) &  0.157 & 89033-90860, 92315-94133 & 157  & 0.03 & 110 & 1.18   & 10  \\
CS(2-1) $^b$ &  0.60  &  92500-100280  & 209 & 0.02 & 80 &  1.19  & 7  \\
\hline                                             
\end{tabular}
\tablefoot{Note that to maximize S/N in the lines of
the low-resolution spectra, we combined them with the high-resolution data.  The
rms listed is thus the rms of the combined spectrum.
$^a$ The setup also covers the $^{13}$CO line. $^b$ Only low resolution data. All observations were obtained on 2014 March 5-6. We include the
average sky opacity ($\tau_{ave}$) during the observations, the average system
temperature, the ratio between forward efficiency and beam efficiency \citep[F$_{eff}$/B$_{eff}$;][]{kramer13}, and the average rms for each configuration.}
\end{table*}

\begin{figure*}
\centering
\includegraphics[width=0.9\linewidth]{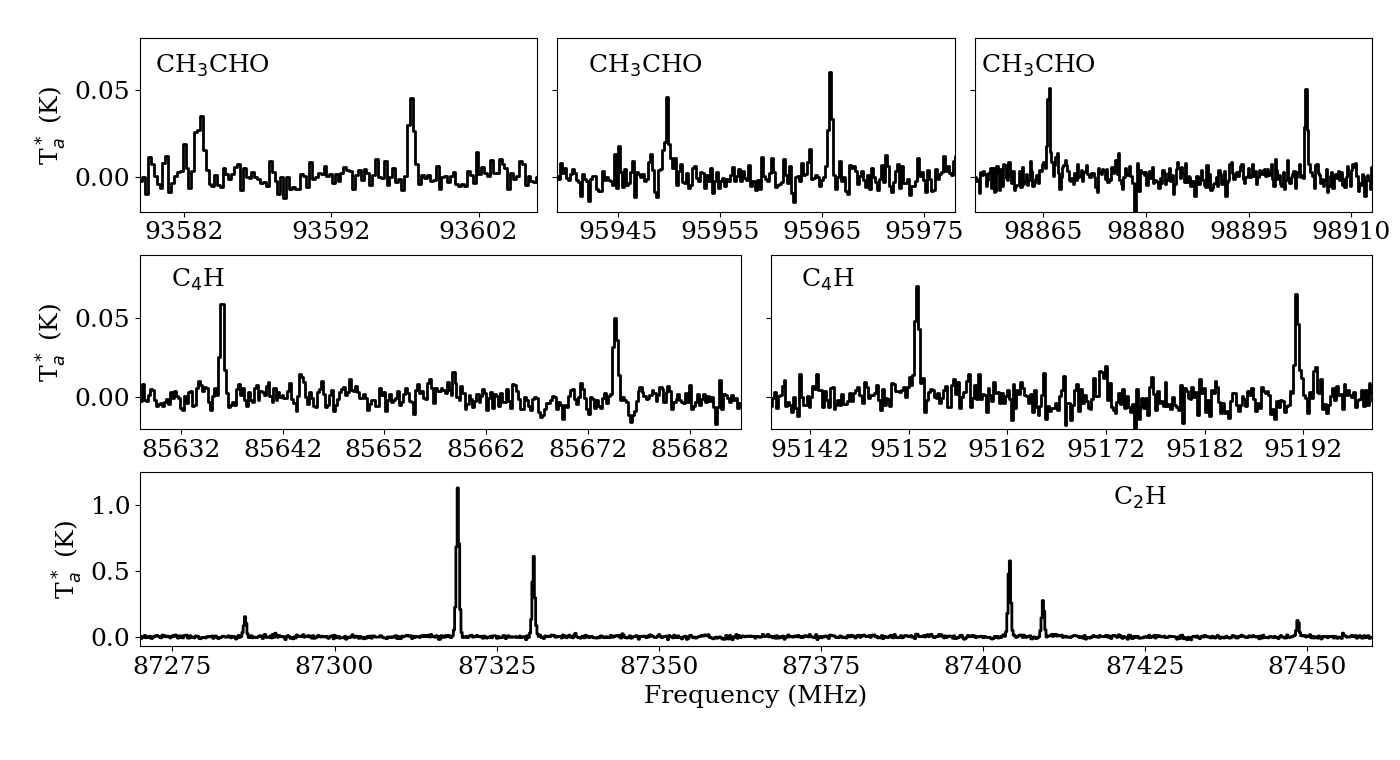}
\caption{Some of the complex molecules and carbon chain lines detected towards the region. Top: CH$_3$CHO. Middle: C$_4$H. Bottom: C$_2$H. 
\label{examplelines-fig}}
\end{figure*}

To obtain the molecular-line data, we observed with EMIR E0/E2 parallel mode
on 2014-03-05 and 2014-03-06.  The beam sizes for the two EMIR frequency ranges were $\sim$12" (for E2)
and $\sim$27" (for E0).
To optimize the observing times, we mapped only the region around the Class 0 source including
the bright rim behind the ionization front. For each setup, OTF maps with uniform coverage in an area of
$\sim$1.4'$\times$1.4' ( which results in a mapped area around 2.2'$\times$2.2' with lower
S/N towards the edges) were obtained. Each map consisted of a scan and perpendicular
cross-scan map to minimize the instrumental signature. The initial plan was to use the low
resolution backends for EMIR, which result in a velocity resolution of 0.25 km/s at 230 GHz
or 0.65 km/s at 90 GHz, together with a large frequency coverage (see Table \ref{obs-table} for
details). Since the source was brighter than expected, we switched to the high-resolution
backends after the first set of observations, obtaining velocity resolutions of
0.06 km/s at 230 GHz, and 0.16 km/s at 230 GHz. Therefore, we have both 
maps with a large frequency coverage, and detailed maps with high-velocity-resolution data on
selected lines.
The EMIR parallel mode was used to map three E2/E0 configurations,
including $^{12}$CO(2-1) and HCO$^+$(1-0),
C$^{18}$O(2-1) and N$_2$H$^+$(1-0), and $^{13}$CO(2-1) and CS(2-1). The last configuration
was only observed at low resolution, since the high resolution mode included both
the C$^{18}$O(2-1) and the $^{13}$CO(2-1) lines.  Planets were used as primary focus
calibrators, and a nearby bright source was used for
focusing before starting the maps. The atmospheric opacity for the EMIR observations
was determined using the chopper wheel calibration.
that were repeated regularly depending on weather conditions. As for the continuum maps,
OFF positions were taken in locations that are clean from nebular
emission. The detailed observing conditions
and exposure times are listed in Table \ref{obs-table}. Since there is partial overlap
between different instrumental configurations, 
all observations that cover the relevant frequency range were combined 
to maximize the signal-to-noise
ratio (S/N) in the line analysis. This means that, for the analysis of low-resolution
data, both low- and high-resolution data were combined, as well as the regions
between 214920-221860 MHz (covered by both the $^{13}$CO and C$^{18}$O
setups), 87680-91490 MHz (covered by the N$_2$H$^+$ and HCO$^+$ setups)
and 92500-97475 (included in the N$_2$H$^+$ and CS setups).
There was no overlap for the high-resolution data, which were reduced and
analyzed independently. 

Although most of the strong lines are
detected towards extended parts of the globule, some of the higher-density
tracers reveal very compact emission. Without having a constrain on the
emitting structure of the source, it is hard to estimate to which extent
these lines may be affected by beam dilution. In particular, DCO$^+$ and N$_2$D$^+$ are
only observed towards the Class 0 source. Due to their strenghts relative to
the non-deuterated species in comparison to what is observed towards other
protostars in similar environments \citep[e.g.][]{pety07}, we deduce that the emitting region should not be much different in size from the  $\sim$11" beam. In fact,
the 160$\mu$m observations reveal that the envelope size is probably
comparable or larger than the PACS 160$\mu$m PSF ($\sim$11"$\sim$12" for the medium scan speed of 20"/s\footnote{https://www.cosmos.esa.int/documents/12133/996891/\\PACS+Observers\%27+Manual}, which is very similar to the IRAM beam at 230 GHz). Further higher resolution
observations will be needed to determine the source
structure.

The molecular-line data were reduced using the GILDAS/Class software 
(Bardeau et al. 2006\footnote{http://www.iram.fr/IRAMFR/GILDAS}).
The data were calibrated following standard IRAM procedures using the $MIRA$\footnote{https://www.iram.fr/IRAMFR/GILDAS/doc/html/mira-html/mira.html} package at
the telescope to account for atmospheric corrections. The beam efficiencies are the standard values for the telescope \footnote{https://www.iram.fr/GENERAL/calls/w08/w08/node20.html}
and have been estimated with observations of the Moon and planets \citep{kramer13}. We reduced the data 
extracting each strong line individually. 
A constant baseline, measured over a small frequency range around each line,
was subtracted locally. For the final maps, the data were
regridded using a pixel spacing 8" and a convolving kernel 11.9" (for the 
E2 data) or a pixel spacing 16" and a convolving kernel 19.6" (for the E0 data). 
Maps with resolution 16" (for E2) or 32" (for E0) where then constructed
using Class $xy\_map$ task. The regridded maps are the starting point for the momenta and bitmap
analysis to explore the spatial origin of the emission.  Due to the large size of the beam,
some leakage across the map occurs for several lines, but these regions are excluded from 
the analysis (see Appendix \ref{velo-app} for details).
The resulting maps show strong line emission towards the globule and, in particular, the Class 0 source,
with a systemic velocity around $-$7.8 km/s, in agreement with \citep{patel95}.

All the lines identified in the region with the low-resolution data
are listed in Appendix \ref{app1}, Table \ref{lowreso-table}. The presence of
multiple long-carbon chains (e.g. c-C$_3$H$_2$, C$_3$H, C$_4$H) other molecular species
typical of protostars at the edge of a HII region (e.g. HCOOH, CH$_3$OH, CH$_3$CHO)
suggest that the source contains a hot corino with warm carbon chain chemistry \citep[WCCC;][]{watanabe12} 
illuminated by UV radiation.  Further high-resolution observations will be needed to 
confirm this hypothesis. With the low resolution and relatively low sensitivity for faint lines, it is hard to distinguish the emission from the ionization front and the source without further interferometric data. We leave the complex chemistry analysis for a subsequent paper,
concentrating here on the gas dynamics and the structure of the region.
The lines that are strong enough for a
detailed high-velocity resolution analysis are listed in Table \ref{hireso-table}.
We also include in this table the CS(2-1) line since it conveys important information
about the photodissociation region, even though it was only observed in the low-velocity resolution mode.
Figure \ref{examplelines-fig} shows some of the complex molecular lines detected.

\begin{table}
\begin{footnotesize}
\caption{Strong lines detected in the high-resolution spectra. }              
\label{hireso-table}     
\centering                                     
\begin{tabular}{l c l c  }       
\hline\hline                        
Frequency  & Species	& Transition 	& Int. Intensity \\
 (MHz)     &		& (Quantum Nr.)			&  (K [T$_a^*$] km/s)	  \\
\hline                                  
85338.89	& c-C$_3$H$_2$	& 2(1,2)-1(0,1)      & 5.28$\pm$0.01  	 \\
86054.96	& HC$^{15}$N	& 1-0			& 2.39$\pm$0.01  	 \\
86093.95	& SO		& 2(2)-1(1)		&  4.36$\pm$0.01  	 \\
86340.18$^m$	& H$^{13}$CN	& 1(2)-0(1)		&  1.87$\pm$0.01 	 \\
86670.76	& HCO		& 1(0,1,2,2)-0(0,0,1,1) & 2.88$\pm$0.01  	 \\
86677.46	& HCO		& 1(0,1,1,1)-0(0,0,1,1) & 1.37$\pm$0.01  	\\
86708.36	& HCO		& 1(0,1,2,1)-0(0,0,1,1) & 1.29$\pm$0.01  	\\
88633.94$^m$	& HCN		& 1(0)-0(1)		& 10.73$\pm$0.06  	\\
89487.41	& HCO$^+$	& 1-0			& 7.02$\pm$0.04  	\\
90663.59	& HNC		& 1-0			& 12.15$\pm$0.02  	\\	
92494.31	& $^{13}$CS	& 2-1 		&  2.29$\pm$0.01 	\\
93173.70	& N$_2$H$^+$	& 1-0			& 1.93$\pm$0.01  	\\
93870.11	& CCS		& 7(8)-6(7)		&   0.661$\pm$0.004 	\\
96412.94	& C$^{34}$S	& 2-1 		& 3.63$\pm$0.01  	\\
97171.84	& C$^{33}$S	& 2-1 		& 0.759$\pm$0.002  	\\
97980.95	& CS$^*$	& 2-1 		&   24.32$\pm$0.01 	\\
218222.19 	& H$_2$CO	& 3(0,3)-2(0,2)     & 0.80$\pm$0.01  	\\
219560.36	& C$^{18}$O	& 2-1 		&  6.4$\pm$0.1  	\\
220398.68	& $^{13}$CO	& 2-1 		& 26.8$\pm$0.4  	\\
226875.90	& CN		& v=0,1; 2(0,3,2)-1(0,2,1)	& 8.54$\pm$0.03  	\\
230538.00	& CO		& 2-1			&  117$\pm$1  	\\
216112.58	& DCO$^+$	& 3-2			&  2.31$\pm$0.02 	\\
231321.67$^m$	& N$_2$D$^+$	& 3-2			& 2.78$\pm$0.01  	\\
215220.65	& SO		& 5(5)-4(4)		& 2.06$\pm$0.03  	\\
\hline                                             
\end{tabular}
\end{footnotesize}
\tablefoot{Only the lines with enough S/N
to analize their detailed velocity are listed here. The listed intensities correspond to the
average, velocity-integrated T$_a^*$ intensity over the whole region. 
Multiplets are labeled with $^m$
and only the central line is listed, while the integrated intensity of multiplets
includes all hyperfine structure components. $^*$The CS line was only observed in the low-resolution
mode, but given its imporance and that it is included in the spatial and velocity
analysis, we also list it in the table here. All the wavelengths and quantum numbers are taken from the
JPL database \citep{picket98}.}
\end{table}

\subsection{Ancillary data \label{ancillary}}

A wealth of existing optical and IR 
data are used to obtain a complete (multi-phase, dust and gas, temperature range from stellar photospheres to tens of K)
view of the globule and its embedded population. 
Our \emph{Herschel}/PACS data at 70 and 160~$\mu$m\footnote{Open Time proposal "Disk dispersal in Cep OB2", OT1\_asicilia\_1, PI A. Sicilia-Aguilar, AORs 
1342259791 and 1342259792.}
 \citep[for further details regarding observations and data
reduction, see][Paper I]{sicilia15} 
are particularly useful for the characterization of the dust content in the globule and of IC1396A-PACS-1. 
Our [S\,II] narrow-band imaging, obtained with CAFOS on the 2.2m telescope in Calar Alto 
\citep{sicilia13} allows us to characterize the edge of the photon-dominated region (PDR) 
and the impact of the embedded
population in the cloud.
Finally, \emph{Spitzer} IRAC and MIPS data \citep{reach04,sicilia06a,morales09}, together with optical high-resolution spectroscopy
\citep{sicilia06b} complete the characterization of the low-mass cluster members in the region and the radial-velocity picture obtained from the molecular lines, respectively.

\begin{figure}
\includegraphics[width=0.98\linewidth]{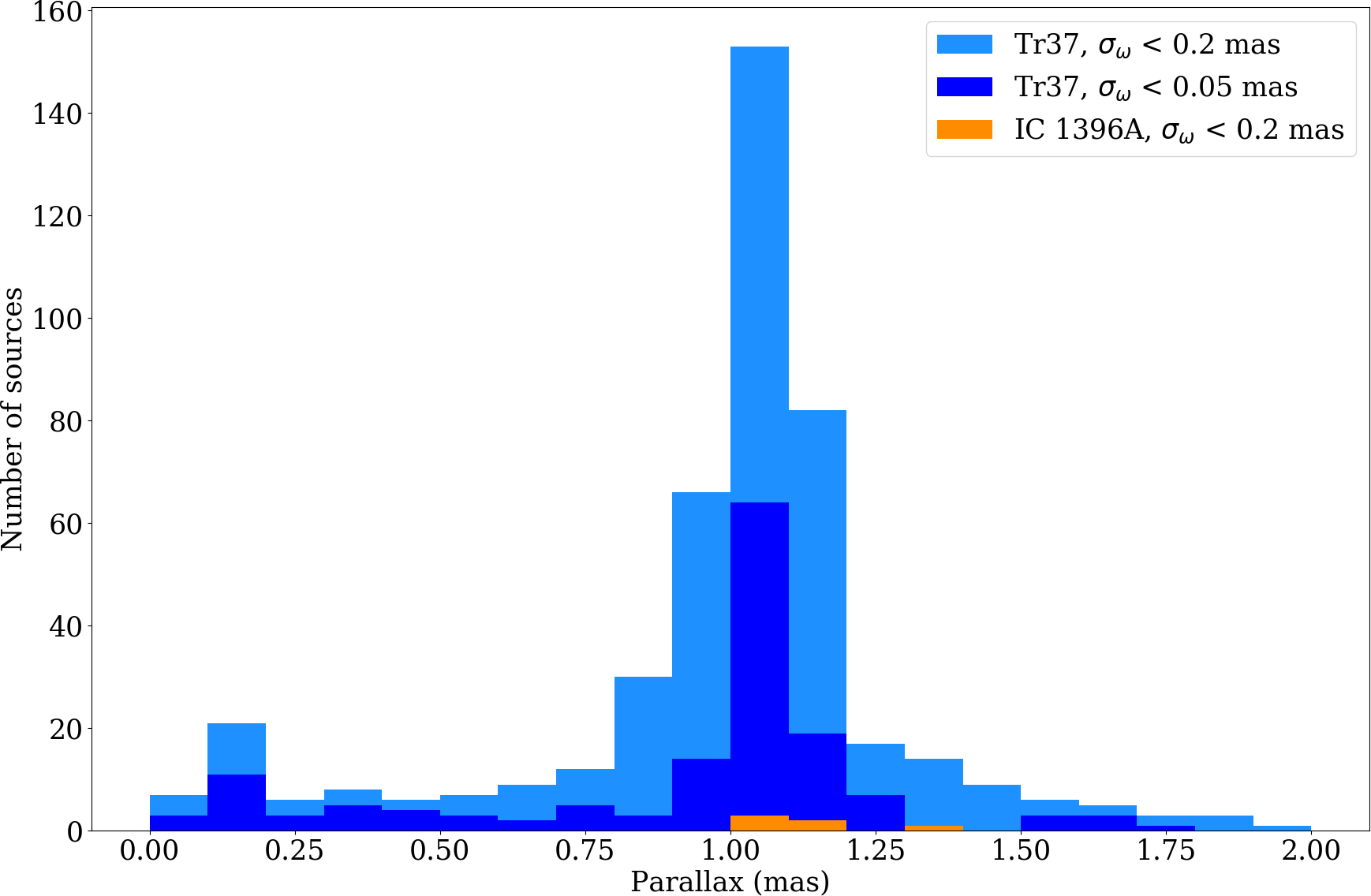}
\caption{Histogram of the parallaxes of the known Tr 37 cluster members with good Gaia 
data (see text). The stars associated with IC\,1396A are marked in yellow. \label{gaia1-fig}}
\end{figure}

In addition, we use Gaia DR2 data \citep{gaiamission16,gaiadr218} available
through Vizier \citep{gaiavizier18} to explore the velocities of the stars associated with IC\,1396A
and the Tr\,37 cluster in connection with the molecular gas observations.  Gaia has been
successfully used to identify cluster structure in other young clusters \citep[e.g.][]{roccatagliata18,franciosini18}, and
can help to obtain a 3-dimensional picture of the region.
We compiled the list of cluster members in Tr\,37 and the IC\,1396A 
region based on spectroscopically-identified members \citep{contreras02,sicilia05,sicilia06b,sicilia13},
Spitzer identifications \citep{reach04,sicilia06a,morales09}, H$\alpha$ search for young stars \citep{barentsen11}, and X-ray surveys \citep{mercer09,getman12}. This produced a list of over 800 members
detected with Gaia, among which 354 had low errors (matching radius $<$0.5 arcsecs,
relative parallax error $\sigma_\omega$/$\omega <$0.1, proper motion errors below 2 mas/yr). 
Among these, 6 sources are associated with IC\,1396A, including
V390 Cep, which is known to be physically associated with the globule thanks to the signs of interaction 
within the eye-shaped hole.  A histogram with the Gaia parallaxes for Tr~37 and IC\,1396A is
shown in Figure \ref{gaia1-fig}.

For a cluster at a relatively large distance and composed of mostly low-mass, faint stars, the errors from Gaia DR2 are often non-negligible, which results in biased distances if the parallaxes are simply 
inverted. Because of this, we follow the Bayesian inference methods of \citet{bailerjones15,astraatmadja16a,astraatmadja16b} to estimate distances and their asymmetric errors.
The distance to the cluster members is obtained assuming an exponentially decreasing density prior,
which is the preferred one for DR2 \citep{bailerjones18}, with a characteristic length l=1.35 kpc
\citep{astraatmadja16a,astraatmadja16b}. 
Following \citet{bailerjones15}, the 
prior for an exponentially decreasing density can be written as a function of the distance r and the characteristic length l
\begin{equation}
P^*_{r^2 e^{-r}}=\begin{cases} 
	\frac{1}{2l^3}r^2 e^{-r/l} &\text{if $r>0$},\\
	0 &\text{otherwise}.
\end{cases}
\end{equation}
For this, the unnormalized posterior is a function of the parallax $\omega$ and the parallax error $\sigma_\omega$
\begin{equation}
P^*_{r^2 e^{-r}}(r | \omega, \sigma_\omega)=\begin{cases}
	\frac{r^2 e^{-r/l}}{\sigma_\omega} exp[-(\omega-1/r)^2/2\sigma^2_\omega] &\text{if $r>0$},\\
	0 & \text{otherwise}.
\end{cases}
\end{equation}
The best estimate of the distance is calculated as the mode of the unnormalized posterior, which can be
obtained from equating to zero the derivative of the posterior \citep{bailerjones15}.
Using the 354 stars with good Gaia data, we find that the average distance to the cluster
is found to be 945$^{+90}_{-73}$pc, where the errorbars mark the 5-95\% confidence
intervals. This is consistent with the previous value of 870pc \citep{contreras02},
especially as we take into account that Gaia DR2 seems slightly biased towards larger distances
\citep{stassun18}, which will be corrected in future data releases. The stars associated with IC\,1396A
are consistent with the cluster distance, as expected from the evident physical relation between the globule and HD\,206267 (see Figure \ref{gaia1-fig} left).

\section{Data analysis \label{analysis-sect}}

\subsection{Dust temperature and  column density maps \label{dust-TNH}}

Although optical and \emph{Spitzer} images suggest a relatively uniform globule with a
hole around the position of V 390 Cep, \emph{Herschel} unveiled
denser, colder material behind the ionization front (Paper I).
The NIKA data confirms the \emph{Herschel} results, showing a sharp rise in intensity
behind the ionization rim, with an intensity varying by over a factor of 40 at both 1.3 and 2~mm between
the maximum point at the Class 0 source and the lower density structures to the west
(see Figure \ref{nika-fig}).

\begin{figure}
\includegraphics[width=0.95\linewidth]{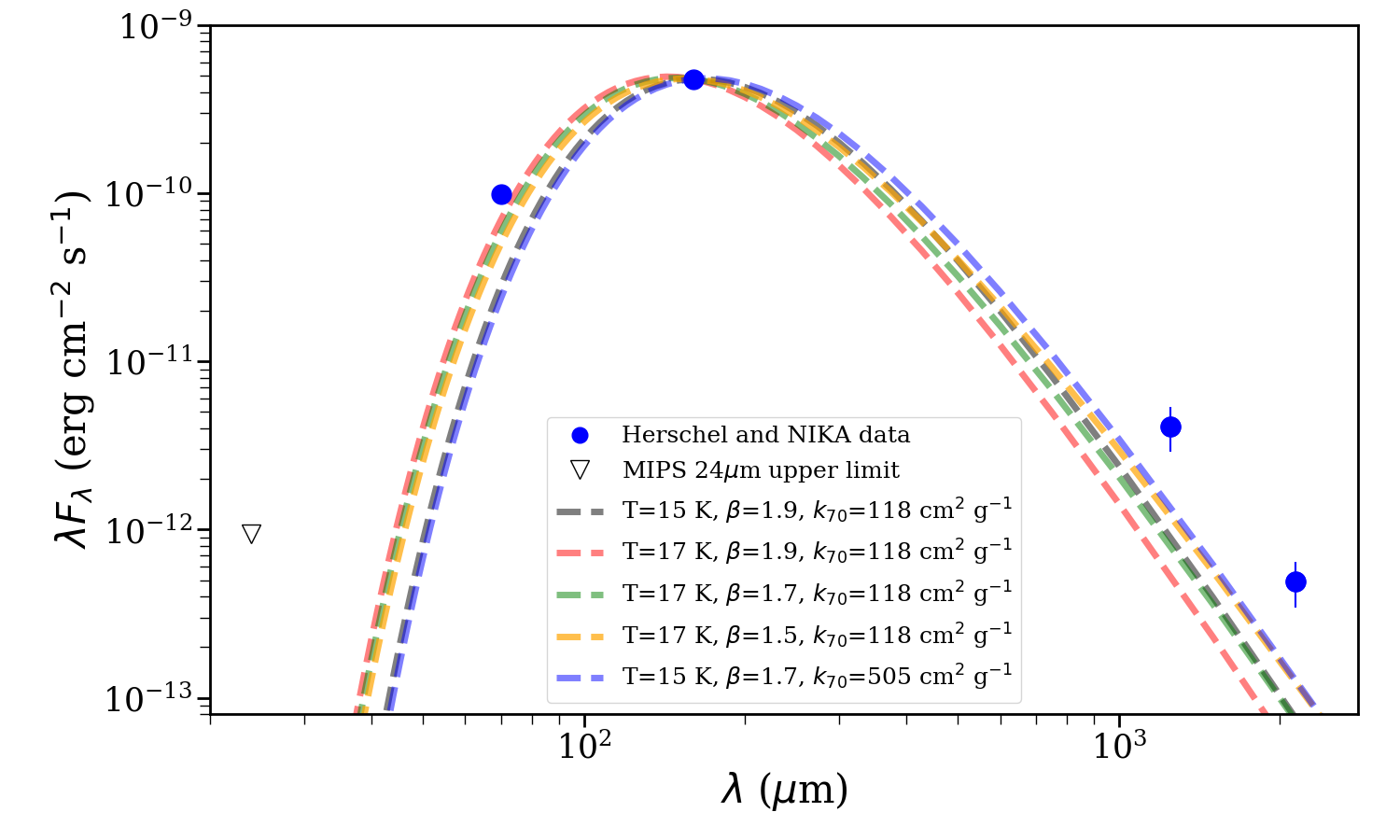}
\caption{SED of the Class 0 source including the \emph{Herschel} and NIKA detections and the MIPS 24$\mu$m upper limit (non-detection) compared to the emission of several modified black bodies with temperatures 15 and 17K, $\beta$=1.9, 1.7, and 1.5, and $k_{70}$=118 and 505 cm$^{2}$g$^{-1}$ (see discussion in text). Error bars are smaller than the dots for the \emph{Herschel} data. \label{SED-fig}}
\end{figure}

The temperature/column density
map from \emph{Herschel} data alone \citep{sicilia15} is highly uncertain due to the use of only two 
wavelengths. Here, we revise the spectral energy distribution (SED) of the Class 0 object, together with the dust temperature
and column density maps around IC1396A-PACS-1 using the NIKA data in combination with
Herschel/PACS. For the object SED, we combine the \emph{Herschel} data  (see Paper I) 
with the integrated
NIKA flux at the position of the Class 0 source. We define the source limits based on the
background emission around the extended structure, which is 0.09 and
0.04 Jy/beam at 1.3 and 2mm, respectively. The errors in the NIKA fluxes depend not only
on the calibration and filtering uncertainties (see Setion \ref{observations-iram}) but also
on the uncertainties defining the source limits in a region with highly variable background,
for which we adopt a conservative estimate of $\sim$30\%. The SED is displayed in Figure \ref{SED-fig}.
If we assume that the dust is optically
thin and that the emission is dominated by a single temperature, the flux emission at a given frequency $\nu$ can be written as
\begin{equation}
	F_{\nu} = \Omega B_{\nu}(T) \tau_{\nu} =  \Omega B_{\nu}(T) k_{\nu} \Sigma. \label{eqfnu}
\end{equation}
Here, k$_\nu$ is the mass absorption coefficient, $\Omega$ is the solid angle subtended by 
the emitting region, $\Sigma$ is the mass column density, 
and B$_\nu$(T) is the black-body emission for a temperature T at the
frequency $\nu$. The frequency dependence can be further simplified assuming that the dust mass absorption coefficient k$_\nu$ varies as a power-law with 
frequency, with values that are typically around 2 \citep[e.g.][]{schneider10,juvela12b,preibisch12, roccatagliata13}. 
Such a power law also offers a good fit to more detailed
dust models in the far-IR and millimeter range \citep{ossenkopf94}.

Although the large NIKA beam includes part of the extended structures 
detected with \emph{Herschel} 70$\mu$m data, and thus the interpretation of the emission needs to be
regarded with care, the NIKA data confirm that the source is dominated by grey-body-like
emission as  it had been suggested by \emph{Herschel}. The best-fitting temperatures are in the
range of 15-17 K and thus do not
significantly change with respect to our previous results based on the two \emph{Herschel} data points (Paper I). We now take advantage
of the NIKA data to derive further constraints on the dust model.  The zero-point of this power law fit
depends on the dust properties, including grain sizes, composition, and the presence of icy mantles \citep{ossenkopf94}, all of which are likely to vary throughout a molecular 
cloud, especially in the surroundings of young objects.

Figure \ref{SED-fig} shows the effect
of modifying the dust mass absorption coefficients at 70$\mu$m and the power law exponent of the frequency
dependence of the dust mass absorption coefficient, $\beta$. Although the typical gas densities in the
cloud are expected to be low (see Section \ref{Tr37-past}) compared to those required for substantial 
dust coagulation \citep{ossenkopf94}, the densities are likely much higher in the source, and dust coagulation and
the presence of thick ice mantles are a possibility. The temperature is relatively well-constrained independly of the dust model used,
even though the data suggest a range of temperatures between $\sim$15-17 K in the source. 
A larger dust mass absorption coefficient would result in lower column densities
and lower source masses, even though the value for grains with thin ice mantles of $k_{70}$=118 cm$^2$g$^{-1}$ 
\citep[derived from model 1.b in Table 1 in][]{ossenkopf94}
provides a very good fit to the data. The choice of a larger $k_{70}$=505cm$^2$g$^{-1}$ for a 
model with thick ice mantles \citep[derived from model 1.c in Table 1 in][]{ossenkopf94} does not appear to be
justified by the data. A lower value of $\beta$ down to 1.7-1.5, as would be expected from grain growth,
offers a better fit, even though it is very hard to distinguish a lower $\beta$ from
the effect of a slight variation of temperature along the line-of-sight of a couple of degrees \citep[as it has been noted in other regions, e.g.][]{juvela12a}.

The best-fitting mass column density $\Sigma$ from Equation \ref{eqfnu} can be used to derive a
mass for the envelope of the Class 0 object. If we assume a gas to dust ratio (R$_{gas/dust}$=100) and take
into account the mass of the hydrogen atom (m$_{\rm H}$) and the mean molecular weight ($\mu$ = 2.8),
the hydrogen number column density  (N$_{\rm H}$) can be estimated as
\begin{equation}
	N_{\rm H}= \frac{2 \Sigma R_{gas/dust}}{m_{\rm H} \mu}, \label{eqnh}
\end{equation}
which can be integrated over the size of the object to derive a total mass. The limits of
the source are uncertain, with the object appearing as a compact source at 70$\mu$m and
envelope being resolved at 160$\mu$m (Paper I). Assuming a size similar to the NIKA 1mm beam, we obtain a mass
in the range of 7-18 M$_\odot$ for typical models with $k_{70}$=118 cm$^2$g$^{-1}$ and $\beta$=1.5-1.9. Given that
the lower masses result from the fits that better adjust to the \emph{Herschel} and that the source
is likely optically thick at 70$\mu$m, higher masses are likely more representative of the Class 0 envelope mass.
The NIKA 2mm datum probably contains part of the cloud emission, since the source envelope is resolved at 160$\mu$m and appears
smaller than the NIKA 2mm beam. A 
larger dust mass absorption coefficient would result in a lower source mass, and extending the source 
limits to the NIKA 2mm
beam would result in a higher mass by a factor of 2.

To derive the temperature and column density maps, we extend the assumption of single dust temperature and
optically thin material to the whole cloud \citep[see][for further details on this approximation]{roccatagliata13}. The
first assumption breaks down if the cloud has a temperature structure along the line-of-sight, which is usually expected. 
In case of regions with
different temperatures, the emission will be dominated by the highest temperature on the line-of-sight.
If emission from a hot point-source (such as a star or protostar) is significant, then the temperature will 
be also biased towards higher values. In our case, the only protostars with significant emission at 70$\mu$m (compared to
the background) in the region are IC1396A-PACS-1 itself and 21364660+5729384 \citep[also called $\alpha$ and located out of our EMIR field;][]{reach04,sicilia14}. 
The assumption of optically thin material may break down  for the 70~$\mu$m emission in 
the very dense parts of the cloud, in particular, around IC1396A-PACS-1.

\begin{figure*}
\centering
\begin{tabular}{c}
\includegraphics[width=1.0\linewidth]{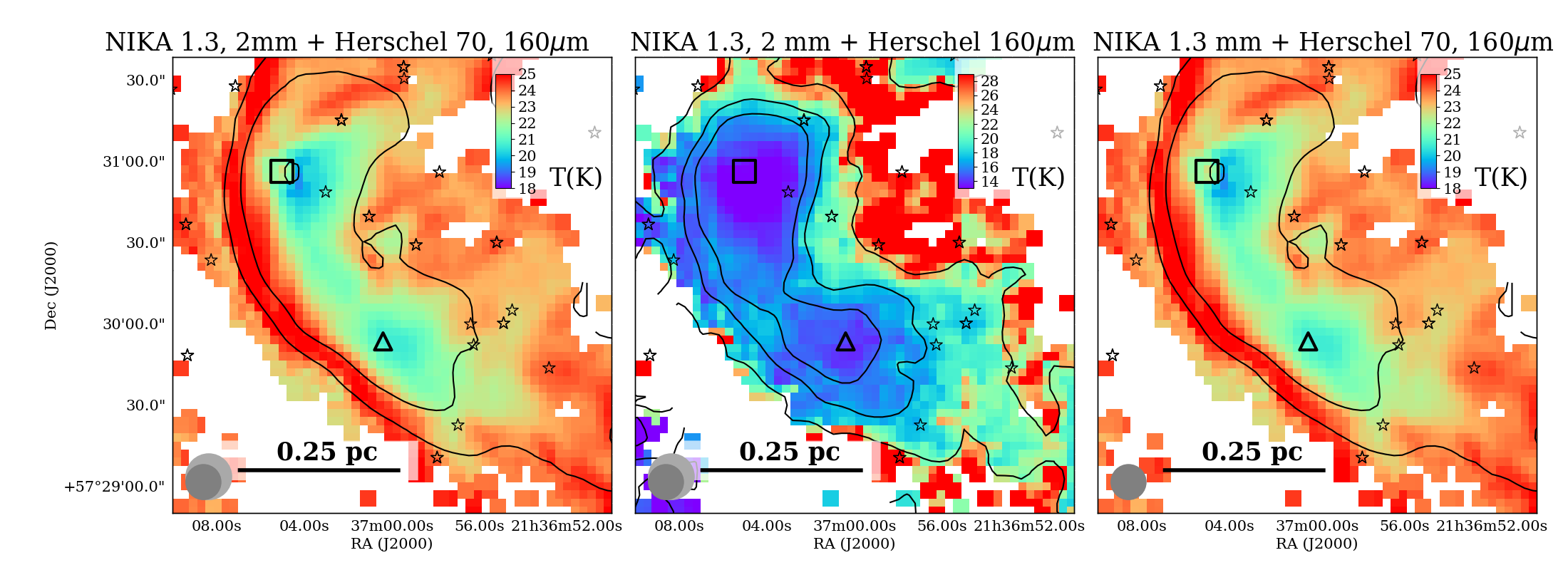}\\
\includegraphics[width=1.0\linewidth]{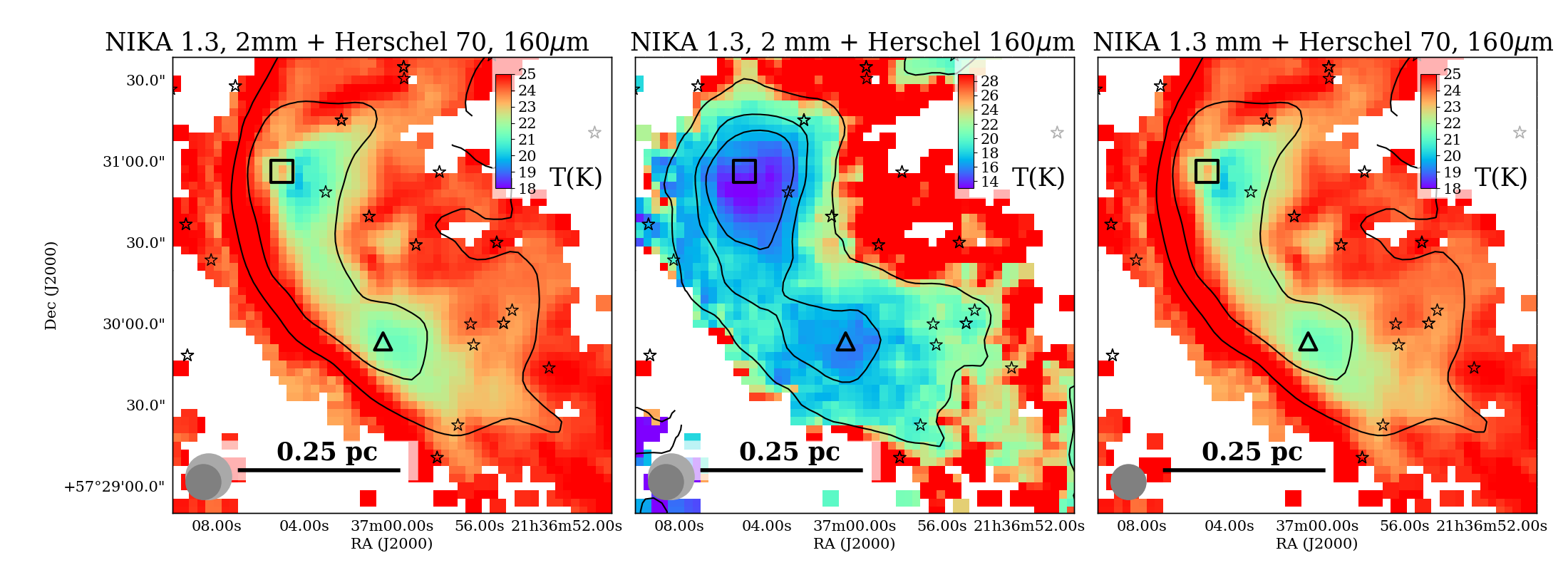}\\
\end{tabular}
\caption{Temperature (color scale) and N$_H$ (contours) in the IC1396A region, derived from all four datapoints (\emph{Herschel} 70 and 160~$\mu$m and the NIKA
data; left),
the \emph{Herschel} 160~$\mu$m and the NIKA data (middle), and the \emph{Herschel} 70 and 160~$\mu$m and the NIKA 1.3mm data (right), using a dust model with $k_{70}$=118 cm$^2$g$^{-1}$, $\beta$=1.9 (top),
and $k_{70}$=118 cm$^2$g$^{-1}$, $\beta$=1.7 (bottom). The 70$\mu$m position of the Class 0 source is marked with a black square,  and NIKA S is marked with a triangle. The beams of the two NIKA bands are shown, together with a size 
indicator. The temperature scales are adapted in each case to show the full range of
temperatures with as much detail as possible. Note that there are some biases towards smaller or larger temperatures depending on the wavelengths used, as described in the text; relative
values are more accurate. The long-wavelength fit is better at determining the global cloud temperature and
structure, while the result from the inclusion of the 70~$\mu$m temperature is biased toward point-like stellar 
contributions  especially near intermediate-mass protostars
such as the objects north of V 390 Cep and the protostar $\alpha$). The temperature and column density are derived only in the pixels with emission
larger than 3$\sigma$ over the background at all wavelengths. The column density contours mark the 5e21,1e22,5e22,1e23 cm$^{-2}$ levels, noting that the NIKA + \emph{Herschel} 160$\mu$m maps start at 1e22 only as they are noise-dominated below this threshold.
 \label{T_NH-fig}}
\end{figure*}

The dust column density and temperature structure 
can be derived by fitting, point-by-point, the multiwavelength continuum data using Equation \ref{eqfnu}
to obtain a mass column density, $\Sigma$, and temperature, and Equation \ref{eqnh} to derive the hydrogen
number column density. For this exercise, we take $\beta$=1.9 \citep[the typical choice for star-forming clouds,][] {roccatagliata13,sicilia15},
and for comparison, $\beta$=1.7.
We regridded the NIKA and \emph{Herschel}/PACS maps to the same pixel size (3"$\times$3", corresponding
to the sampling of our 160~$\mu$m PACS maps) and fitted
Eq. \ref{eqfnu} on a pixel-by-pixel basis to obtain the local temperature and column density.
Note that the spatial resolution of the \emph{Herschel} data is significantly higher than that of
the IRAM data, so spatial structures at scales smaller than the IRAM beam  are not significant. The pixel-to-pixel variations provide information on the uncertainties of the procedure.
For each choice of $\beta$, we constructed three separate maps: one including all four \emph{Herschel} and NIKA bands, a second one including
only the  160 $\mu$m PACS band and the two NIKA channels, and a third one including both \emph{Herschel}/PACS
bands plus NIKA 1.3mm channel. The first has the main limitation that the 70$\mu$m band has substantial emission from the Class 0 point source itself, which breaks down the assumption of
optically thin material on the source location. The second one is better
to characterize the column density and temperature in the densest parts of the cloud (around IC1396A-PACS-1), while the third map offers a better view of the extended, less dense cloud (see Figure \ref{T_NH-fig}).

A further limitation of the maps is that, due to the filtering applied to NIKA data, we are underestimating the emission from the low-density parts of the cloud. This results in lower emission coming from regions with extended emission, which would bias the results towards higher temperatures (and lower densities) along the line-of-sight. To estimate what these loses mean in terms of mass and column density, we
can use the differences observed between the three above listed maps together with the limits in column density detected in our maps, which can be compared to the column density limits observed with \emph{Herschel} data only in the region \citep[Paper I,][]{sicilia15}.  The lowest background column density detected is of the order of 1$\times 10^{21}$  cm$^{-2}$  for maps including \emph{Herschel} data (which is similar to what one would expect for extinction along the line of sight for a region at $\sim$900 pc). Integrating this over the area of the globule, we obtain $\sim$10 M$_\odot$ of low-density material that could be missing in the whole globule. In addition, the globule tail has no significant
NIKA emission, and is therefore not included in our mass estimate. Further discussion on this aspect is
presented in Section \ref{rotd} regarding gas-inferred masses.

Despite these uncertainties, relative values along the line-of-sight are significant. The temperature and column density maps confirm that the Class 0 source is located in the
coldest and densest part of the globule. The NIKA data trace the coldest and densest material 
more accurately than our previous PACS-only maps (Paper I). In particular, the lack of (proto-)stellar emission
and emission associated with the rim behind the photoionization front in the NIKA
data compared to PACS allows to get a more accurate picture of the cold envelope around IC1396A-PACS-1. 
NIKA also confirms the presence of a low temperature but lower density region to the south of
PACS-1 (from now on, NIKA S, see Figure \ref{T_NH-fig}), where we do not detect any evidence of ongoing star formation.

The peak of the hydrogen column density at the Class 0 source position
is in the range 0.5-10$\times$10$^{23}$ cm$^{-2}$, for a temperature 14-18 K (depending on the wavelengths used to derive N$_{\rm H}$
\footnote{Due to the temperature-column density relation, fits based on shorter wavelengths tend to
have lower column densities and higher temperatures,  which thus leads to lower mass estimates.}). Even in the relatively dense NIKA S clump, the densities drop by a factor of 2-6, with temperatures about 2-3 degrees higher
than around IC1396A-PACS-1. The differences in density are more marked in the long-wavelength-based maps. Beyond these dense areas, the densities drop to 1-9$\times$10$^{21}$ cm$^{-2}$
in the parts of the globule that still have some millimeter emission, and below 3$\times$10$^{20}$ cm$^{-2}$ in the parts with \emph{Herschel}
emission only \citep{sicilia15}.

\begin{figure*}
\centering
\begin{tabular}{ccc}
\includegraphics[width=0.31\linewidth]{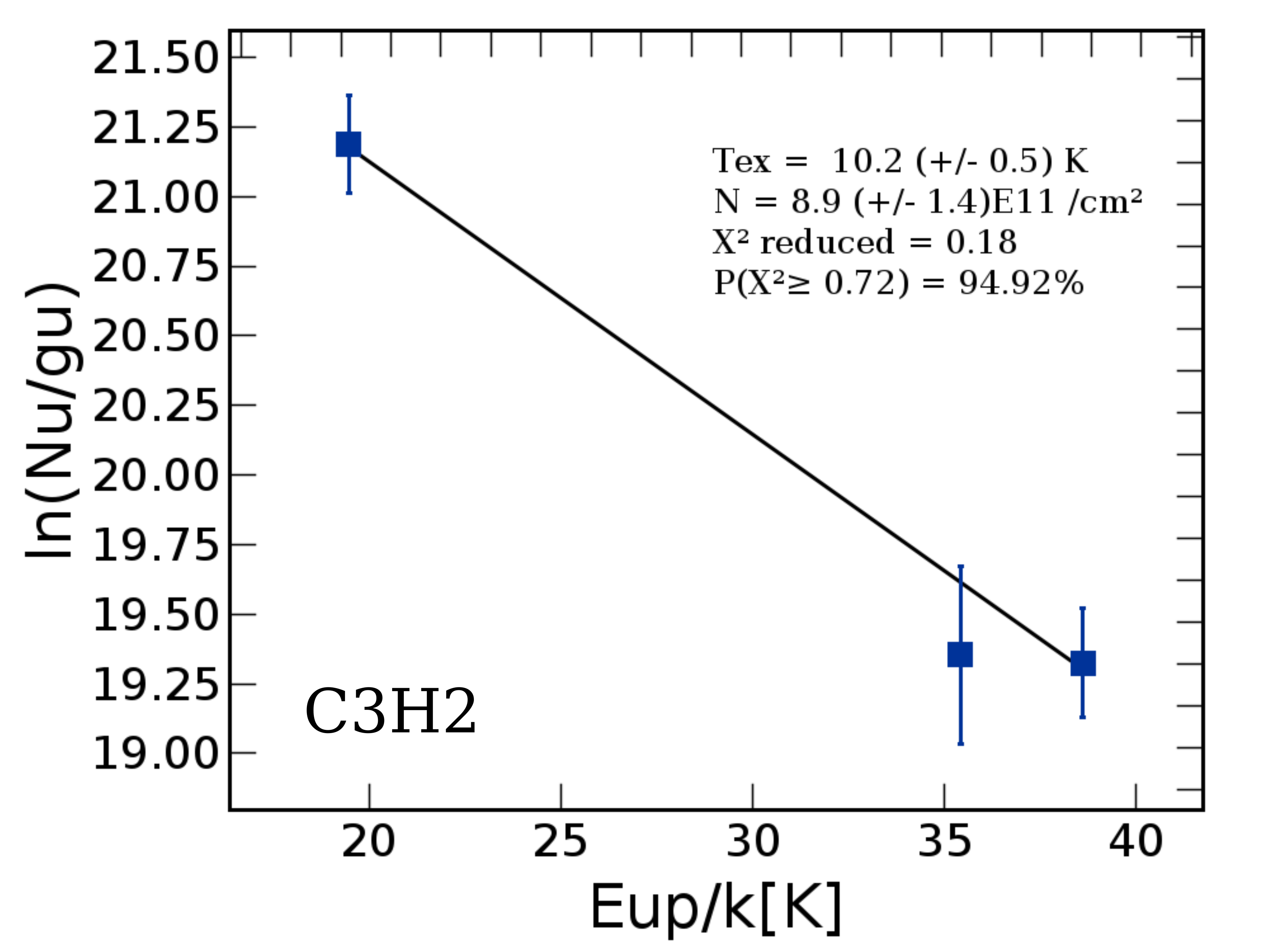} &
\includegraphics[width=0.31\linewidth]{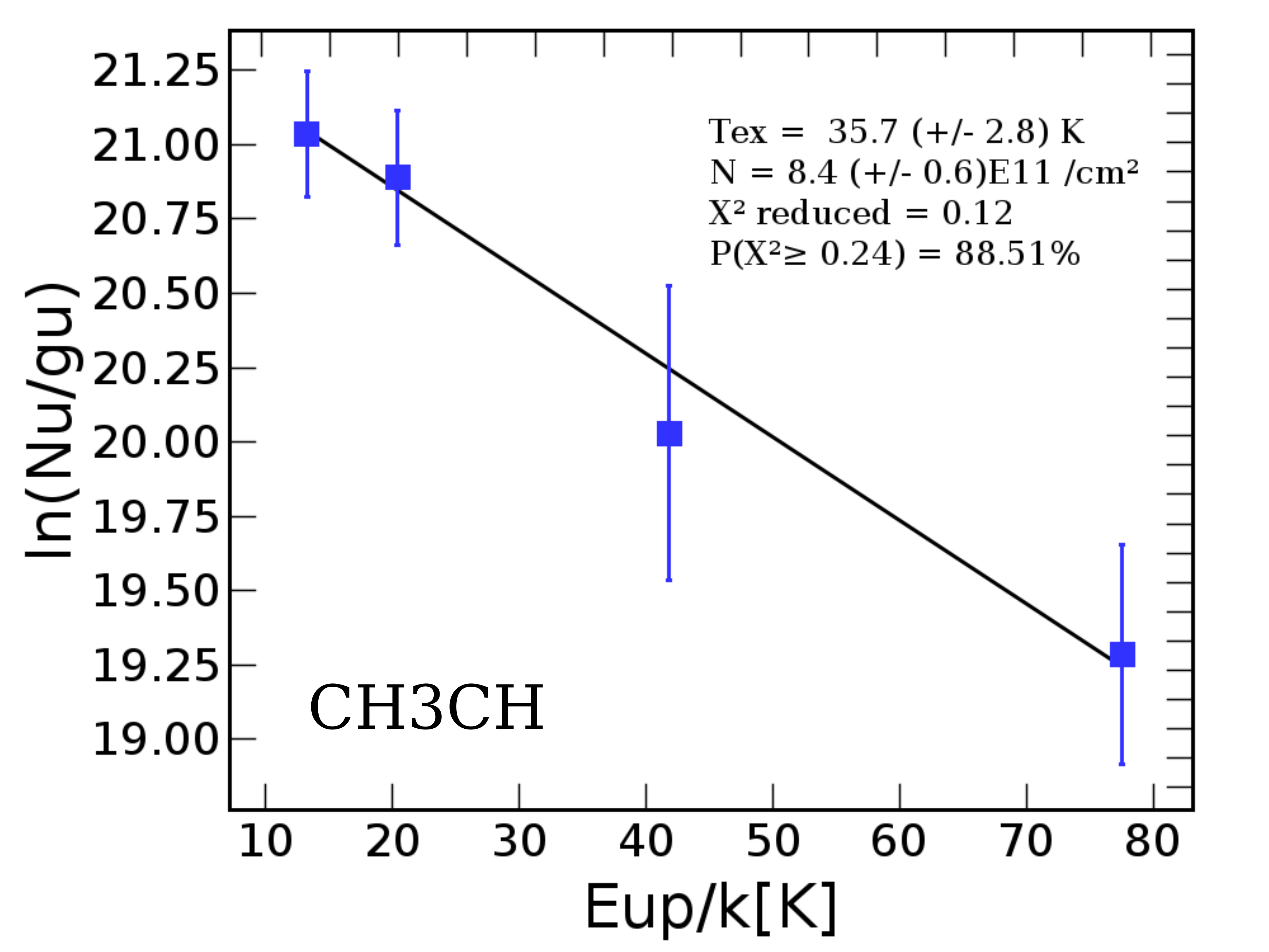} &
\includegraphics[width=0.31\linewidth]{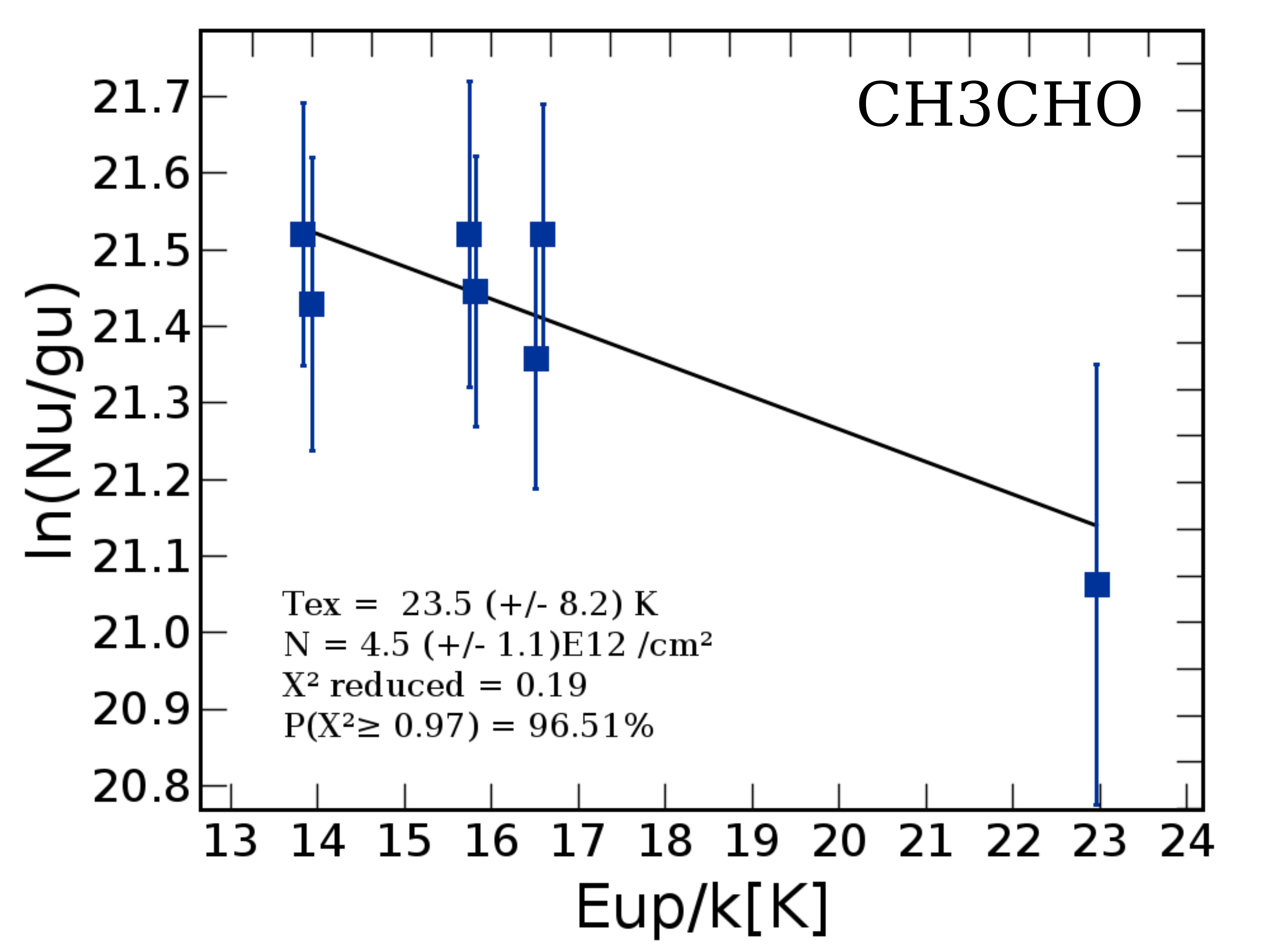} \\
\includegraphics[width=0.31\linewidth]{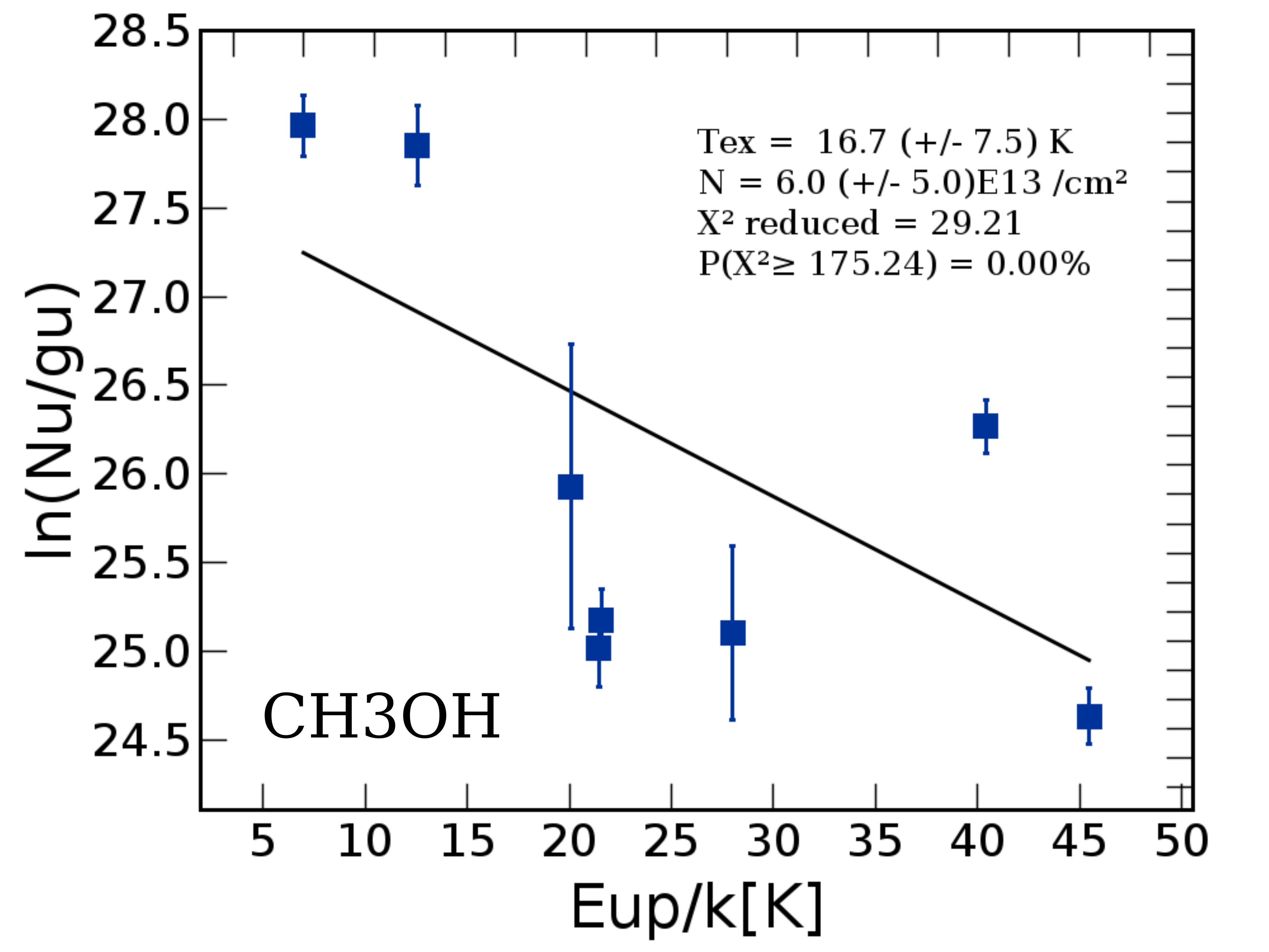} &
\includegraphics[width=0.31\linewidth]{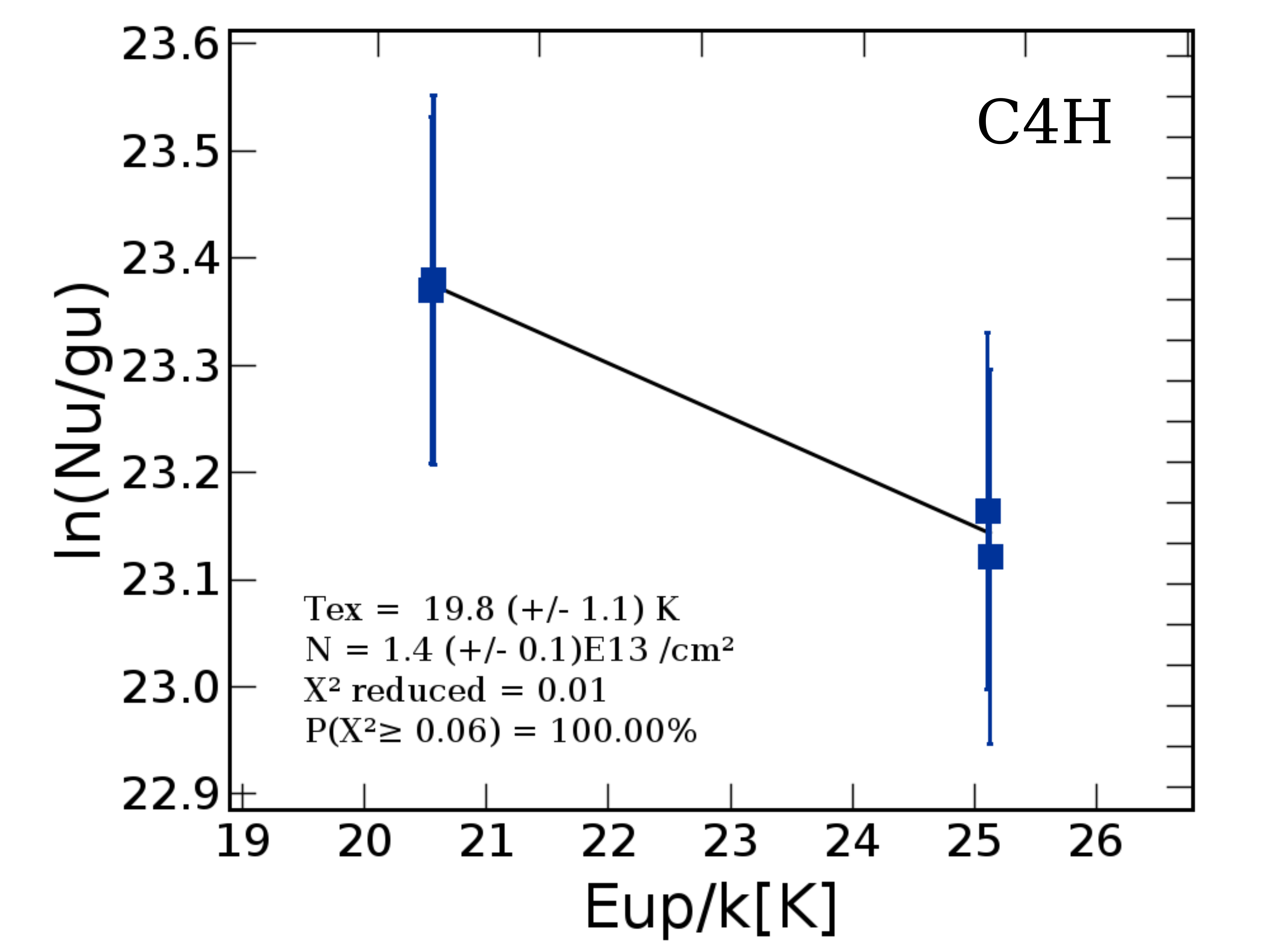} &
\includegraphics[width=0.31\linewidth]{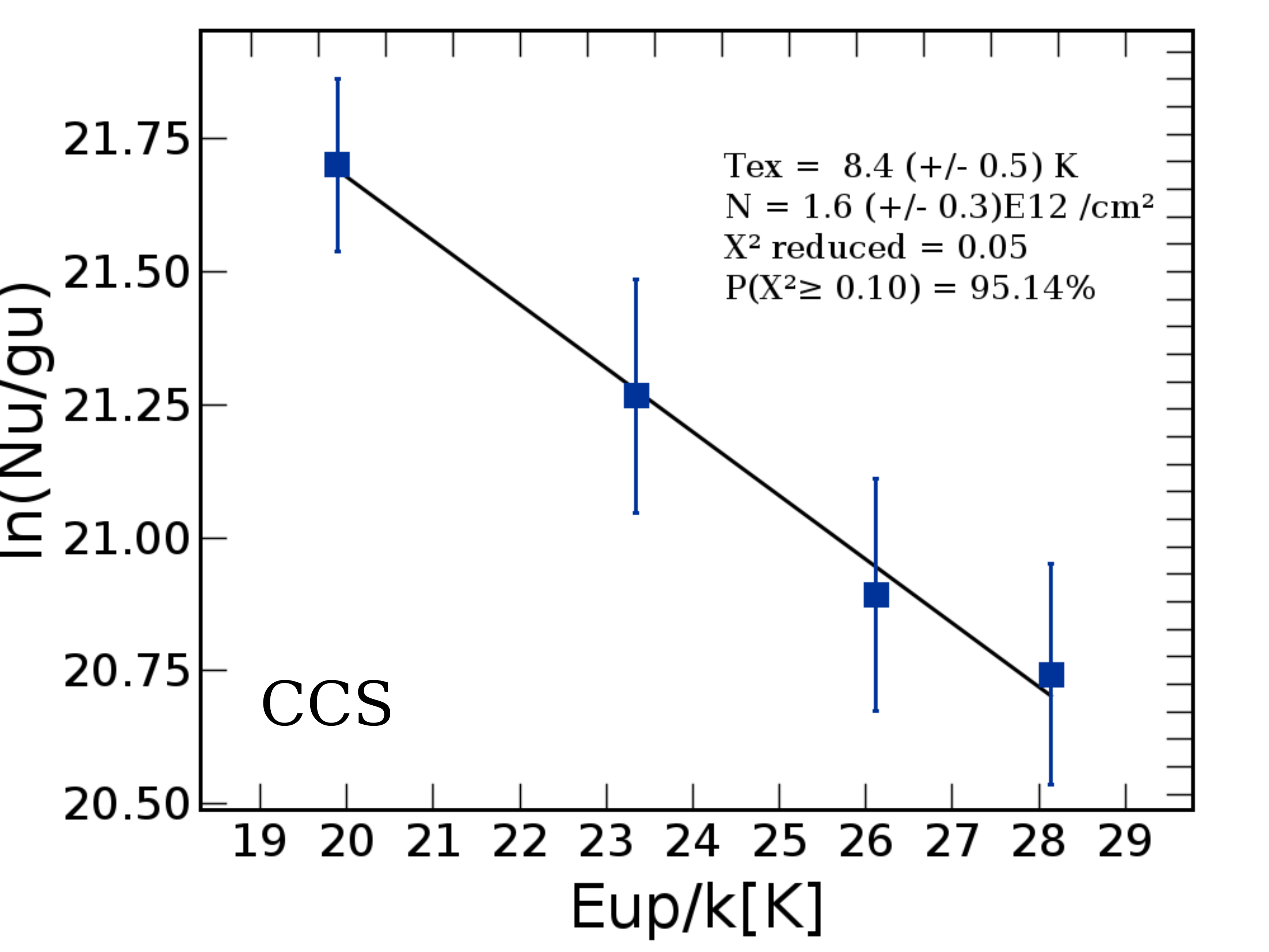} \\
\end{tabular}
\caption{Rotational diagrams showing the best fits to the observed molecular lines  \label{rotd-fig}}
\end{figure*}

Integrating the  column density maps pixel-by-pixel, we can estimate the total mass in different parts of the globule,
keeping in mind that because of the optimization for source extraction in the NIKA maps, the low-column-density dust
emission may be slightly underestimated.
The differences between the maps including NIKA and both (or only one) of the PACS images  allow us to estimate the errors
in a more acurate way.
The total mass in IC\,1396A within regions with column densities higher 
than  our 1$\times$10$^{21}$cm$^{-2}$ threshold varies depending on the model
used and on whether we give more or less weight to the longer or shorter wavelengths. As in Figure \ref{SED-fig}, models that include the shorter wavelengths tend to have higher temperatures and thus lower masses than
models that include the longer wavelengths.
A shallower $\beta$=1.7 tends to produce masses that are nearly a factor of 2 lower than the
masses derived with $\beta$=1.9, although lower $\beta$ values may be hard to justify in the less-dense
parts of the globule.

Depending on the choice of parameters and datasets, the total mass of the globule 
can vary by nearly an order of magnitude between 35 and 270 M$_\odot$. A larger value is more consistent 
with the complete gas maps obtained for the region \citep{patel95}, although we do not include the whole globule tail. 
Note that since the extinction towards the Tr37 cluster is A$_V$=1.67 mag,
with a typical variance between objects of around 0.5 mag 
\citep[attributed to thin cloud material all around the region plus extinction from circumstellar material][]{sicilia05}, there is likely a substantial mass
of low density material that is not detected in our maps. 
The mass in the denser 28" region around IC1396A-PACS-1 ranges between 8-80 M$_\odot$, but with the
most accurate mass derived from SED fitting, larger masses are favoured. 
For comparison, the NIKA S clump contains between 1-16 M$_\odot$ (with the large range here motivated by
the fact that the limits of the clump are not well-defined in flux nor temperature/column density), which is only a small fraction of
the mass surrounding the Class 0 source. Therefore,  independently of the dust model assumed, the head of IC1396A that contains the Class 0 source also contains
about 1/4 of the total mass of the globule and is significantly denser than the rest of the structure. We also note that
the largest difference in mass, temperature, and column density is not due to small variations in the dust model adopted, but rather on whether the map is more or less strongly weigthed towards the shorter or longer wavelengths, with the \emph{Herschel-}based maps giving higher temperatures and substantially lower masses. To improve these results, a multi-temperature fit with additional far-IR and submillimeter data would be required, which is beyond the scope of this work.

\subsection{Gas temperatures and column densities \label{rotd}}

\begin{table*}
\begin{footnotesize}
\begin{center}
\caption{Results of the rotational diagrams obtained with CASSIS. }              
\label{rotd-table}     
\centering                                     
\begin{tabular}{l c c c c c l  }       
\hline\hline                        
Species		& $\lambda$ range & E$_{up}$ range 	&  Nr. of Lines  & T$_{ex}$ 	& N 		 	&   Notes \\
		& (GHz)	          &     (K)      	&              	 & (K)    	& (cm$^{-2}$)		 &  Molecule ID, $\chi_{red}^2$ \\
\hline                                  
c-C$_3$H$_2$ 	& 87.28-87.41	  &  19-39	& 3 		 &  10.2$\pm$0.5  & 8.9$\pm$1.4 e11   	& 38002, $\chi_{red}^2$=0.18\\
CH$_3$CN 	& 91.98 	  & 13-78   	& 4 		 &  35.7$\pm$2.8 & 8.4$\pm$0.6 e11 	& 41001, $\chi_{red}^2$=0.12 \\
CH$_3$CHO 	& 93.58-98.90 	  & 13-23	& 7 		 &  24$\pm$8    & 4.5$\pm$1.1 e12 	& 44003, $\chi_{red}^2$=0.19 \\
CH$_3$OH 	& 96.75/97.58 	  &  7.5-45  		& 8 		 &  17$\pm$8  & 6$\pm$5 e13    		& 32003, $\chi_{red}^2$=29.21\\
CCS 		& 86.18-227.14 	  &  19-28  	& 4 		 &  8.5$\pm$0.5  & 1.6$\pm$0.3 e12  	& 56007, $\chi_{red}^2$=0.05 \\
C$_4$H		& 85.63-85.67  	  &  20-25	& 8		 &  19.8$\pm$1.1 & 1.4$\pm$0.1 e13	& 49003, $\chi_{red}^2$=0.01 \\
\hline                                             
\end{tabular}
\end{center}
\end{footnotesize}
\tablefoot{Only the lines that are strong enough, span a large enough range of upper level energies (E$_{up}$)
and have no significant velocity structure are included. The notes include the reduced chi square of the fit ($\chi_{red}^2$) and
the identification number for the molecular model used in CASSIS,
taken from the 
JPL database\footnote{https://spec.jpl.nasa.gov/}.}
\end{table*}

For the lines that are narrow and strong enough and span a large enough range of 
upper level energies (E$_{up}$), we use a rotational
diagram  to estimate their temperatures and densities. 
Given the complexity of the region, at this stage we
concentrate on narrow lines observed towards the map center, dominated by a single component, given that
complex lines (such as the HCO transitions and the broad CN lines)  would need to be first decomposed in their various velocity and density contributors
to obtain a meaningful fit. 
The complete chemical analysis, including the various velocity components, will be presented in a 
second paper. The main limitation for this exercise is that the lines are weak and the beam is
too large to be able to distinguish the spatial distribution of the line emission, so some of the lines
explored could be in part related to PDR emission and in part to the Class 0 source, as for instance is seen for the strongest c-C$_3$H$_2$ line. Future interferometric data will be required to give a more detailed picture.

We used the CASSIS software\footnote{http://cassis.irap.omp.eu}to extract all the lines in the spectra, fitting a baseline plus Gaussian model to every individual line for a given molecule,
examining the fits and the lines for any inconsistent velocities that may represent a line misidentification. 
For the rotational diagram fit, we assume an uncertainty of 15\% in the flux calibration.
The HCOOH lines have large errors and lead to no meaningful excitation temperature, 
which suggests that some of the transitions are either misclassified, 
contaminated by other species, or that the emission originates in different regions with various temperatures. The transitions of c-C$_3$H$_2$, CH$_3$CHO, CH$_3$CN,CH$_3$OH, and CCS produce 
good rotational diagrams and reveal the temperature structure of the gas in the region, although the fit for the
CH$_3$OH line has a very large uncertainty. A total of 4 transitions identified as C$_4$H produce bad fits regarding
velocity and/or S/N, but excluding them results in 8 well-fitted lines for the rotational diagram. An opacity correction to the fits does not significantly
change the results, since the errors are dominated by the S/N of the lines. 
Table \ref{rotd-table} summarizes the results, which are displayed in Figure \ref{rotd-fig}.

The rotational diagrams reveal gas temperatures and column densities consistent with the dust observations,
assuming typical abundances in star-forming regions. Considering that all the lines are strongest towards
the Class 0 source, we can compare the observed temperatures and column densities with those derived for the dust in 
Section \ref{dust-TNH}. The density peak around IC1396A-PACS-1 has a column density of 2$\times 10^{22}$ to 1$\times 10^{24}$ cm$^{-2}$ and
a temperature around 15-17 K (depending on the wavelengths used to derive the maps). The c-C$_3$H$_2$, CCS, and CH$_3$OH have
excitation temperatures significantly lower than the dominant temperature, which can be due to the fact that the temperature 
derived from the continuum images is dominated by the highest temperature along the line-of-sight. CH$_3$CN and
CH$_3$CHO track material at a higher temperature than the rest, consistent with observations of WCCC. 
For CH$_3$OH we find a column density of the order of 1$\times 10^{14}$ cm$^{-2}$,
which compared to the hydrogen column density of  2$\times 10^{22}$-1$\times 10^{24}$ cm$^{-2}$ suggests a ratio CH$_3$OH/H$_2$=1$\times 10^{-8}$ to 1$\times 10^{-10}$, in line with values found in
envelopes of cold cores in an early evolutionary stage \citep{vandertak00,kristensen10,oberg14}. The higher
temperature of CH$_3$CN is consistent with an origin in a deeper and warmer region of the core \citep{oberg14}, although the
relative abundance with respect to CH$_3$OH ($\sim$0.02) is lower than expected, which may be due to beam dilution.
CCS is usually found in early-stage star formation, and would suggest an age $<$10$^5$ yr for IC1396A-PACS-1 \citep{suzuki92,gregorio06},
placing it among the youngest YSOs known. The abundance of c-C$_3$H$_2$ is similar to what is found in other low-mass star-forming regions,
although the low temperature is in contrast with the usual origin of the molecule in the WCCC region \citep{sakai10}. Some contamination from c-C$_3$H$_2$ from the PDR region is expected, due to the large beam and to the fact that some c-C$_3$H$_2$ emission is detected towards the densest parts of the PDR as well (see Section \ref{emline}), which may be also the reason of the discrepant temperature values.

A further constraint on the gas mass can be obtained from the integrated line intensity for optically
thin lines \citep{scoville86}. Our main limitation is the lack of a reliable measure of the excitation temperature
since we did not observe several transitions for the same molecule. We follow the procedure in \citet{pineda10} to estimate the excitation temperature from the optically thick CO line\footnote{Note that this has the strong limitation that the $^{13}$CO and C$^{18}$O lines trace much deeper material than the CO line, so their excitation temperatures are likely different.}. We use their relation beween the corrected main beam temperature ($T_{mb,c}$) and the excitation temperature ($T_{ex}$),
\begin{equation}
T_{mb,c}= T_0 \Big[ \frac{1}{e^{T_0/T_{ex}}-1} - \frac{1}{e^{T_0/T_{bg}}-1} \Big] (1-e^{-\tau}), \label{eqtx}
\end{equation}
where $T_0 = h \nu/k$ for the line frequency, $T_{bg}$=2.73 K is the background temperature, anffd $\tau$ is the
line opacity. The main beam temperature is related to T$_a^*$ by the telescope efficiencies, which gives a factor of 1.559 for the CO(2-1) frequency, and 1.522 for the C$^{18}$O(2-1) transition \citep{kramer13}. For a large-enough source (which may be applied to CO since it is extended rather uniformly over the whole field), the beam filling factor can be taken to $\sim$1. Since CO is optically thick, we can derive its excitation temperature from the line peak, obtaining
39.8 K on the Class 0 source, for which T$_a^*$=22 K for the CO(2-1) line. For comparison, the value for the globule average is 36.6 K (T$_a^*$=20 K) and for the NIKA S clump (T$_a^*$=18 K) we find 33.5 K. Assuming the same excitation temperature for all the CO lines, we can use Equation \ref{eqtx}
to derive the optical depth of the $^{13}$CO and C$^{18}$O lines. 

We find that $\tau_{13CO}\sim$1 in all regions, thus marginally optically thick. For C$^{18}$O, the assumption that the emission comes from a region
that is very large compared to the beam breaks down. While it is true that the emission is strongly peaked at the position of the Class 0 source, there is some significant emission towards NIKA S
and all around the globule. We thus estimate a filling factor of 0.7 for the compact sources \citep[assuming a size comparable to the NIKA 2mm beam for the C$^{18}$O emission, see for instance][]{shimajiri14}.
This gives us $\tau_{C18O}$=0.31 in the Class 0 source, and  $\tau_{C18O}$=0.16 in the NIKA S clump. With these values, the C$^{18}$O column density (N$_{C18O}$) can be derived \citep[see e.g.][]{scoville86,ao04,pineda10}. Assuming that the excitation temperature is much higher than the background 
temperature and using the beam-averaged opacity $\tau$, we obtain
\begin{equation}
N_{C18O} = \frac{3 k^2}{8 \pi^3 B \mu^2 h \nu} \frac{e^{hBJ(J+1)/kT_{ex}}}{(J+1)} \frac{T_{ex} + hB/3k}{e^{-T_0/T_{ex}}} \frac{\tau}{(1-e^{-\tau})} \int T_{mb,c} dv,
\end{equation}
where $h$ and $k$ are the Planck and Boltzmann constants, respectively, J is the quantum number of the lower level, $B$ is the rotational constant (54.891 GHz for C$^{18}$O\footnote{https://physics.nist.gov/PhysRefData/MolSpec/Diatomic/Html/Tables/ CO.html}), and $\mu$ is the electric dipole moment (0.1098 Debye for C$^{18}$O). 

To calculate the mass of the sources, we need to take into account the abundance of C$^{18}$O with respect to H$_2$. There is a substantial uncertainty in this, which moreover depends on the type of region observed (e.g. PDR regions vs YSO vs HII regions), with some authors suggesting lower \citep{areal18} or higher \citep{shimajiri14} C$^{18}$O vs H$_2$
abundances that can lead to significantly different results. We adopt the calibration of \citet{frerking82} for high-density enviroments in $\rho$ Ophiuchi. The resulting $H_2$ column densities range from 8.5e+23 cm$^{-2}$ on the Class 0 source, to 4.1e+23 cm$^{-2}$ in  NIKA S, and  7e+22 cm$^{-2}$ for the globule average within the EMIR map, which are roughly comparable to the values derived from the 160$\mu$ plus NIKA 1mm, 2mm column density map, and as in this case, more weighted towards lower temperatures and higher densities. Using the same mean molecular weight 2.8 and taking into account the size of the sources, we estimate a mass for the Class 0 source around 45 M$_\odot$, and 22 M$_\odot$ for the NIKA S clump, which are 
roughly consistent with the higher estimates based on dust emission. Given the uncertainties in the C$^{18}$O abundance, the possibility of some degree of C$^{18}$O being frozen onto the grains and, to a lesser extent, the uncertainties in the excitation temperature, the above given masses are highly uncertain. In any case, the results are in good agreement with the amount of mass expected  around a forming intermediate-mass star.

Since we did not observe the whole globule with EMIR, it is not possible to derive a full-globule mass.
The mass derived from the gas observations towards the mapped globule tip suggests
that the total gas mass may be a few times higher than the globule estimate from dust mass. Although the dust-based
and the gas-based mass estimates for the Class 0 source agree within a factor of few for the lower temperature estimate, the gas-derived mass of NIKA S is significantly higher, being about half (instead of about 12-20\%) of
the mass associated to the Class 0 clump. This may be due to a combination of gas freezing in the colder regions
\citep[despite this not being detected in the maps nor in the SMA maps][]{patel15}, filtering of extended emission in the NIKA maps, uncertainties in the C$^{18}$O/H$_2$ ratio, 
and variations of the excitation temperature and the C$^{18}$O/H$_2$ ratio throughout the different parts
of the globule.

\subsection{Emission line analysis: exploring the velocity structure of IC1396A\label{emline}}

\begin{figure}
\centering
\includegraphics[width=0.99\linewidth]{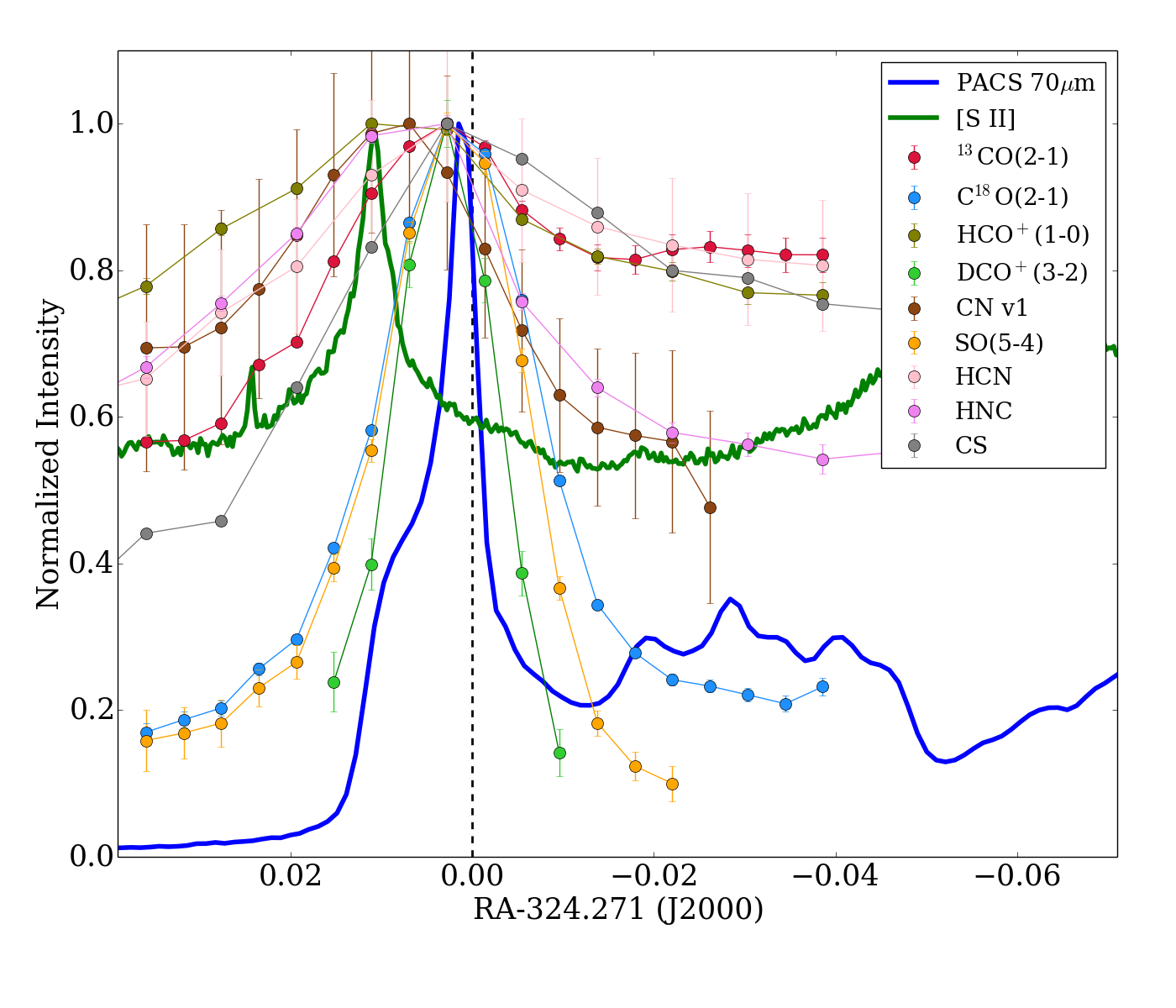}
\caption{Intensity of various tracers along a cut across the tip of the globule along the line between HD~206267 and IC1396A-PACS-1.
The normalized flux in  [S II] and PACS/70~$\mu$m data along the same region is also shown. 
The position of IC1396A-PACS-1 is marked as a vertical dashed line.
HD~206267 would be located at RA=324.74007, thus at $\Delta$RA=+0.46907 degrees along the same line.
The physical scale for a distance of 945 pc is 0.178 pc per $\Delta$RA=0.02 degree.
 \label{slice-fig}}
\end{figure}

The global structure of IC1396A is well-characterized by a combination of multi-wavelength,
multi-species data. Figure \ref{slice-fig} shows a cut through IC1396A-PACS-1, starting in the
outer part cleared by HD\,206267, and including the edge
of the PDR and along the globule. The \emph{Herschel}/PACS data clearly show the onset of the dusty
globule and the density enhancement where IC1396A-PACS-1 is located. Narrow-band [S\,II] data \citep{sicilia13} reveals the
peak of the PDR, behind which the dust density rapidly increases. The peak flux of various
tracers also reflect their nature, associated with the PDR and/or with the dense
globule.  The profile across the globule rim  and the 
projected distance bewteen the ionization region (marked by the [S II] peak)
and the maximum density (shown by both the \emph{Herschel}/PACS data and the molecular
high-density tracers in Figure \ref{slice-fig}) is $\sim$0.09 pc, which is similar
to what it would be expected for RDI in a relatively small but massive and
dense cloud \citep{miao09} about 0.4 Myr after the onset of significant
exposure to ionizing radiation.

A portion of the CN, CS, HCO+, HNC and HCN appears associated with the photodissociation rim, while highest-density tracers
 (C$^{18}$O, N$_2$D$^+$, DCO+, SO) peak at the position of the source and the \emph{Herschel} dust rim,
suggesting that the molecular line emission originates from dark regions protected from the UV front.
The HNC and HCN data can be used as a temperature tracer, since the HNC/HCN ratio is larger than
unity in cold regions, and below unity in warmer regions \citep[e.g.][]{tennekes06}. Figure \ref{slice-fig} shows how the HNC/HCN ratio
decreases towards the west of the globule, compared to the higher value of HNC/HCN at the source position,
revealing a higher temperature towards the less dense parts of the globule, as expected for external heating.

\begin{figure}
\centering
\includegraphics[width=9.0cm]{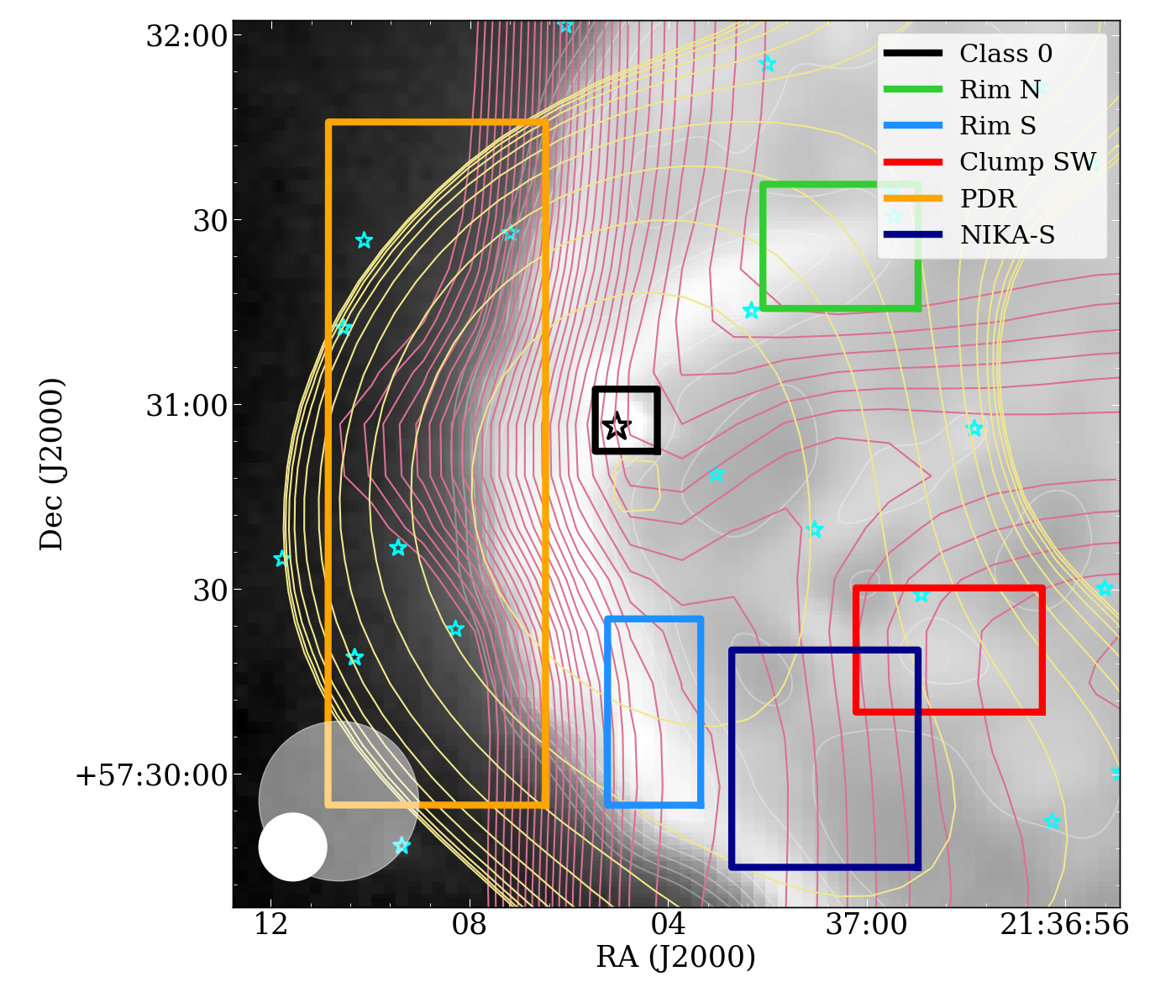}
\caption{Selected areas in IC1396A marked by colored boxes. The background grey image with white contours is the
Herschel/PACS 70~$\mu$m map. The yellow contours mark the NIKA 1.3mm emission as in
Figure \ref{nika-fig}. The violet contours display the $^{13}$CO emission (23 linear contours
in the range 8-30 K (T$^*_a$) km/s). 
The Class 0 object is marked with a large black star, the rest of the
cluster members are marked with small cyan stars. The beams for the two frequency settings 
are also displayed.
\label{regions-fig}}
\end{figure}

To explore the velocity structure in the region, we first extracted integrated velocity maps of 
the lines observed with good S/N in the high velocity 
resolution mode (see Appendix \ref{velo-app}). 
Due to the high velocity resolution of the data and the large number of channels (e.g. over 50 for the $^{12}$CO main component, more than 30 for C$^{18}$O), plotting the data for individual channels is unpractical. We thus present the velocity-integrated line
intensity calculated in nine 0.5 km/s velocity bins, centered from -9.25 to -5.25 km/s, which 
corresponds to the total velocity span observed for all lines. These maps are the first step to visualize and investigate
the velocities associated
with the different structures within the globule. Based on the velocity maps, on the
\emph{Herschel} data, and on the  column density/temperature maps derived from \emph{Herschel} and NIKA data, 
we extracted the part of the spectrum corresponding to
several distinct structures within IC\,1396A, which include:
\begin{itemize}
\item The Class 0 object IC1396A-PACS-1 (labeled as ``Class 0" from now on).
\item The edge of the PDR region (``PDR"). 
\item The $^{13}$CO clump to the south-west of the region (``Clump SW"), which appears globally redshifted.
\item The arc-shaped structure to the north of the Class 0 object (``Rim N").
\item The arc-shaped structure to the south of the Class 0 object (``Rim S").
\item NIKA S, the dense and cold region to the south of the Class 0 object that shows strong NIKA 1mm emission above mentioned. 
\end{itemize}
All the regions are selected to be isolated to avoid contamination by nearby ones, although
due to the beam size, some contamination is unavoidable.
For each region, we estimated the average line
emission using GILDAS/Class, which is then used for the velocity structure analysis
of the cloud. Figure \ref{regions-fig} shows the location of the various components compared to the 
diverse features in continuum and line emission. 
 These regions appear clearly distinct
in their molecular emission (velocity, intensity, line profile) and also in their general physical properties,
although due to projection effects and to the 3D structure of the region,
we can expect some degree of contamination from other structures in all of them\footnote{For instance,
the whole region shows PDR-related lines, probably arising from the illuminated
globule behind the dark structures we observe in the optical.  Also note that the regions
with potential leaks across the scan and cross-scan directions are excluded from the 
analysis (see Appendix \ref{velo-app} for details).}.

\begin{figure*}
\centering
\begin{tabular}{ccc}
\includegraphics[width=0.3\linewidth]{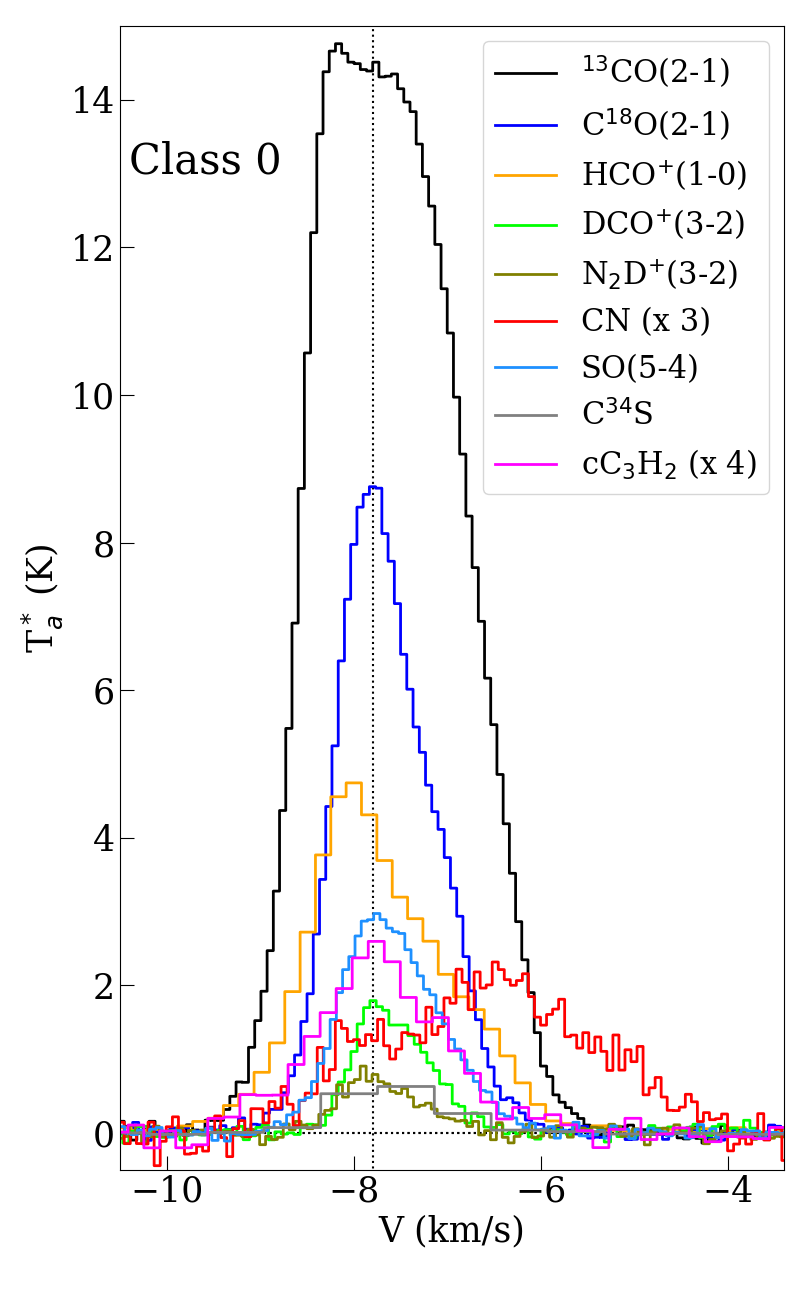} &
\includegraphics[width=0.3\linewidth]{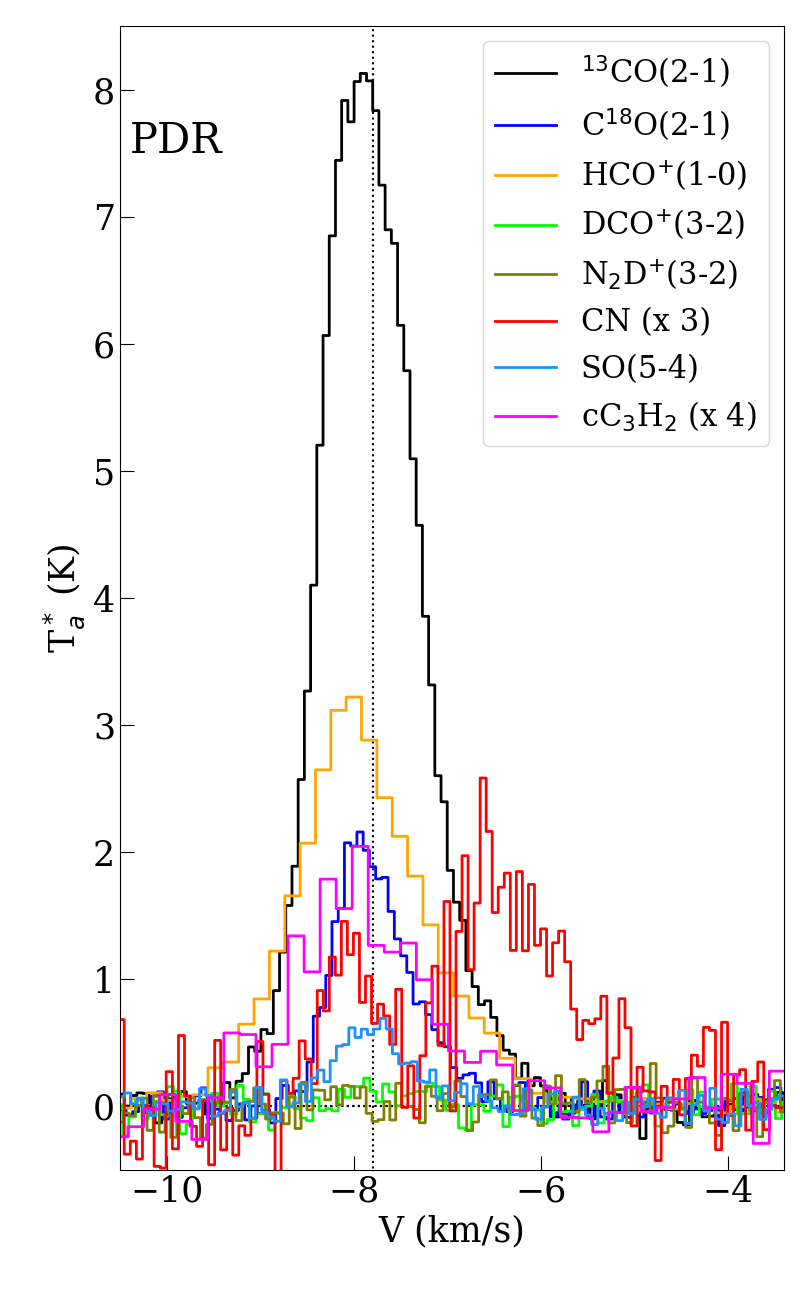} &
\includegraphics[width=0.3\linewidth]{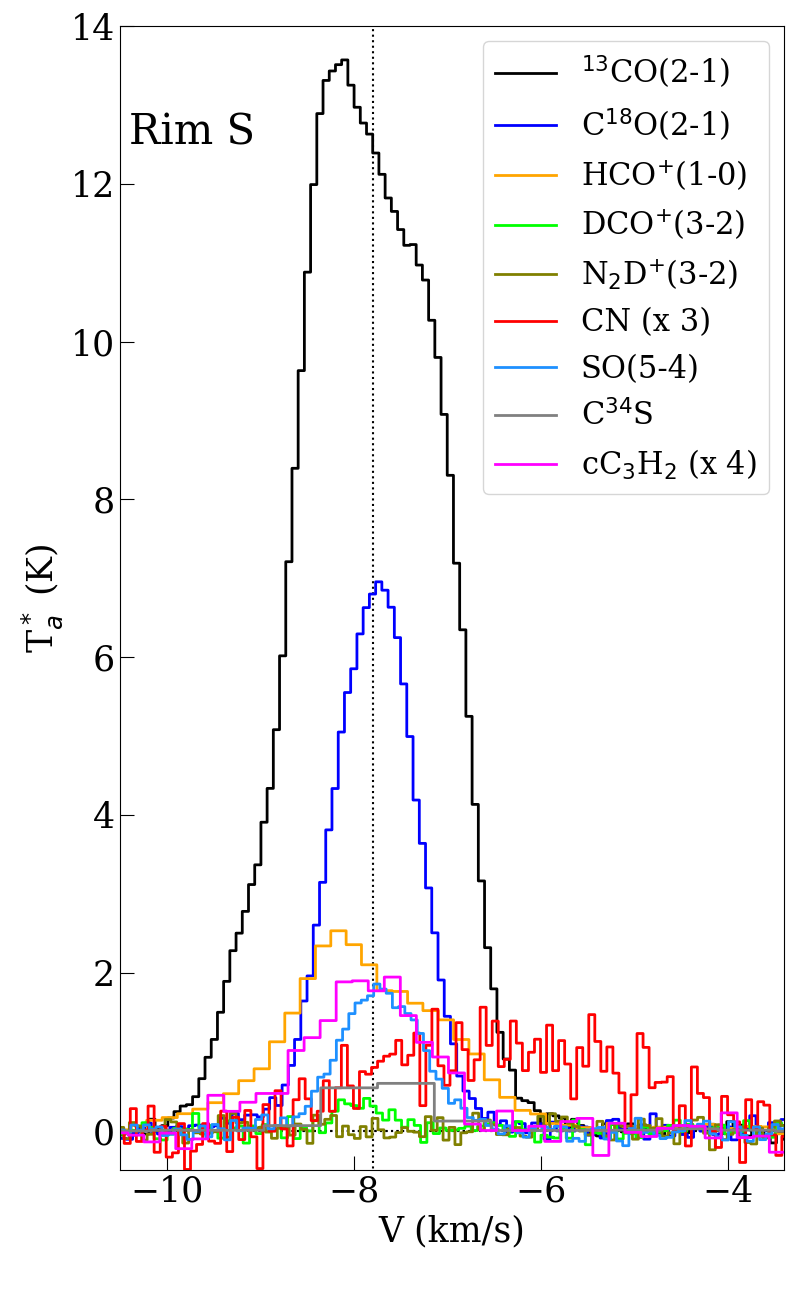} \\
\includegraphics[width=0.3\linewidth]{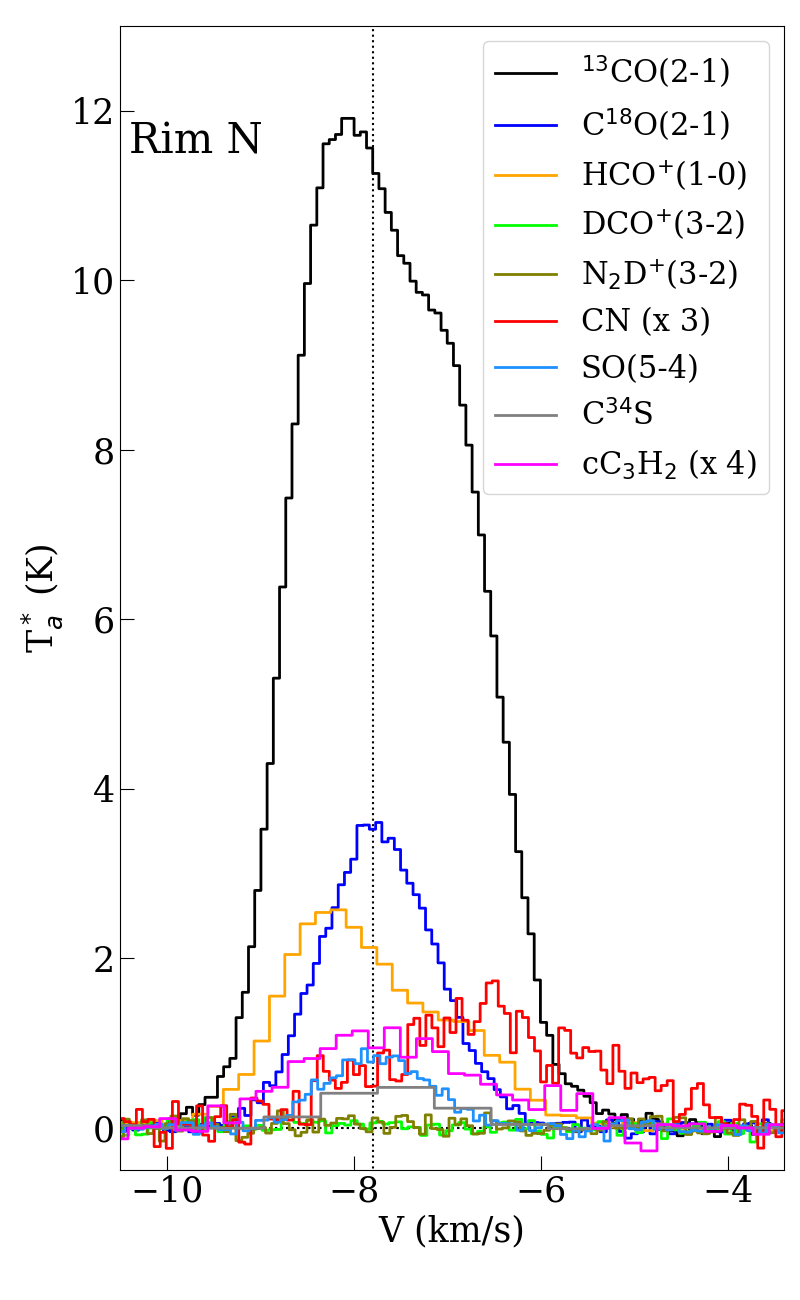} &
\includegraphics[width=0.3\linewidth]{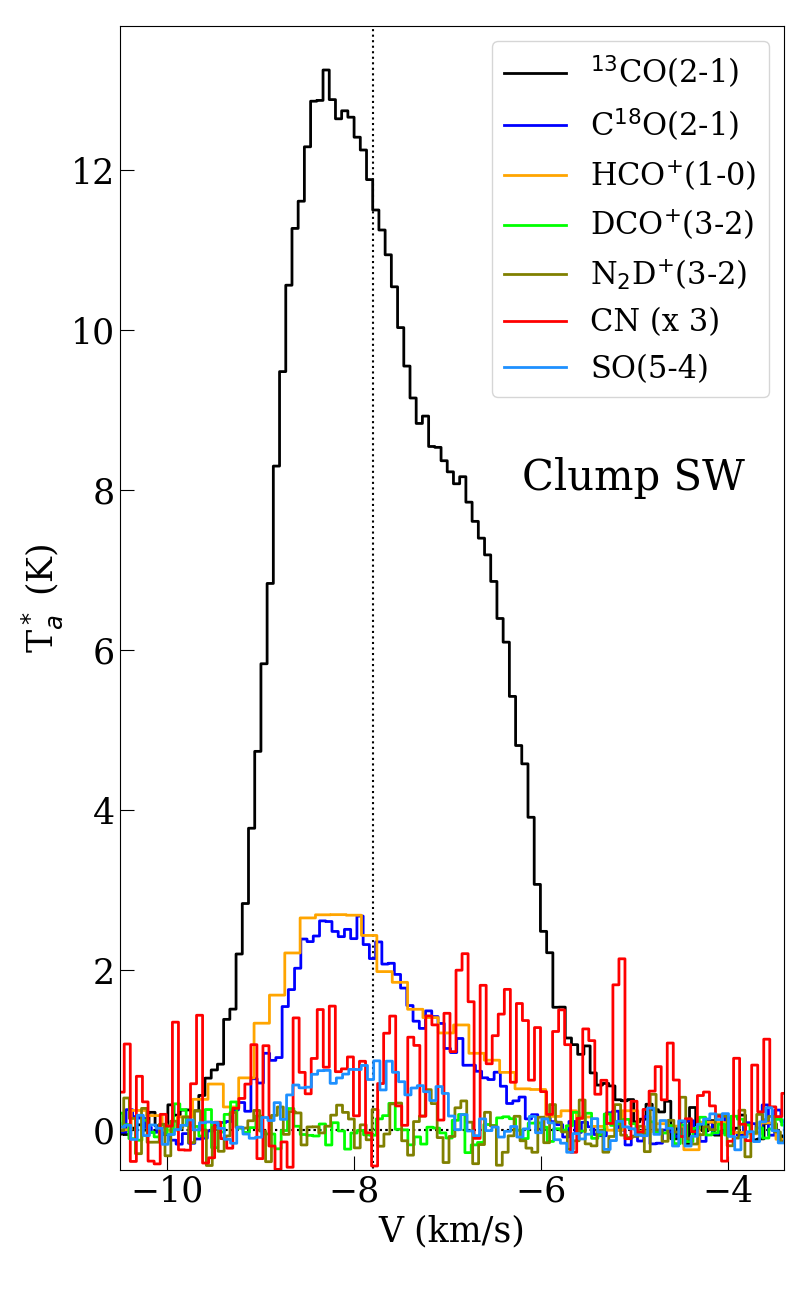} &
\includegraphics[width=0.3\linewidth]{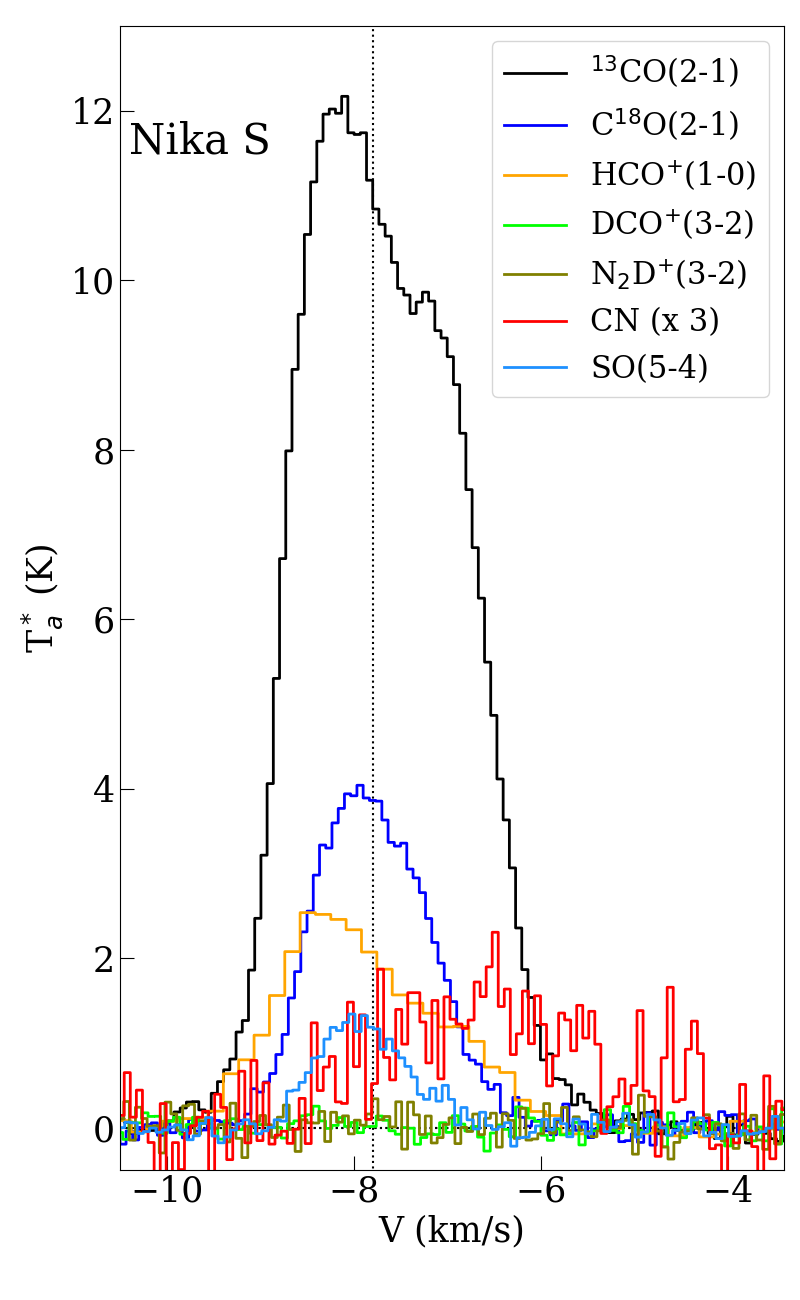} \\
\end{tabular}
\caption{Average emission lines detected towards the various regions marked in Figure \ref{regions-fig}. Note the difference in line profiles and strength between the Class 0
core and the rest of the regions. The Rim S region also appears denser than the rest, although
it is clearly less dense than observed towards the class 0 source. 
High-density tracers (N$_2$H$^+$, DCO$^+$) are observed only 
towards the densest regions. The CN line (which is presented here multiplied by a factor of 3 because of
its relative weakness) is strongly associated with the edges of the cloud, as expected if
dominated by the photoevaporating cloud interface. 
 \label{lineregions-fig}}
\end{figure*}

\begin{figure}
\centering
\includegraphics[width=0.9\linewidth]{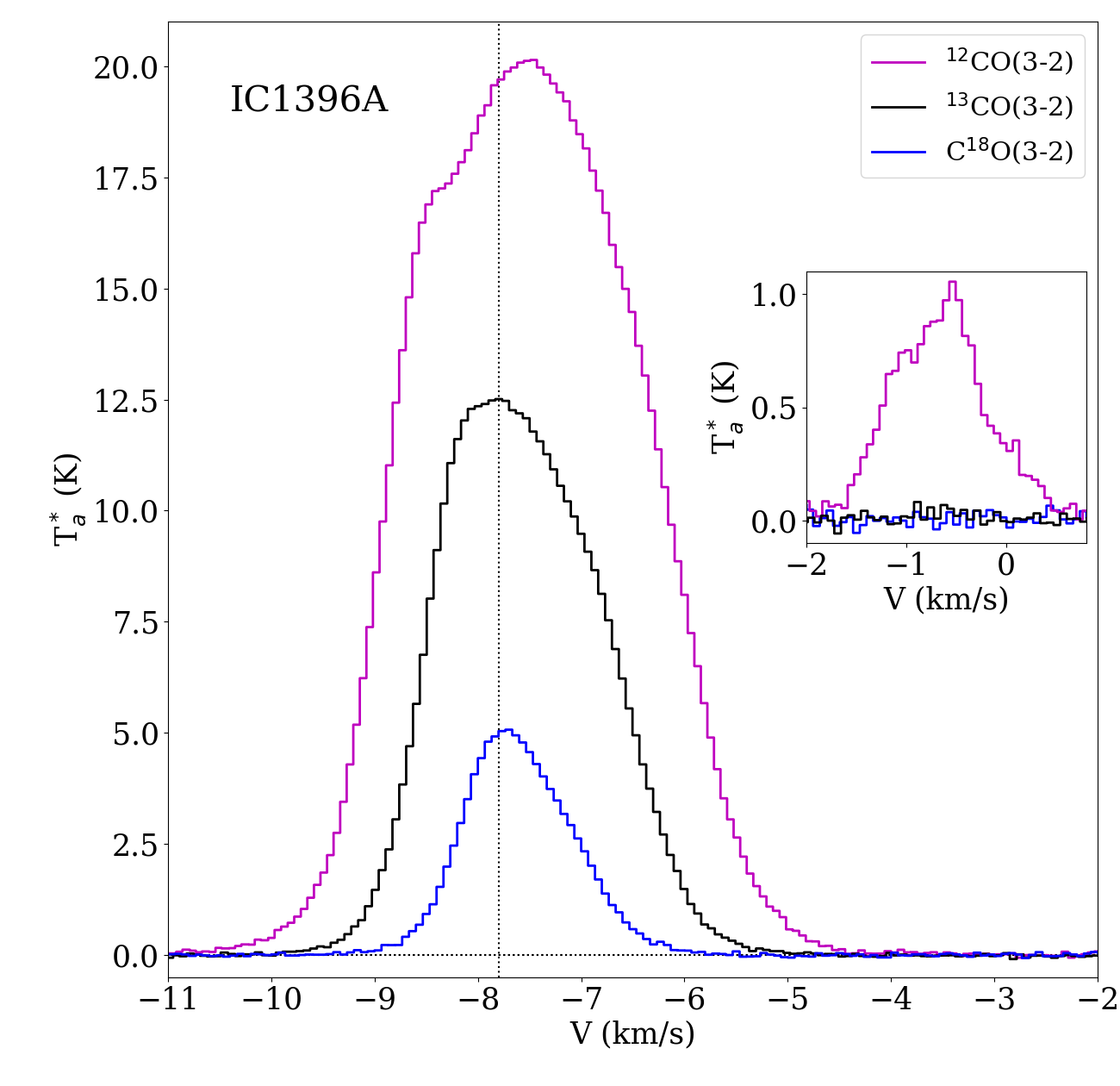}
\caption{ Averaged $^{12}$CO and  $^{13}$CO emission towards IC1396A. The inset shows the faint component at velocity -0.7 km/s. Note that there is no detectable $^{13}$CO emission
at this velocity.
 \label{faint-fig}}
\end{figure}

The global structure of the low-density tracers is very complex, as the line
profiles can include different velocity components along the line-of-sight even
when we integrate over different (projected) spatial locations. 
The $^{12}$CO line is strongly saturated on the globule around the systemic velocity and thus does not
offer much information. Nevertheless, the line wings can be used to estimate the limits of the maximum velocities observed in low-density
gas along the line-of-sight, as we discuss in Section \ref{Tr37-past}. In addition, a faint $^{12}$CO
component at $-$0.7$\pm$0.1 km/s is detected throughout the entire mapped region (see Figure \ref{faint-fig}). 
There is no evidence of gas at this velocity in $^{13}$CO nor any other of the tracers, 
which is a signature of
low density and of the line being optically thin. The intensity of the faint $^{12}$CO component is quite uniform, increasing towards the west of the region.
Its line wings extend up to $\pm$2 km/s, thus more than observed in other optically thin lines.

Higher-density tracers
reveal the density enhancement inside the globule, and even the asymmetry between the 
less-dense northern part and the denser southern side of IC1396A.
If we leave aside the lines that are highly saturated and have distorted profiles
($^{12}$CO and $^{13}$CO) and
concentrate on lines that are observed towards all five regions (such as C$^{18}$O, HCO$^+$, and HNC;
Figure \ref{lineregions-fig}),
we also observe that the line width increases off-source and that the Class 0 object
is systematically blueshifted by 0.5-1 km/s with respect to the surrounding nebula. 
The blue-shifted asymmetry points towards collapse, while the increased line width is
consistent with increased turbulence and the bulk motions in the surrounding clump. 
The off-source line profiles are also asymmetric but less sharply-peaked than
on-source. They are blue-dominated and thus suggestive of collapse or, in case
of a globule that is being photoionized on the far side from our line-of-sight, 
it could be a sign of generalized RDI. 
The broad-but-asymmetric profiles
of the lines in the less-dense parts of the globule are consistent with the
gas being disrupted and removed from the globule by the effect of HD~206267.

Only the region around IC1396A-PACS-1 has significant emission in high-density
tracers (such as DCO$^+$, N$_2$D$^+$, and H$_2$CO; see Figure \ref{lineregions-fig}).
There is weak DCO$^+$ emission associated with Rim S, although it
is one order of magnitude weaker than the emission associated with the IC1396A-PACS-1
core. This indicates that the Class 0 protostar
is forming in the densest parts of the globule, and is consistent with the factor of 5-10
higher  column density around IC1396A-PACS-1, measured by the continuum observations. The $^{13}$CO line
is saturated at the position of the source, while the C$^{18}$O presents a blue-asymmetric
profile. 
There is no evidence of CO depletion in the source, despite the
detection of nitrogenated species and the potential disparity between gas-based and dust-based masses, 
which could be an effect of the large beam and the complexity of the 
source  to be explored with higher resolution observations (see Patel et al. in prep).  
The C$^{18}$O line profile could be interpreted as infall, athough the proximity of the 
PDR and the photoevaporative velocities associated with it and the fact that the $^{13}$CO presents the
same blue-dominated profile towards the rest of the cloud suggest that the profile could be also affected
by global cloud motions and photoevaporation.
The rest of emission lines from high-density tracers, especially for those detected with
high S/N (DCO$^+$, N$_2$D$^+$) are the best indicators of the properties and velocity 
of the Class 0 source, and 
in this case they are also found to be asymmetric, with a blueshifted peak, suggestive
of infall in the densest parts of the core.

We also find that the SW clump, besides being redshifted, has systematically larger line widths
than the rest of the structures. In general, the
cloud positions have a significantly stronger extended red tail, compared to the object.
Both the Class 0 source and the PDR lack these red tails, which is a further point suggesting
the association of IC1396A-PACS-1 and the ionization front. Detailed inspection of the spectra reveals 2 components in
several of the lines (CN, HCN), centered at $\sim$-8 and $\sim$-6 km/s. 
The redshifted component could be a sign of
photoevaporation in the outer parts of the globule facing the O star, 
while the blueshifted component would correspond to the material 
associated with IC1396A-PACS-1. The presence of a redshifted tail and broader
lines within the SW clump suggests a higher range of velocities and turbulence, 
a signature of mass loss and dispersion along several directions over the line-of-sight, 
compared to what is observed towards the Class 0 source and
PDR rim.  Note that other relatively massive protostars
such as $\alpha$ are too far from this redshifted clump for it to be caused
by outflows, and that the embedded globule population is mostly composed
of late-type Class I and Class II objects that are not expected to drive 
such powerful and broad outflows as to explain the SW redshifted emission.

The next step was to analyze the pixel-by-pixel structure in the different line tracers for which we have
high-resolution data. Since the lines are highly complex and often self-absorbed
we use a multi-Gaussian approach to create a model-independent, non-parametric way of describing the line strength, velocity, and width. Although
multi-Gaussian fits have been successfully used on large scales \citep[e.g.][]{hacar13}, 
the environment around IC1396A-PACS-1 is highly
complex and the multi-Gaussian fits are strongly degenerate. Therefore, in an analogy to complex optical emission lines \citep[e.g.][]{sicilia17}, 
we derive instead several line parameters, including line peak and peak velocity, 
integrated flux, line width, and line asymmetry (blue vs red components, for both the flux and the velocity). As occurs in the optical, molecular emission lines can be extremely complex and thus a
simple geometrical fit does not have a direct physical interpretation, especially in regions where 
saturation and/or self-absorption occur. The advantage of the fit is that it allows us to derive
line-based parameters  that take into account the global shape of the line, thus enabling us to systematically explore emission, velocities, and line asymmetry on a 
pixel-by-pixel scale. In this way, we can visualize and detect changes of the structure that are not seen by other means such 
as channel maps, especially in cases where the structure is very complex and the velocity resolution very high.

For this exercise, the lines are first interactively fitted with a multi-Gaussian model containing 1 to 3 Gaussian components, 
selected according to the line shape and $\chi^2$ of the fit. This fit is then used to derive the line
parameters including flux, peak velocity, width, and asymmetry in flux and in velocity (see  more details in Appendix \ref{gaussfit-app}). 
As long as the Gaussian model profiles provide a good fit ($\chi^2\leq$1) to the line, the derived 
parameters have the advantage
that they are not significantly dependent on the particular choice of Gaussian components, 
so that the line parameters are model-independent, circunventing the intrinsic degeneracy of the fit.
After this exercise, we can explore the position-velocity, position-width, position-intensity, 
and position-asymmetry
diagrams to derive information about the region.
Since the lines originate in gas with different temperatures
and densities, the velocities and velocity dispersion
in the different gaseous lines give us a 3D dynamical picture of the region,
whose details are discussed in Section \ref{3d-sect}.

\subsection{Tangencial velocities from Gaia DR2}

Further information regarding the velocity of the globule can be
obtained from the analysis of the Gaia DR2 data \citep{gaiamission16,gaiadr218}. 
The cluster proper motion can be calculated by weighted average of the 
individual proper motions
for the 354 objects with good Gaia data (see Section \ref{ancillary}), 
being $\mu_{RA}$=$-$2.5$\pm$1.5 mas/yr and 
$\mu_{Dec}$=$-$4.6$\pm$1.3 mas/yr, respectively. The uncertainties given correspond to the typical spread in proper motion between confirmed cluster members,  estimated as the
standard deviation for the 354 cluster members.

\begin{figure*}
\centering
\includegraphics[width=0.7\linewidth]{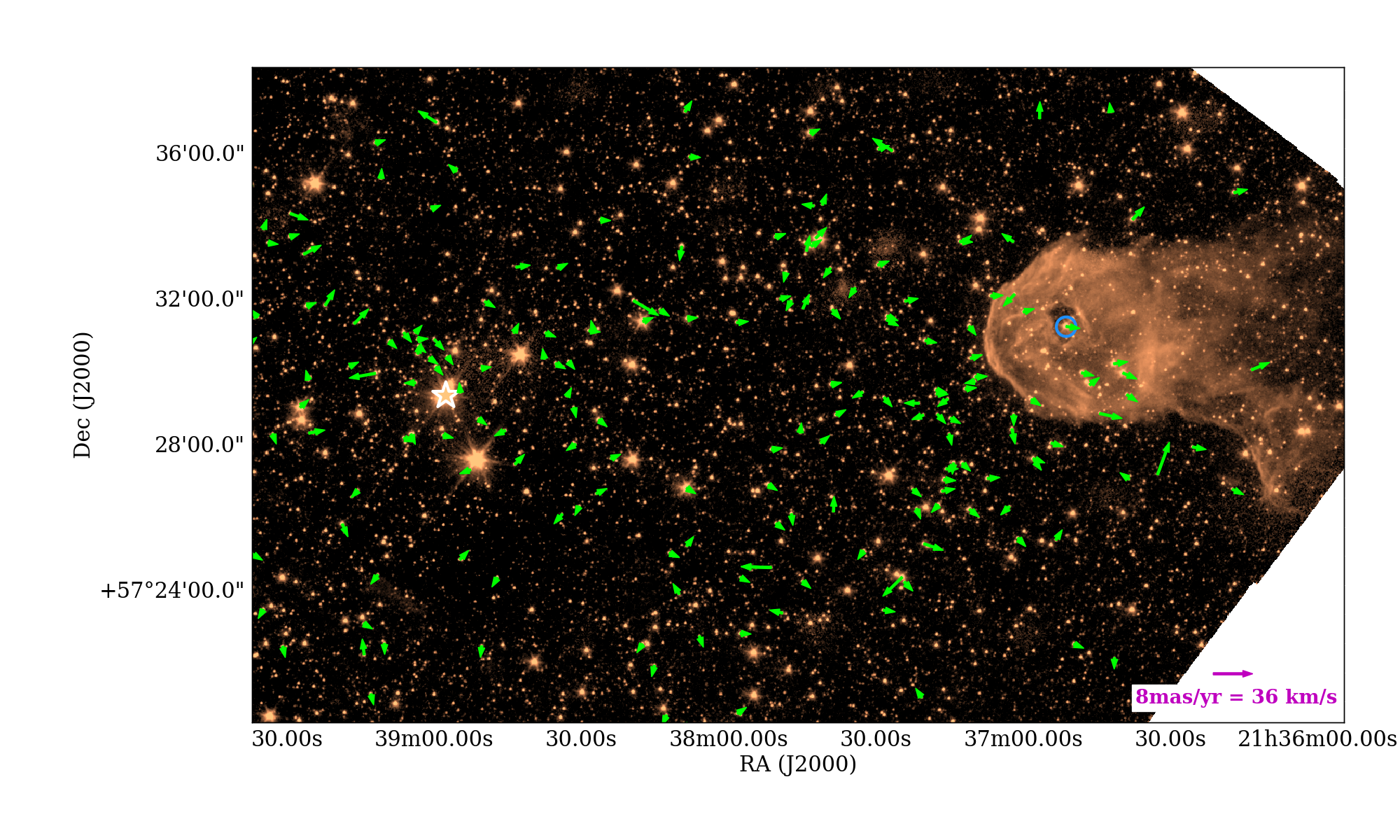}
\caption{Proper motions of the Tr 37 cluster members between HD\,206267 (marked with a large white star) and IC 1396A with parallax errors below 0.2 mas and proper motion errors below 2mas/yr, shown over the IRAC 1 3.6$\mu$m/Spitzer map of the region.
The arrows represent the proper motion with respect to the mean cluster proper motion pm$_{RA}$=$-$2.5$\pm$1.5 mas/yr, pm$_{Dec}$=$-$4.6$\pm$1.3 mas/yr. The stars associated 
to IC 1396A, including V390 Cep  (marked with a blue circle), 
have proper motions consistent with the cluster, with no evidence of
strong acceleration in the plane of the sky.  \label{gaia-fig}}
\end{figure*}

Figure \ref{gaia-fig} shows that the  spread of velocities in the plane of the sky of IC\,1396A members are not too different from what is observed elsewhere in
the Tr\,37 custer. The proper motions for the globule star V390 Cep,
for which its interaction with the surrounding material offer a clear signature of
association with the globule, are $\mu_{RA}$=$-$3.53$\pm$0.07 mas/yr and $\mu_{Dec}$=$-$4.78$\pm$0.07 mas/yr. There are 5 more objects seen in projection against the
globule and having good quality Gaia DR2 data. Including them together with V390 Cep, we
derive the weigthed mean proper motions for the globule to be  $\mu_{RA}$=$-$3.4$\pm$0.5 mas/yr and $\mu_{Dec}$=$-$4.8$\pm$0.5 mas/yr, where the errors reflect the standard deviation
of all sources found towards the globule. These values are essentially identical
to the velocity of V390 Cep, and all together suggests a tendency for V390 Cep and the globule to move systematically westwards (away from HD\,206267,  see Figure \ref{gaia-fig}).  The proper motion difference is
significant in RA, for which $\Delta \mu_{RA}$=-0.9$\pm$0.1 mas/yr, while the
proper motion difference in the Dec direction is essentially consistent with
zero, $\Delta \mu_{Dec}$=-0.2$\pm$0.1 mas/yr. For a distance of 945 pc, 
this is equivalent to 4 km/s westwards on the plane of the sky (in RA) and up to 0.9 km/s northwards (in Dec).

\section{Discussion: The structure and formation history of IC\,1396A and Tr\,37 \label{discussion-sect}}

\subsection{Gas dynamics in IC\,1396A \label{3d-sect}}

The pixel-by-pixel line component analysis reveals the velocity structure on the plane of the sky
with unprecedented resolution. Putting together the various lines, we can trace
the cloud at different depths around the Class 0 source. Many processes 
(e.g. velocity fields, infall, outflows, depletion, self-absorption) can
affect the shape of the line and thus the line parameters, which means that 
a single line parameter is unlikely to provide much information on physical processes
or structure.
Neverthelesss, by combining them all gives us a powerful way to explore the 
velocities and velocity gradients throughout the cloud (using the peak velocity for lines that are symmetric and
have no signs of self-absorption), detect relative expansion and contraction in higher density
tracers (which induce shifts in the
observed peak velocity and line and flux asymmetries with a dominant blue or red part, respectively), 
or identify the presence of more than one component (e.g. by checking
line peaks vs peaks of individual Gaussian components and the line width).

\begin{figure*}
\centering
\begin{tabular}{ccc}
\includegraphics[width=0.3\linewidth]{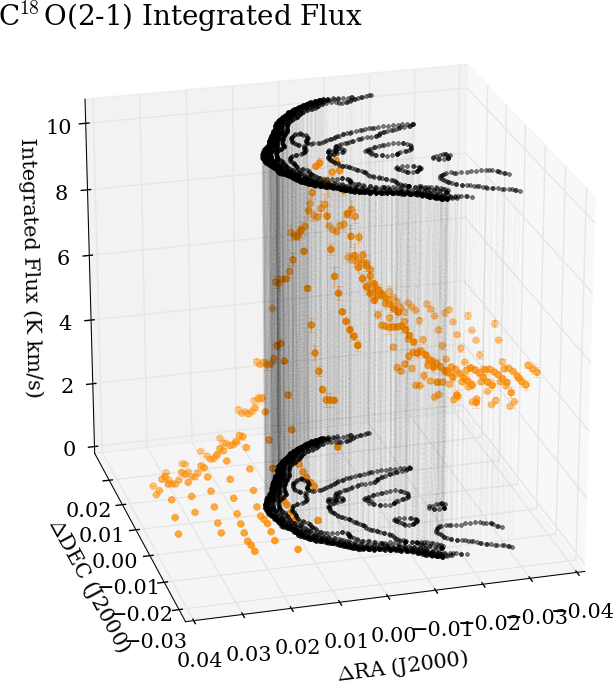} &
\includegraphics[width=0.28\linewidth]{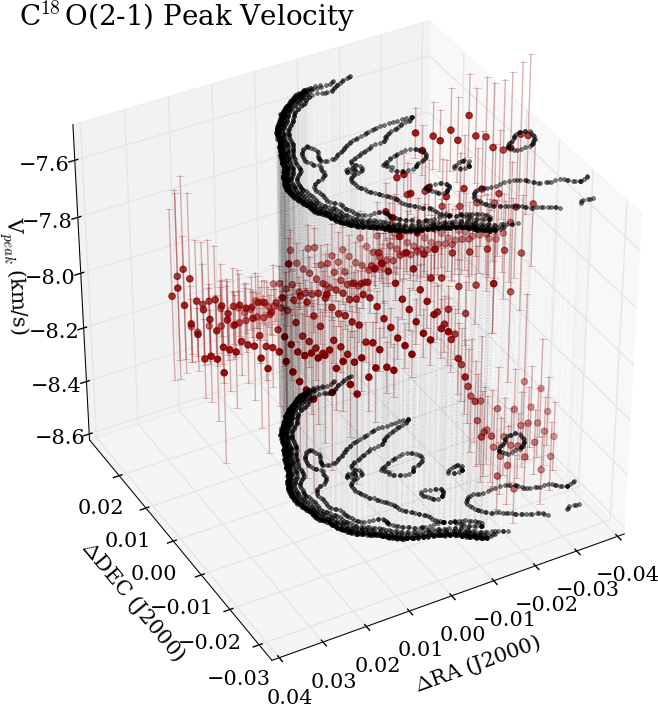} &
\includegraphics[width=0.3\linewidth]{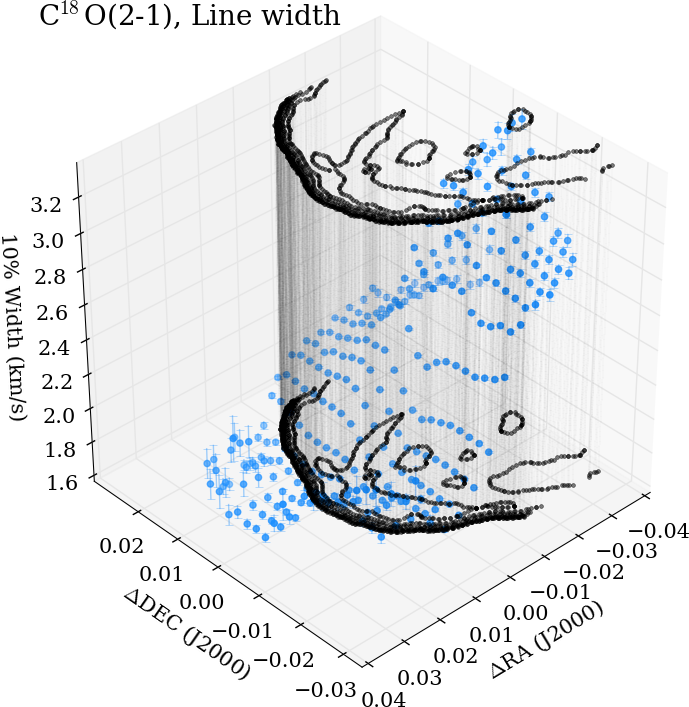} \\
\includegraphics[width=0.3\linewidth]{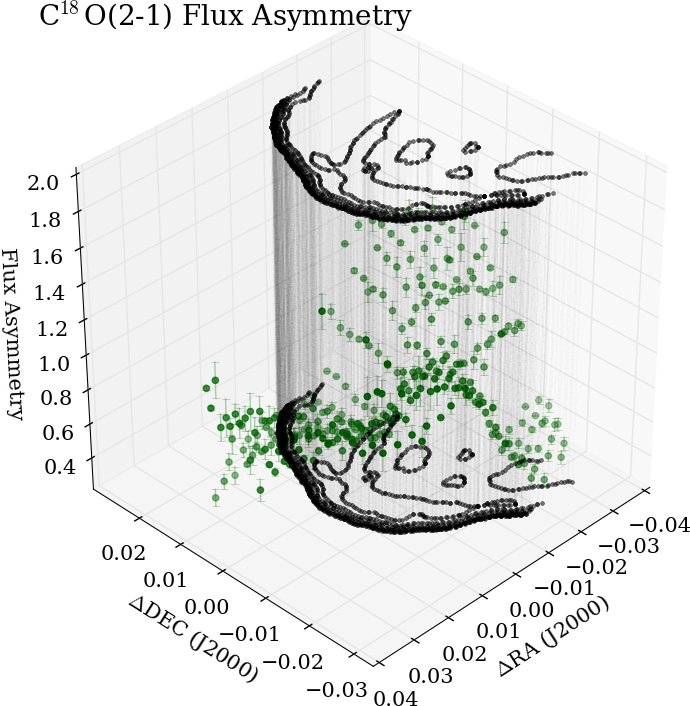} &
\includegraphics[width=0.3\linewidth]{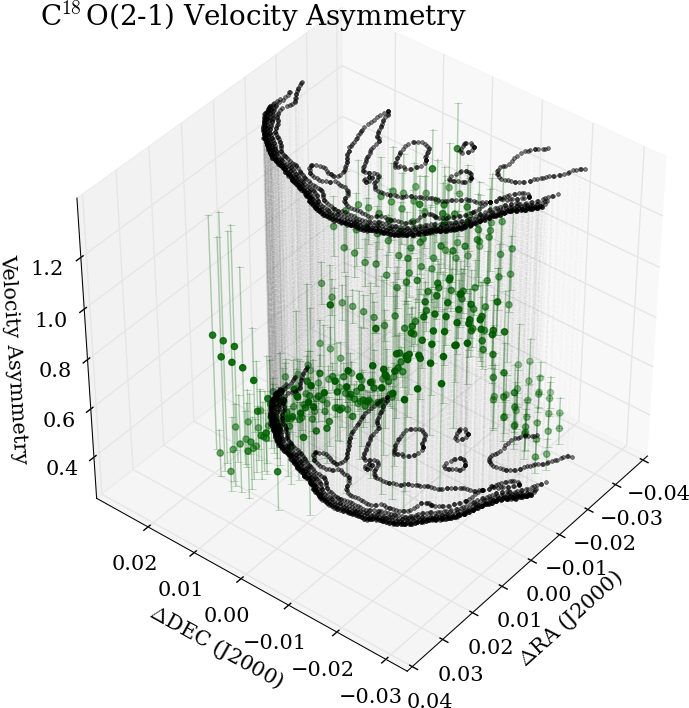} \\
\end{tabular}
\caption{Pixel-by-pixel rendering of the integrated flux, peak velocity, line width,
flux asymmetry, and velocity asymmetry (colored dots) for the C$^{18}$O(2-1) line. The black contours mark the
structures seen in the globule with \emph{Herschel}/PACS 70~$\mu$m. \label{C18O_3D-fig}}
\end{figure*}

The $^{13}$CO emission rises sharply at the globule rim, and is mostly saturated towards the region around IC1396A-PACS-1,
so that the $^{13}$CO line parameters do not tell much about the structure of the cloud.
The C$^{18}$O(2-1) line (see Figure \ref{C18O_3D-fig}) clearly reveals the location of the
density peak behind the cloud rim, showing no signs of CO depletion despite the presence of nitrogenated species, probably due to the large beam. The C$^{18}$O peak velocity shows a strong gradient towards the SW
clump, marking the 3D structure of the globule. The line width also increases in the same direction,
as do the line asymmetry for velocity and flux,
indicating a clear change in the velocity pattern and a distinct behavior, compared to
the main part of the globule. The velocity asymmetry and, to a lesser extent, the flux
asymmetry, also reveal more symmetric, less turbulent lines towards the densest parts of
the region. The tendency to find blue-shited asymmetry in the lines could be an
indication of ongoing RDI collapse. The difference in line asymmetry between the inner and the outer part of the globule (Figure \ref{C18O_3D-fig}) suggests that the globule is being eroded mostly in the outermost parts. The increased width towards the south-west and the fact that the line peak shifts by about half a km/s in this direction also points to higher velocities (probably due to evaporation of the near-side of the globule) in this region. 

Several PDR-related lines are detected
within our EMIR field. This includes a very weak SiO line
and CN emission. The SiO line is very weak,
but detected towards the PDR, the Class 0 object, and the extended 
southern rim (Rim S). The CN line is remarkably broad and globally redshifted, compared to all other lines,
having a typical central velocity of $\sim$-6.5 km/s and a 10\% width in the 3-4 km/s range. 
A global redshift is typically observed in optical lines towards the tips of photodissociated pillar-like
structures \citep{mcleod15}, which is observed in the multi-Gaussian analysis of the CN line.
The central velocities of all other strong lines, appearing around -7.8 km/s, 
are instead tracking denser material inside the globule.
The structure of the CN line is also quite
stable throughout the globule, being usually well-fitted by two individual
Gaussian components, although since
the line is weak, there is a substantial uncertainty in the line parameters. The first component peaks at
$\sim-$6.4 km/s, while the second peaks at $\sim-$8.1 km/s. The blueshifted component
dominates towards the Class 0 source and the northern side of the PDR rim. It is also the
narrowest component ($\sim$0.6 km/s), and its width
increases towards the globule rim. The redshifted component
is very variable in intensity and line width ($\sim$0.9-1.2 km/s), without showing any discernible 
pattern except for being stronger around the PDR region. Both components can be
interpreted as the redshifted and blueshifted sides of a photoevaporation flow, where the most distant side of
the globule would be more photoevaporated, as expected if the far-side receives more illumination by
HD\,206267. The velocity difference compared to the rest of lines would be about 0.6 km/s.

\begin{figure*}
\centering
\begin{tabular}{ccc}
\includegraphics[width=0.3\linewidth]{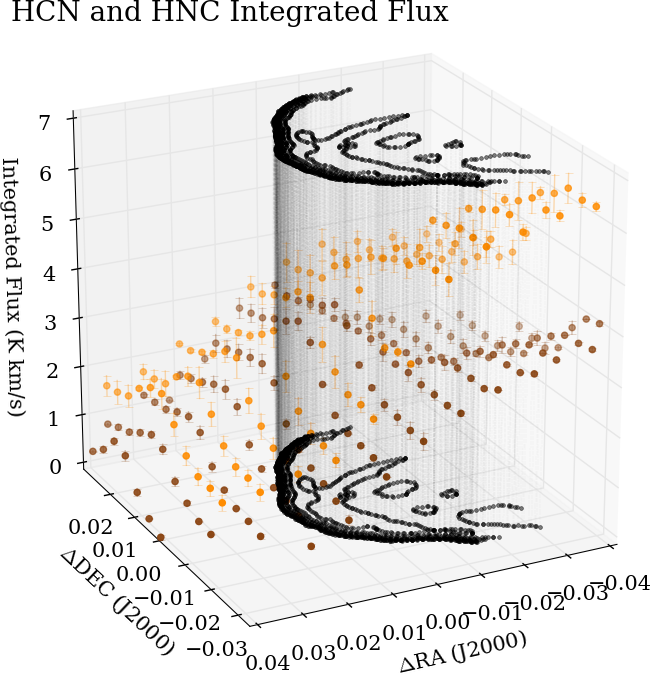} &
\includegraphics[width=0.3\linewidth]{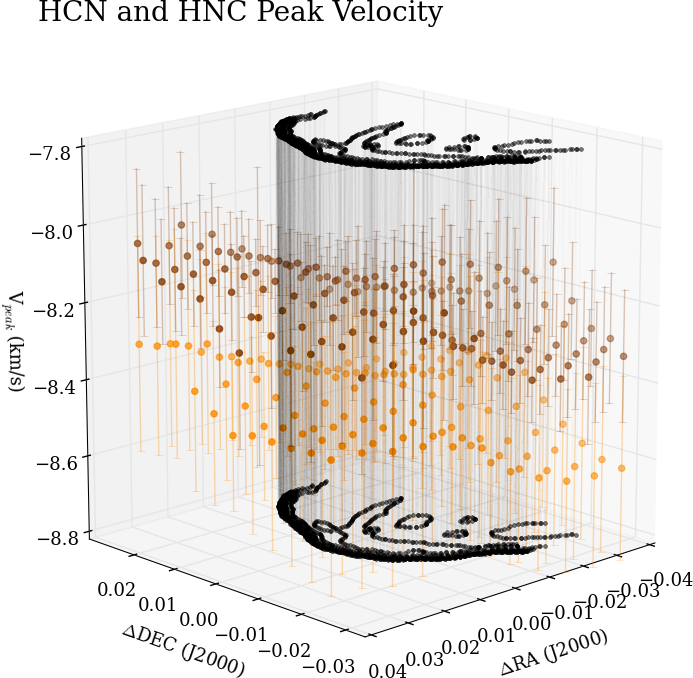} & 
\includegraphics[width=0.3\linewidth]{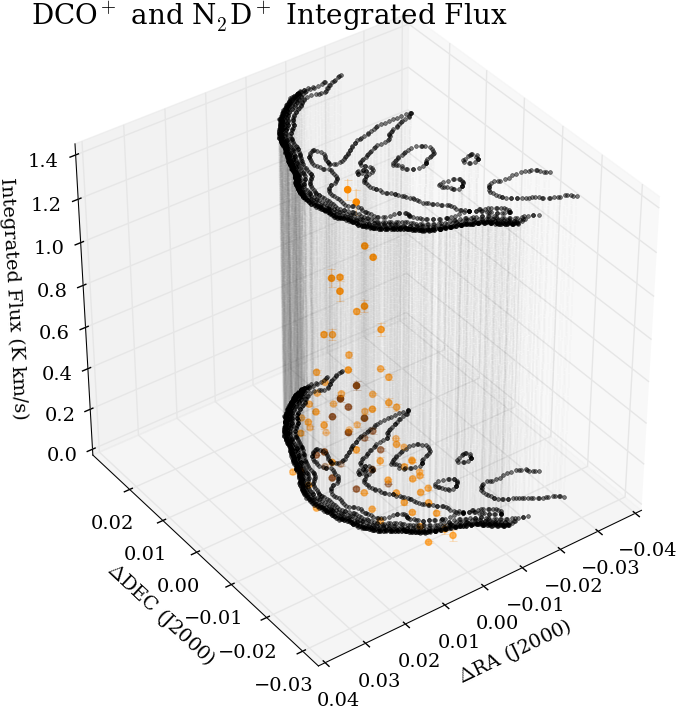} \\
\end{tabular}
\caption{Pixel-by-pixel rendering of the integrated flux (left) and peak velocity (middle) of the 
HCN(1-0) (yellow dots) and HNC(1-0) (brown dots) lines, and the integrated flux for the DCO$^+$(3-2) 
(yellow dots) and N$_2$H$^+$(3-2) (brown dots) lines (right). The black contours mark the
structures seen in the globule with \emph{Herschel}/PACS 70~$\mu$m. Note that only the densest parts of the cloud produce significant emission
in the high-density tracers N$_2$H$^+$ and DCO$^+$. \label{HCN_HNC_3D-fig}}
\end{figure*}

The HCN and HNC lines have similar line profiles, with the difference in flux (related to the temperature)
pointed out previously. The 3D maps show that HCN is quite uniform over the globule, while HNC is clearly stronger 
towards the rim, as expected (see Figure \ref{HCN_HNC_3D-fig} left). Both lines have a slight blue-dominated asymmetry, and
their peak velocity does not vary much over the mapped area, but shows a relatively
constant offsets of about 0.3-0.4 km/s (see Figure \ref{HCN_HNC_3D-fig} middle), likely an optical depth effect. HCN is more blueshifted, with a velocity similar to the blueshifted CN component
(which may hint also a photoevaporation origin),
and it tends to be slightly broader than HNC.

The HCO$^+$ line intensity increases steeply towards the globule and and also shows a trend to
redder velocities as we move to the western part of the globule, which could be a 
sign of material loss 
and globule evaporation.
The peak is at the rim, as expected for a high density tracer.
Higher-density lines such as DCO$^+$ and N$_2$D$^+$ are detectable only in the densest part of the globule (Figure \ref{HCN_HNC_3D-fig} right).
They have blue-dominated profiles characteristic of infall, and a peak velocity of $-$7.8 km/s (DCO$^+$)
and -8.0 km/s (N$_2$D$^+$). The lines are relatively narrow and weak, making it difficult to analyze the line parameters in detail.

\subsection{The velocity history of IC\,1396A within Tr\,37 \label{Tr37-past}}

IC1396A-PACS-1 lies at the interface beween a dense cloud and a
PDR as shown in Figure \ref{slice-fig}. 
From existing narrow-band images,
IC\,1396A is a dark globule \citep{osterbrock89,barentsen11}, illuminated by 
HD~206267 mostly from the east. 
The [S II] images also reveal that although the rim is significantly
bright, the [S II] emission from the globule is not significantly different from
what is observed towards the H II region 
\citep[see Figure \ref{slice-fig} and][]{sicilia13}.
This, combined with the thin rim observed in the \emph{Herschel} continuum data (Paper I), 
suggests that the pair HD~206267 and IC~1396A are {at a low angle with respect
to the plane of the sky}. As mentioned in \ref{3d-sect}, both 
the velocity and the line width of the CN line are suggestive
of an origin in the photoevaporated material around the edge of the globule.
The distance between the globule tip and the massive star must be at least equal to the
projected distance of 4.9 pc (considering the revised cluster distance of 945 pc). This distance is
significantly larger than typically observed towards other photoevaporated globules \citep[such as the Pillars of Creation, at $\sim$2 pc;][]{mcleod15}.

The velocities of the clouds around the Tr~37/IC~1396 region are very
diverse. 
\citet{wilson53} measured a radial velocity of $-$7.8 km/s for HD~206267.
This result was later revised by \citet{stickland95}, who obtained
a (highly uncertain) systemic velocity around $-$24.8 km/s and signatures of spectroscopic
binarity. Velocities derived for the CO molecular line
emission of the nebular structures in the whole region
by \citet{patel95} range between V$_{LSR}$=+5 to $-$9 km/s, with IC~1396A 
having V$_{LSR}$=$-$7.9 km/s. The velocity we derive for IC~1396A is fully consistent
with that value, V$_{LSR}$=$-$7.8 km/s on average, and with previous estimates of the velocity of
the globule \citep{morgan10}. Lines with various optical
depths show sligthly different velocities, suggesting a small variation by $\sim$0.3 km/s
throughout different depths. In particular, lines associated with
the photodissociation region are clearly redshifted compared to  high-density 
tracers, consistent with the surface of the cloud being eroded.

The radial velocities of the parental cluster Tr~37 are significantly different
from those of IC\,1396 by about 7 km/s.
\citet{sicilia06b} used optical spectroscopy to measure the
radial velocities of T Tauri stars (and their spread) in Tr~37, obtaining a typical radial velocity 
cz= $-$15.0$\pm$3.6 km/s  (V$_{LSR}\sim -$1 km/s), clearly distinct from that of IC\,1396A. 
Molecular-line $^{12}$CO emission has been also detected towards a star
with a remarkably massive disk \citep[GM~Cep;][]{sicilia08},
confirming the  V$_{LSR}$=$-$1$\pm$2 km/s, in agreement with the optical
mean velocity of the Tr~37 cluster.
The $^{12}$CO weak component 
centered at $-$0.7 km/s (Figure \ref{faint-fig}), distributed relatively uniformly over IC1396A
is thus consistent with a diffuse component tracing low density
remnant material of the original cloud that gave rise to Tr\,37.
Combined with the velocity in the plane of the sky measured with 
Gaia, we obtain a velocity difference of about 8 km/s in magnitude between IC\,1396A and Tr\,37.
This distinct velocity suggest different origins for both the main
Tr~37 cluster and IC~1396A within parts of the many clouds that constitute the Cep~OB2 region
\citep{patel98}. 
With this in mind, the connection between Tr~37 and IC\,1396A has to be revised.  Exploring
the causes of this velocity offset is a first step in this direction. 

Gravity is unlikely to provide the observed velocity difference.
If we consider the approximate mass of the main Tr~37 cluster to be around 1000 M$_\odot$\footnote{Based 
on the known members and considering that the region is essentially
devoid of gas now \citep{sicilia06a,sicilia13,barentsen11}}, the velocity
expected if IC~1396A were being gravitationally pulled by the main cluster would be 
of the order of 1 km/s. Even if we assume a 20$\times$ larger mass to account for the gas
that is now dispersed (for a star formation efficiency of 5\%), the gravitational 
pull would not exceed 4.5 km/s. This is clearly insufficient to explain the disparate
velocities of Tr~37 and IC~1396A. 
If we consider infall towards the larger mass of the L1149 and L1143 clouds,
located 50 pc to the east of Tr~37 and with a total mass of 25200\,M$_\odot$ \citep{patel98},
the maximum infall velocity expected would only average $\sim$1.5 km/s.

The natural expansion of H \,II regions can provide
velocity differences between massive star clusters and their surrounding clouds.
Considering the expansion of a Str\"{o}mgren sphere \citep{mckee84,osterbrock89}, the
Str\"{o}mgren radius R$_{St}$ is given by
\begin{equation}
R_{St}=67 (S_{49}/n_{m}^2)^{1/3} \,pc,  \label{Rstromgren-eq}
\end{equation}
where S$_{49}$ is the rate of ionizing photons emitted by the star in units of 10$^{49}$ photons/s
\citep[$\sim$1.5 for an O6.5 star like HD~206267;][]{sternberg03} and n$_m$ is the mean number
density of the cloud.
For an isothermal sound speed of c$_s$=10 km/s, this translates to an expansion time of
\begin{equation}
t_{St}= R_{St}/c_s = 6.5\times10^6 (S_{49}/n_m^2)^{1/3} yr. \label{tstromgren-eq}
\end{equation}
For IC~1396A/HD~206267 and a typical number density of 10 cm$^{-3}$ \citep{patel98}, 
these values correspond to $\sim$16.5 pc in about 1.4 Myr, which is consistent
with the $\sim$15 pc ring of bright-rimmed clouds observed around HD~206267 \citep{patel95,barentsen11}.
Denser environments would result in smaller radii. For instance, for a distance of 2-2.5 pc,
a minimum density of 200 cm$^{-3}$ would be needed to keep the ionization front from propagating inside
a globule, which is well below the estimated density in the thickest parts of IC~1396A.

Rocket acceleration \citep{elmegreen76b} of the globule
by the effect of the radiation from HD~206267 could induce a velocity away
from the ionizing source in a globule like IC\,1396A. The magnitude of the imparted velocity
can be up to several tens of km/s as predicted by some models
of BRC evolution \citep{miao06,miao09} and observed in other globules \citep{mcleod15}.
The global velocity observed for IC\,1396A is much smaller than expected
for  sustained rocket acceleration,
although some degree of  rocket acceleration cannot be excluded.
To acquire such velocity, globules usually need to
be at most at 2-2.5 pc distance from the ionizing star. 
One issue with strong rocket effect acceleration is that it predicts line-of-sight
velocities on opposite sides of the globule tail and body comparable 
to the bulk motion of the globule \citep{miao06,miao09}. Such velocity spreads have
been observed in photoevaporated globules \citep{mcleod15}, but
they are inconsistent with the small velocity spread observed for IC\,1396A in molecular
lines. Our EMIR data detect 
variations up to $\sim$0.3 km/s for lines with different optical depths, up to 2 km/s between 
the red and blue peaks of
the CN line, and up to $\pm$4 km/s for low-density gas according to the line wings of CO (see Section \ref{3d-sect} and Figure \ref{faint-fig}),  which would suggest 
that rocket acceleration accounts for up to a few km/s at most.
In this respect, the simple expansion models for H\,II regions from \citet{patel95},
provide velocities more in agreement with the current observations, 
with a rapid acceleration that would quickly stabilize after 1-2 Myr
around a value of $\sim$4-6 km/s \citep[see][]{patel95}. This value is similar
to the velocity on the plane of the sky measured with Gaia, but on the lower
side for the observed bulk velocity difference of 8 km/s. The
angle between the velocity vector and the plane of the sky is $\sim$60 degrees, 
pointing to the west and towards the observer. 

If the observed velocity is exclusively caused by rocket acceleration
with respect to HD\,206267, the angle between the observer, HD\,206267, and
the globule would be about 30 degrees. This would result in a current
distance to the star of nearly 10 pc, which is close to the limit at which the O6.5
star can supply enough ionizing radiation to significantly affect the
globule \citep[$\sim$11 pc for an O6.5 star with an ionizing flux
of 1.5$\times$10$^{49}$ photons s$^{-1}$; ][]{sternberg03,bisbas11}. 
In fact, the usual requirement of a ionizing photon flux at least of 
1$\times$10$^9$ photons cm$^{-2}$s$^{-1}$ 
\citep{bisbas11,miao09} imposes a minimum angle of about 30 degrees between IC\,1396A
and the line-of-sight (LOS) towards HD\,206267 for the massive star to have
a significant effect on the globule. If IC\,1396A had been moving
at a constant rate, the observed velocities would place it at only 1.1 pc
from HD\,206267 1 Myr ago. But for such a close distance, the total
rocket acceleration expected would be rather of the order or tens of km/s \citep{miao06,miao09}
instead of the 8 km/s observed. The age of HD\,206267 and
the Tr37 cluster are 3-4 Myr\citep{sicilia05,getman12},  
which poses an additional problem to the idea that the globule is being radially
accelerated. A very close (1-2 pc) globule near a massive star would not only be subjet to strong
acceleration and would
lose matter at a rate of at least several tens of solar masses per Myr \citep{mcleod16},
halving the mass of a globule like IC\,1396A in about 0.3-0.4 Myr \citep{miao06,miao09} for a ionizing photon flux of $\sim$1.5e11 cm$^{-2}$s$^{-1}$, a value that is reached at about 1 pc distance from HD\,206267.
This means that if the velocity is entirely radial, it cannot have been constant
in time (scenario 1), or that the
velocity observed cannot be exclusively due to rocket acceleration in the radial direction
away from the star (scenario 2). 

While the expansion of the H II region is expected to slow down with time, 
rocket acceleration is expected to increase in time, with the velocity saturating when the
HII region is about 10 pc and $\sim$2 Myr old \citep{patel95}. In scenario 1,
IC\,1396A would be currently at about $\sim$10 pc from the ionizing source ($\sim$30
degree angle with respect to the LOS towards HD\,206267). This  
distance is comparable to the models for the expansion of an H\,II region for 
an age between 2-3.5 Myr for
Tr\,37 \citep[see Fig 11 in][]{patel95}, but the bulk velocity of 8 km/s
is about a factor of 2 higher than expected, so 
a short time at the observed velocity would bring IC\,1396A
too close to HD\,206267 for the velocity to be so low compared to the
predictions of rocket acceleration.

For scenario 2 and with the angle limitation imposed by the 
current minimum ionizing flux mentioned above, 
the only possibility would be sto have a larger angle so that IC\,1396A would be 
closer to the plane of
the sky and thus currently closer to HD\,206267. A closer current distance is also required to 
cause the observed overpressure \citep{morgan04}. 
 This would help understanding
the H$\alpha$ and [S II] emission, and would reconcile
the velocity in the plane of the sky with the expansion velocities of the H II region. 
Nevertheless, scenario 2 also requires an extra velocity component in the 
radial direction with an origin other than rocket acceleration and/or expansion
within the H\,II region. 

\begin{figure}
\centering
\includegraphics[width=0.99\linewidth]{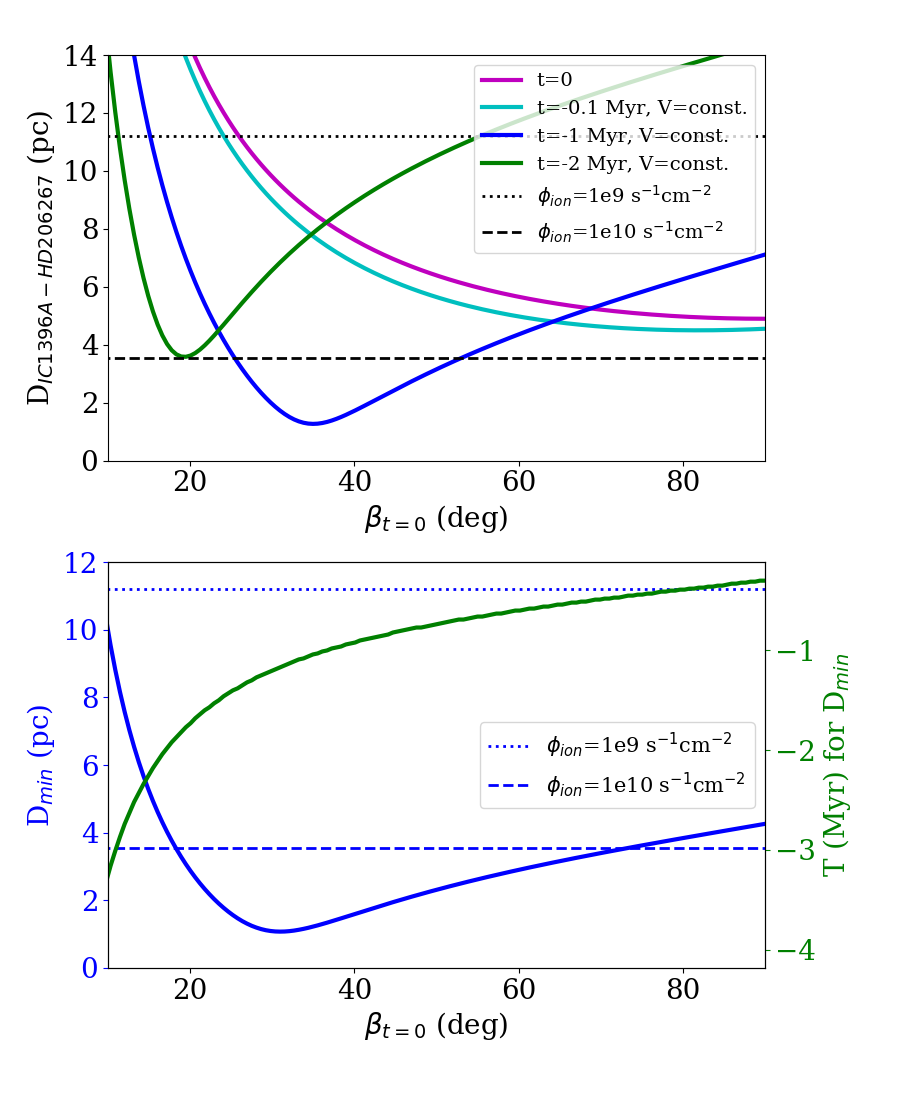}
\caption{Top: Distance between IC\,1396A and HD\,206267 as a 
function of the current angle with respect to the line-of-sight ($\beta_{t=0}$).
Bottom: Minimum distance between IC\,1396A and HD\,206267 (blue) and time at which 
the minimum distance was reached (green) as a function of the current angle with respect to 
the line-of-sight ($\beta_{t=0}$). The time is given in Myr with the present time being
set at 0 Myr, which makes all quantities negative. The velocity is considered
as constant and equal to the velocity observed in the plane of the sky and radial
directions, which is a good approximation for the past 2-1.5 Myr (see text). The
distances at which the
ionizing photon flux is equal to 1e9 and 1e10 cm$^{-2}$s$^{-1}$ are 
marked as horizontal dotted and dashed lines, respectively. \label{dmin-fig}}
\end{figure}

Although the expansion velocity is not expected to be constant, 
H\,II region expansion velocities change slowly after the first 1-1.5 Myr \citep[see][ Figure 11]{patel95}.
For a cluster age between 3-4 Myr,
this means that we can consider the velocity constant in the past 1.5-2 Myr, which
are the most relevant for the formation of the population associated with IC\,1396A
since their ages are estimated to be $\sim$1 Myr \citep{sicilia05, getman12}.
For each possible angle between the line-of-sight, HD\,206267, and IC\,1396A, we can
thus calculate the current distance and extrapolate the distances back in time
within the last $<$2 Myr. The minimum distance to the ionizing source, together with
the time at which this minimum distance was reached, are also easy to estimate.
Figure \ref{dmin-fig} shows the results.

The distance to the ionizing source is also a critical parameter to estimate
the time needed to disrupt the cloud and whether this time allows for
substantial star formation, and depends on the position of the cloud (see Figure
\ref{dmin-fig}, bottom). For example, for the minimum angle of 30 degrees with respect to the line-of-sight,
 a minimum distance of 1.1 pc would have been reached some 1.2 Myr ago
(consistent with the Class I/II population), but the distance would have only increased
since then, reaching values at the present time that are nearly incompatible with the
formation of the Class 0 source.

\subsection{Triggered, sequential, and multi-episode star formation}

The revised position and velocity history affects the triggering scenario by
affecting the location of the globule with respect to the ionizing source
in time.  Scenario 1, in which IC\,1396A had been initially much closer to the massive
star, results in the problem that the velocities observed are
only slightly higher than the expansion velocities of the H II region, which
is unexpected from rocket acceleration and from what is observed in
other BRC \citep[e.g.][]{mcleod15}. Scenario 1 also would suggest that
the globule was much closer to the ionizing source 1-2 Myr ago at the time when the most evolved population inside IC\,1396A 
formed \citep[V390 Cep, 14-141, and the Class I/II objects associated with the globule, 
$\sim$1-2 Myr ago;][]{reach04,sicilia05b,sicilia06a}.

Although IC\,1396A is significantly more massive than most of the globules used
in RDI simulations, the distance behind the ionization rim at which stars are forming 
is expected to increase with decreasing ionizing flux \citep{bisbas11}. 
Since the Class 0 source is formed closest to the ionization rim, 
it is likely that the closest distance between IC\,1396A and HD\,206267 was
reached at about the time of the formation of IC1396A-PACS-1, thus $\leq$0.1 Myr. 
With all associated uncertainties, this suggests an angle over 70 degrees,
which would place the line between  IC\,1396A and HD\,206267 nearly on the
plane of the sky. Therefore, if we consider that the Class I population in the globule is further away
from the globule rim than the Class 0 source, we would expect that the 
globule was located further away when the 1-2 Myr old population formed.
This would result in an angle with respect to the line-of-sight of at least
37 degrees (assuming a 2Myr age for this first population of stars that are
now Class II and Class I sources) or 68 degrees (assuming
a 1 Myr age for the first triggered population), which may compromise the
possibilities of triggering for the older globule population.

For scenario 2, a velocity component not related to expansion away from HD\,206267
is needed. The relatively low global velocity observed can be explained if
the globule has been in the past at distances from the ionizing source too large
to induce any significant acceleration. This is achieved if the current position
of the line between HD\,206267 and the globule is relatively close to the plane of
the sky ($<$20 degrees with respect to the plane of the sky),
which is in agreement with the appearance in H$\alpha$ and [S II] as
mentioned above. Scenario 2 nevertheless introduces the problem that the globule may have been
too far away from HD\,206267 at the time when the Class I/II population formed, for
triggered star formation to be efficient.
Thus the options remain that either rocket acceleration and RDI triggered star 
formation models are poorly understood, or that the only star for which consistent
evidence of triggered star formation exist is the Class 0 source IC1396A-PACS-1.

These dynamical and age considerations make it difficult to confirm the triggering of the 
oldest populations within the globule. If IC\,1396A were located beyond the $\phi_{ion}$=1e9 cm$^{-2}$s$^{-1}$
line about $\sim$1-2 Myr ago, when V390 Cep and 14-141 formed,
it would have been beyond the ionization front and too far for RDI \citep[e.g.][]{bisbas11}).
Figure \ref{dmin-fig} shows that this is unlikely for 1 Myr at any angle, but it could have 
been the case for angles lower than about 55 degrees with respect to the LOS for stars 
with ages $\geq$2 Myr old.

Considering that the intermediate-mass star V390 Cep, belonging to the most evolved
globule population, is carving its own ionization hole within the globule
without any noticeable external effects, 
star formation seems to have started relatively unaffected by external influences (either because
of a large distance, or because of being very highly embedded).
The second burst of star formation, resulting in the embedded Class I
and Class II sources
\citep{reach04,sicilia06a,getman12} appear distributed behind the head of the globule, which
together with the age difference, suggest a RDI-triggering origin \citep{getman12}, although the
dynamical evidence seems so far elusive. Further, deeper observations targetting the lower
density gas and distributed population in the globule could be used to search for further dynamical clues.
As the globule moved closer to HD~206267, the ionizing front would have started
to erode the less dense parts of the globule, carving it in until it reached the
denser core around the Class 0 object IC1396A-PACS-1. The pressure of the ionization rim
is now acting directly on the core, as the 70~$\mu$m filament shows, and
as the density, temperature, and turbulence tracers indicate.
Therefore, IC1396A-PACS-1 is the only object in IC~1396A for which an unambiguous
sign of triggering exists.

IC\,1396A is comparable to the models of 30\,M$_\odot$
clouds in the presence of a perpendicular UV radiation field by \citet{kinnear14,kinnear15}. 
The aspect ratio and curvature of the models
are very similar to the observed two-arm structure detected with \emph{Herschel} (Paper I) and extending on both sides of
the Class 0 source. The collapse time, once exposed to the UV field, would be well below 1 Myr, 
which is also consistent
with the recent formation of the Class 0 source compared to the age of HD~206267 and the Tr\,37 cluster.
The large mass in the dense core containing 
IC1396A-PACS-1 is consistent with
the higher-density models in \citet{kinnear14}, although symmetric clouds tend to produce two 
cores at the two gravitational 
foci or ends of the linear structure, which is not the case here. Asymmetries in the direction of the
incident UV radiation with respect to the initial dense cloud are also seen in the models if the initial shape 
of the cloud is elongated along
the direction of the incomming UV radiation \citep{kinnear15}. The observed differences between the
dense core and the rest of the cloud (about 3-4 orders of magnitude in column density) are also consistent with 
the models, as well as the relatively lower density towards the center of the globule, away from the
edges, due to gravitational focusing \citep{kinnear15}. The initial
density structure of the globule could also contribute to the North-South density difference at the tip of
IC\,1396A.

\subsection{The future of IC\,1396A}

Considering the current densities and
temperatures in the globule, IC1396A-PACS-1 is also likely to be the last star-forming episode in
IC~1396A. Although the NIKA data reveal several smaller clumps in IC~1396A,
the temperature maps show that the gas there is significantly warmer, less dense,
and contains less mass altogether. Higher-resolution observations will be needed to search for additional
star formation, but considering the lack of \emph{Herschel} counterparts in these regions,
it is highly unlikely that more intermediate-mass stars are forming
in the globule. Formation of faint, low-mass objects cannot be ruled out at this stage
in the denser southern clump and around IC1396A-PACS-1. 
IC1396A-PACS-1
is the predecessor of an intermediate (probably B-type) star \citep[similar to the intermediate-mass
star formation inferred in other BRCs, e.g.][]{morgan04}. If more low-mass stars were
to form within the clump, it could evolve into
an irregularly-shaped mini-cluster, similar to others observed in other
Tr~37, such as those around the binary B star CCDM+5734
\citep{sicilia15}.

The NIKA image reveals high density clumps associated with some of the objects: around
the Class II, M1-type, emission line star 21365947+5731349, around
the IR source behind 14-141 (at the rim of the V390 Cep hole),
on the Class I protostars 21355793+5729099 and 21360798+5726371, surrounding
the protostars $\epsilon$ and $\delta$ ,
and between the Class I protostar 21364596+5729339 and  
the K6 Class II object 21364762+5729540 
\citep[see reference to the objects in][]{sicilia06a,reach04,sicilia13}. 
All these objects are Class I and Class II 
sources without high-mass envelopes, suggesting that the NIKA emission traces extended 
cloud material rather than envelopes. There is a further
faint emission to the west of the K6 diskless star 11-2487, although
lacking velocity information, it may correspond to material in
the background, not associated with 
Tr\,37.
Although the [S II] forbidden-line map revealed several outflows associated with
the low-mass population at the tip of IC\,1396A \citep{sicilia13}, 
we do not find any evidence of molecular line emission associated with any of the rest of Class I protostars
and T Tauri stars within our EMIR field. This is probably due to both a combination of the
complexity of the molecular emission from the cloud and PDR, the large beam of IRAM, and the fact that all
the objects are low-mass, with low envelope masses and relatively low accretion rates.

The fact that the bulk of the Class I and Class II sources are
associated with the less dense material  in the globule suggests that
a substantial amount of cloud mass has been removed from around the objects.
Given that the average ages for objects in the globule is 1 Myr \citep{sicilia06a,getman12},
a substantial gas heating and removal needs to have taken place in a relatively
short timescale.
The connection of the known stars with the warmer parts of the globule
may also suggest small-scale feedback, as observed around V390 Cep, but 
since our temperature estimates are dominated by the highest temperatures
along the line-of-sight, we cannot exclude that far-IR emission from the stars themselves
contributes to artificially increase the temperature estimates in their surroundings.

The approximate mass loss due to photoevaporation can be estimated following a similar method
to \citet{mcleod15}. In their approximation, the total mass loss dM/dt is given by
\begin{equation}
dM/dt = v \rho A,
\end{equation} 
where $v$ is the velocity of the photoevaporated gas, $\rho$ is the mass density
in the photoevaporating regions, and $A$ is the
area of the globule. We consider the velocity of the photoevaporating gas to be
of the order of 0.6 km/s (from the CN two-component data) and up to $\pm$4 km/s for the lowest density
gas with $^{12}CO$ emission only. The typical column density
of the globule in the outermost parts that are subject to
photoevaporation can be estimated to be of the order of 7$\times 10^{21}$ cm$^{-2}$ (from the N$_{\rm H}$ map), measuring it towards the small, well-defined
25 arcsec-radius clump at 21:36:56 +57:31:58\footnote{Choosing other low-density regions does not change the
result by more than 30\%} can be used to estimate an approximate number density
$\rho$=2$\times 10^4$ cm$^{-3}$.  Note that although the number density of the globule
can be significantly larger in the innermost parts, if the photoevaporation is regulated by the photon rate emitted by HD\,206267, the rate will not significantly change once 
the low-density parts of the globule are eroded and denser parts 
are exposed. Summing over the whole globule area of about 1.5 pc$^2$, we estimate a mass loss rate of the
order of 4$\times 10^{-4}$ M$_\odot$/yr, which would result in the evaporation of the whole globule on a timescale below 0.5 Myr. This value is on the high-end
for a globule around a O6.5 star \citep{mcleod15}, being also suggestive of a relatively 
close distance to the ionizing source.

In this calculation, we must note that our dust-derived mass is a factor of 3 smaller than
previous estimates \citep{patel98}, which is probably caused by the fact that our observations are sensitive 
only to the densest regions. A globule mass of $\sim$100-300 M$_\odot$ is only a few times the mass 
in stars, based on the stellar census \citep[most of them low-mass stars, see][]{reach04,sicilia06a,getman12}. Although
some of the sources may be seen in projection, at least half of them show significantly higher extinction and/or are in an earlier evolutionary phase than Tr37, suggestive of association with the globule population. This is an indication
that the original globule must have been significantly more massive than it is currently, to
result in a reasonable star-forming rate, as already noted by \citet{getman12}. 
The fact that substantial mass loss needs to have happened during the life of 
the globule is also a sign that, despite
distance, external photoevaporation must have played an important role in shaping the globule and 
its population.

From all the above considerations, the environment around IC1396A-PACS-1 comprises
the densest and coldest part of the cloud. 
Its high density has likely contributed to keep it cold and isolated from  previous star formation episodes until the
pressure from the expanding H~II region was enough to start triggering the
collapse. The most massive optically-visible star within IC~1396A is V 390 Cep, classified as
an intermediate-mass star \citep[and thus likely 2-4 M$_\odot$,][]{contreras02,siess00}. As an object with 
A/F spectral type, it produces 
a very low ionizing flux compared
to massive stars, but its
action on the cloud is clearly noticeable on small spatial scales, having resulted
in the opening of the eye-shaped hole in the globule (see Figures \ref{nika-fig} and \ref{gaia-fig}). 
The temperature map also reveals local heating and short-scale feedback by the embedded
population.

Local feedback in the cloud can be estimated using the the A/F-type star V390 Cep as example. With the
data on stellar properties and magnitudes \citep[SIMBAD,][]{wenger00}, 
we can estimate its ionizing photon rate to be S$\sim$1.5$\times$10$^{44}$ s$^{-1}$.
For this calculation, we assume the same spectrum as for a white dwarf with the same temperature, and scale
the result to the total stellar luminosity of 2.42 L$_\odot$ \citep{hills73},
which gives us S$_{49}\sim$1.5$\times$10$^{-5}$ s$^{-1}$. 
This flux would result in the opening of a small
Stromgren sphere, about 0.15-0.07 pc in radius (depending on the initial cloud density),
in good agreement with the observed 0.08 pc size of the hole.
The hole is a sign that the local feedback by low-mass stars can be of
importance in a cloud undergoing crowded star-formation, and its associated gas removal and
heating may prevent further star formation on small, nearby scales and significantly 
contribute to cloud heating and mass loss in the absence of (or far away from) massive stars.
The fate of the dense structure along the southern rim and its possibilities of further low-mass
star formation may also depend on this 
small-scale feedback, since the rest of
the globule is warmer and significantly less dense.

\section{Summary and conclusions \label{conclu}}

Our results are summarized below:
\begin{itemize}
\item Our NIKA and EMIR data image the IC~1396A globule at millimeter wavelengths with
significantly increased resolution  compared to previous millimeter
studies. We use the dust and gas data to trace the temperature, 
density, and dynamics of the region, investigating the origin and triggers of the star
formation episodes within the globule. Combining the IRAM data with Gaia DR2 
velocities and proper motions, we complete the 3D picture of the region.
\item Emission suggestive of warm carbon chain chemistry (WCCC) corino is found
towards IC1396A-PACS-1, consistent with the location of the source in an environment with
high UV irradiation, although some contamination from PDR lines cannot be
excluded at present due to the large IRAM beam. 
The observed chemistry and, in particular, the presence of CCS associated 
to the source, place the object
among the youngest protostars known. Further interferometric observations will be needed to confirm the properties and structure of the source. 
\item The head of the globule where IC1396A-PACS-1 is located appears significantly more
massive than the rest of the cloud, containing about 1/4 of the mass inferred from continuum data,
and is significantly denser and colder than the rest of IC1396A.
\item From the temperature, density, and dynamical analysis, we
conclude that the star formation episode that produced IC1396A-PACS-1 is probably
the last one in IC\,1396A, at least regarding intermediate-mass stars, given that 
the region around the Class 0 object is the
last one that appears sufficiently dense, cold, and quiescent. 
\item The dynamics of the cloud and its surroundings reveal a new picture of the region.
The main velocity of the globule is significantly different from the velocity of Tr\,37 ($-$7.8 vs $-$1 km/s), considering
both the radial velocities of the stars and of the surrounding gas. 
We detect a faint $^{12}$CO component at $-$0.7 km/s, which probably corresponds to the remnants of 
the cloud that formed Tr\,37.  The Gaia DR2 stellar proper motions, together with
the gas radial velocities, reveal a total velocity of $\sim$8 km/s for IC\,1396A
with respect to Tr\,37, which is too low compared to expected rocket acceleration
if the globule had been much closer to HD\,206267 in the past.
 Depeding on the angle of the globule with respect to the massive star and the LOS, the 
distance between IC\,1396A and the ionizing source varies, affecting the possibilities of
triggering for the older, 1-2 Myr population. 
 This result prompts us to revise the 
history of triggered and sequential star formation in the region, and also 
demonstrates the power of 
combined radial velocity and Gaia data to understand cluster structure 
and formation history with unprecedented detail.
\item The formation of V~390~Cep and 14-141 seems to have occurred rather undisturbed. RDI triggering is the most likely formation mechanism for the Class 0 source IC1396A-PACS-1, and can be inferred from the blue-dominated profiles of the molecular lines observed towards the source and the globule. 
For the ClassI/II/III population inside IC~1396A, their location is suggestive of 
RDI \citep{getman12}, 
although the large distance of the globule at the time the stars were formed could have been a problem for triggering.
\item Finding several star-forming episodes within a structure as small as the
IC~1396A globule ($\sim$0.5pc in size) suggests that various modes of
sequential fragmentation and star formation can
occur in clouds, even on very small spatial scales. The population emerging from such an
scenario can thus have age differences of 1-2 Myr, which also would include differences
in the evolutionary stage of their disks. Moreover, having various star-forming episodes potentially
triggered by different mechanisms may also result in a variety of initial conditions for neighbouring
protostars, leading to potential differences in disk formation
and affecting their future evolution.
\end{itemize}

Acknowledgments:  We thank the editor, M. Tafalla, for his help during the submission, and the referee for his/her thoughtful comments that contributed to clarify
the paper. We are very grateful to the personnel at the IRAM telescope, for their
help and for making the observation stays so enjoyable. In particular, we are very grateful
to M. Gonz\'{a}lez, N. Billot, and I. Hermelo for their help with the observation preparation and
the observations at the telescope. We also thank Sylvie Cabrit, Ana L\'{o}pez-Sepulcre, 
Nina Sartorio and Ian Bonnell for 
their comments and discussion. 
This work is based on observations carried out under project number 166-13 with the IRAM 30m telescope. IRAM is supported by INSU/CNRS (France), MPG (Germany) and IGN (Spain).
This work has made use of data from the European Space Agency (ESA) mission
{\it Gaia} (\url{https://www.cosmos.esa.int/gaia}), processed by the {\it Gaia}
Data Processing and Analysis Consortium (DPAC,
\url{https://www.cosmos.esa.int/web/gaia/dpac/consortium}). Funding for the DPAC
has been provided by national institutions, in particular the institutions
participating in the {\it Gaia} Multilateral Agreement.
V.R. is partly supported by the European Union's Horizon 2020 research and innovation programme under the Marie Sklodowska-Curie grant agreement No 664931. 
This research has made use of the SIMBAD database,
operated at CDS, Strasbourg, France. This research was carried out in part at the Jet Propulsion Laboratory, 
which is operated for NASA by the California Institute of Technology.
This research includes analysis carried out with the CASSIS 
software and the JPL (http://spec.jpl.nasa.gov/) spectroscopic database. CASSIS has been developed by IRAP-UPS/CNRS (http://cassis.irap.omp.eu). This work makes use of the NIST Diatomic Spectral database (https://www.nist.gov/pml/diatomic-spectral-database).This work is partly based on observations obtained with the \emph{Herschel Space
Telescope} within open time proposal ``Disk dispersal in Cep OB2'',
OT1\_asicilia\_1. \emph{Herschel} is an ESA space observatory with science in-
struments provided by European-led PI consortia and with important
participation from NASA.

\Online
\onecolumn
\begin{appendix}

\section{Line list \label{app1}}

The table below contains the entire list of lines detected in the spectra, listed according to their
frequency and observed towards the center of the field. Note that for very weak lines, it is not
possible to infer the spatial distribution of the emission. The S/N is variable in throughout the dataset, especially considering that some of the regions were
covered by more than one setup, as listed in Table \ref{obs-table}. The presence of multiple carbon 
chains results in the classification of the source as a potential hot corino.

\begin{longtable}{lccl} 
\caption{\label{lowreso-table} Lines identified in the low-resolution spectra as observed in the
region-averaged spectra. 
Multiplets and marginal detections are accordingly labelled. Weak and uncertain lines are 
marked with ':' . } \\                
\hline \hline                        
$\nu_{obs}$ & Species & T$_{peak}$ & Notes \\
(GHz)	    &         &  (K) 	   &    \\
\hline                             
\endfirsthead
\caption{Continued.}\\
 \hline \hline                        
$\nu_{obs}$ & Species & T$_{peak}$ & Notes \\
(GHz)	    &         &  (K) 	   &    \\
\hline                             
\endhead   
{\bf E0}\\
84.411 & $^{34}$SO & 0.02: & marginal \\
84.521 & CH$_3$OH & 0.31  & \\
85.139 & OCS & 0.05  & \\
85.162 & HC$^{18}$O$^+$ & 0.045  & \\
85.339 & c-C$_3$H$_2$(0) & 0.41  & \\
85.348 & HCS$^+$ & 0.095  & \\
85.634 & C$_4$H & 0.07 & \\ 
85.672 & C$_4$H & 0.06 & \\ 
85.926 & NH$_2$D & 0.14 & multiplet \\
86.053 & HC$^{15}$N & 0.11  & \\
86.094 & SO & 0.65  & \\
86.181 & CCS & 0.035  & \\
86.340/.338/.342 & H$^{13}$CN & 0.28/0.16/0.07  & multiplet \\
86.670 & HCO  & 0.24   & \\
86.708 & HCO   & 0.17  & \\
86.754 & H$^{13}$CO$^+$ & 0.52  & \\
86.806 & HCO  & 0.06  & \\
86.847 & SiO & 0.06 & PDR tracer \\
87.091 & HN$^{13}$CO & 0.13  & \\
87.284 & C$_2$H & 0.14  & \\
87.317 & C$_2$H & 1.12 & \\
87.328 & C$_2$H  & 0.55  & \\
87.402 & C$_2$H & 0.57  & \\
87.407 & C$_2$H & 0.27  & \\
87.446 & C$_2$H & 0.11  & \\
87.925 & HNCO & 0.12  & \\
88.631/.630/.634 & HCN & 2.35/1.07/0.80  & multiplet \\
88.646 &  H$^{18}$ONO  & 0.085  & \\
88.866 & H$^{15}$NC & 0.040  & \\
89.045 & C$_3$N & 0.013: & \\
89.065 & C$_3$N & 0.015  & \\
89.188 & HCO$^+$ & 2.60  & \\
89.488 & HOC$^+$ & 0.030  & \\
89.579 & HCOOH   & 0.024  & \\
89.861 & HCOOH  & 0.012 & faint but clear\\
90.664 & HNC & 1.65  & \\
90.686 & CCS & 0.029  & \\
90.979 & HC$_3$N & 0.058  & \\
91.494/.498 &  c-C$_3$H   & 0.044/0.024 &  \\
91.700 &  c-C$_3$H & 0.029  & line at .692 marginal \\
91.980 & CH$_3$CN & 0.024 & faint but detected\\
91.985 & CH$_3$CN & 0.065 & \\
91.987 & CH$_3$CN & 0.068 & some other CH$_3$CN lines are not detected\\
92.494 & $^{13}$CS & 0.170  & \\
93.174/.172/.176 & N$_2$H$^+$ & 0.34/0.29/0.13  & multiplet \\
93.267 & S$^{18}$O  & 0.025  & \\
93.581 & CH$_3$CHO  & 0.041  & \\
93.595 & CH$_3$CHO  & 0.037  & \\
93.870 & CCS & 0.065  & \\
95.150 & C$_4$H & 0.060  & \\
95.169 & CH$_3$OH & 0.020: & marginal\\
95.189 & C$_4$H & 0.065  & \\
95.914 & CH$_3$OH & 0.03 & \\
95.947 & CH$_3$CHO & 0.04 & \\
95.963 & CH$_3$CHO & 0.05 & \\
96.412 & C$^{34}$S & 0.44 & \\
96.632 & CH$_3$CHO & 0.02: & marginal \\
96.755 & CH$_3$OH & 0.04 & \\
97.172 & C$^{33}$S & 0.07 & \\
97.301 & OCS & 0.05 & \\
97.583 & CH$_3$OH & 0.04 & \\
97.715 & $^{34}$SO & 0.25 & \\
97.981 & CS(2-1) & 2.5 & \\
97.995 & C$_3$H & 0.03 & 2 peaks \\
98.012 & C$_3$H & 0.03 & 2 peaks \\
98.260 & $^{13}$CO image  & 0.18 &  \\
98.863 & CH$_3$CHO & 0.05 & \\
98.901 & CH$_3$CHO & 0.04 & \\
99.300 & SO & 2.45 & \\
99.866 & CCS & 0.025 & \\
100.029 & SO & 0.025 & \\
100.094 & CH$_2$CO  & 0.05 & \\
100.193 & CH$_2$CO  & 0.025 & \\
\\
{\bf E2}\\
\\
215.221 & SO & 0.41 & \\
215.839 &  S$^{34}$O & 0.073  & \\
216.112 & DCO$^+$ & 0.36 & \\
216.278 & c-C$_3$H$_2$  & 0.065 & \\
217.237 &  DCN  & 0.02: & marginal, very uncertain\\
217.822 & c-C$_3$H$_2$ & 0.085 & 2 peaks\\
217.940 & c-C$_3$H$_2$  & 0.025: & weak but detected\\
218.222 & H$_2$CO & 1.10 & \\
218.440 & CH$_3$OH  & 0.16 & \\
218.476 & H$_2$CO & 0.17 & \\
218.760 & H$_2$CO & 0.18 & \\
219.560 & C$^{18}$O & 4.25 & \\
219.908 & H$_2^{13}$CO  & 0.055 & \\
219.949 & SO  & 1.3 & \\
220.398 & $^{13}$CO  & 12.3 & \\
225.698 & H$_2$CO & 1.25 & \\
226.314 & CN, v = 0, 1   & 0.060: & marginal, broad \\
226.342 & CN, v = 0, 1   & 0.060: & marginal, CN line at .323 not detected\\
226.360 & CN, v = 0, 1   & 0.13 & \\
226.632 & CN & 0.17 & CN at .616 not detected\\
226.659 & CN & 0.36 & \\
226.664 & CN & 0.15 & \\
226.679 & CN & 0.17 & \\
226.875 & CN & 0.6 & complex, blend .874 and .876 \\
226.887 & CN & 0.14 & \\
226.892 & CN & 0.18 & \\
228.910 & DNC   & 0.10 & \\
230.537 & $^{12}$CO & 20 & \\
231.322 & NND$^+$  & 0.11  & \\
\hline                                             
\end{longtable}

\section{Velocity integrated maps for strong lines \label{velo-app}}

\begin{figure*}
\centering
\begin{tabular}{cc}
\includegraphics[width=0.44\linewidth]{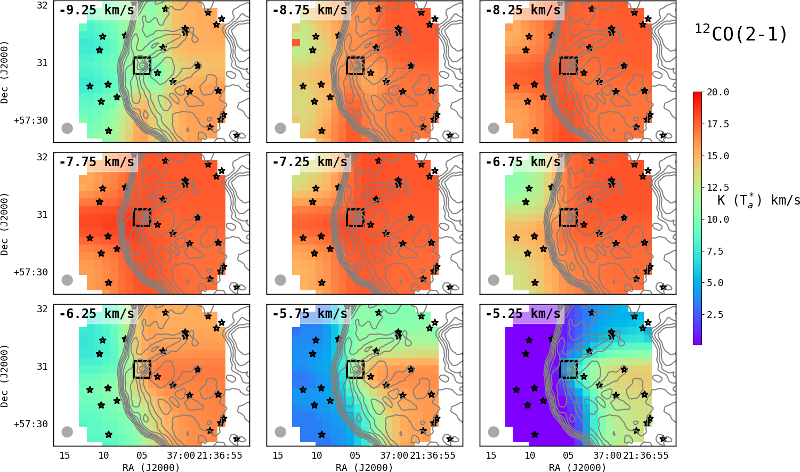} &
\includegraphics[width=0.44\linewidth]{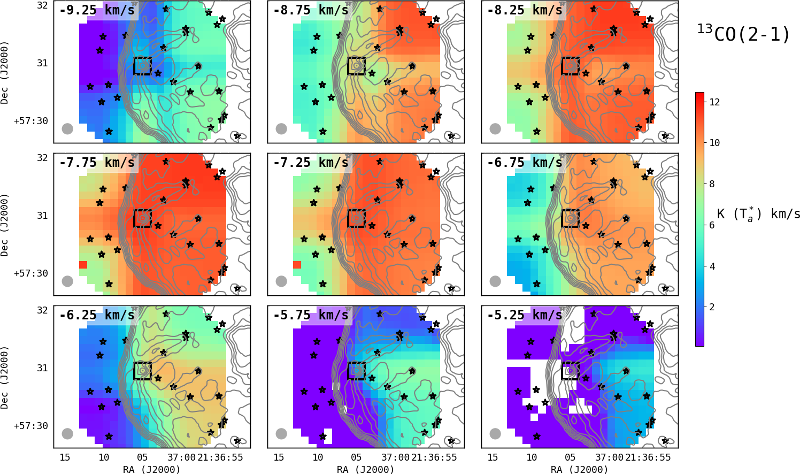} \\
\includegraphics[width=0.44\linewidth]{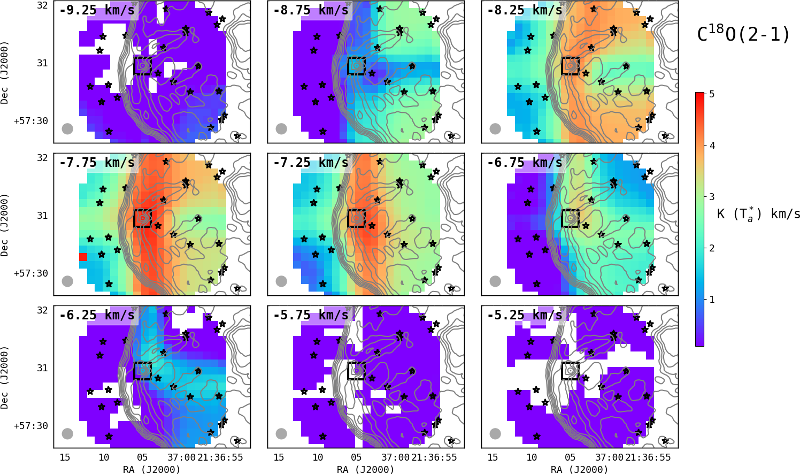} &
\includegraphics[width=0.44\linewidth]{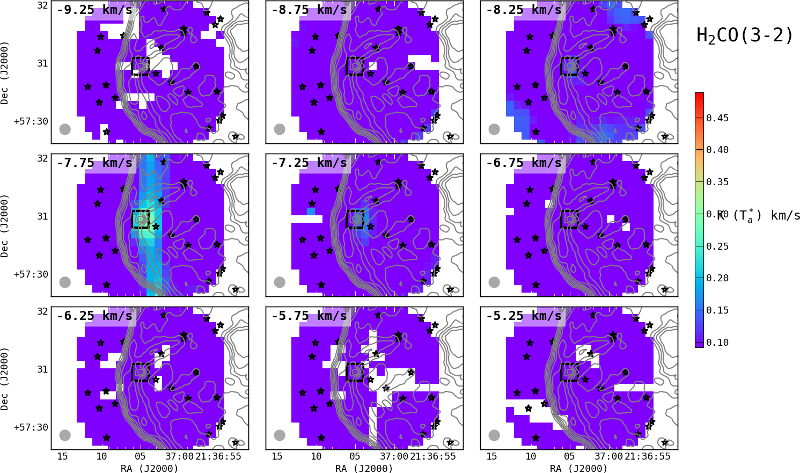} \\
\end{tabular}
\caption{Velocity integrated bit maps. The colors show the line intensity integrated around 0.5 km/s per velocity bin between 3$\sigma$ and the maximum. White is used in case no line emission beyond 3$\sigma$ is detected in the bin (the bottom of the color scale would correspond usually to the noise level). Note that the velocity bins displayed are much larger than the spectral resolution for the sake of space, as most of the lines are detected in between tens to over two hundred channels. The contours mark for reference the \emph{Herschel} 70~$\mu$m emission on a log scale between 0.1 and 2 Jy/beam.
The beam size is shown for each line. Symbols: IC1396A-PACS-1 (large black square), other YSO candidates (small, black star symbols).
From left to right, top to bottom: $^{12}$CO, $^{13}$CO, C$^{18}$O, H$_2$CO.\label{momentCOs-fig}}
\end{figure*}

\begin{figure*}
\centering
\begin{tabular}{cc}
\includegraphics[width=0.44\linewidth]{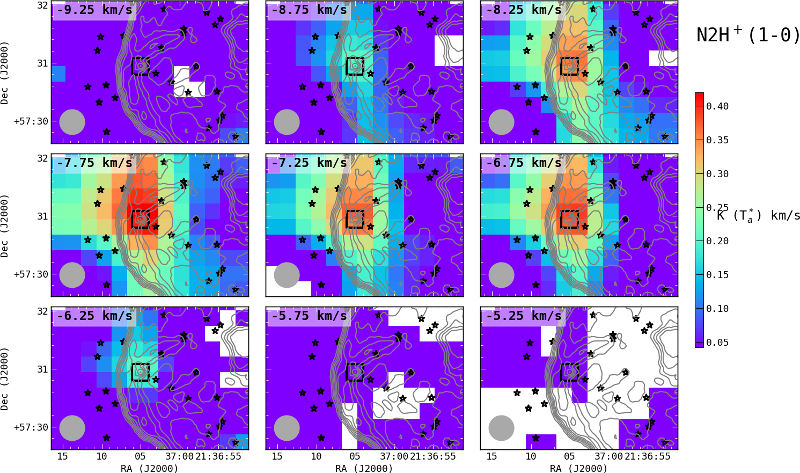} &
\includegraphics[width=0.44\linewidth]{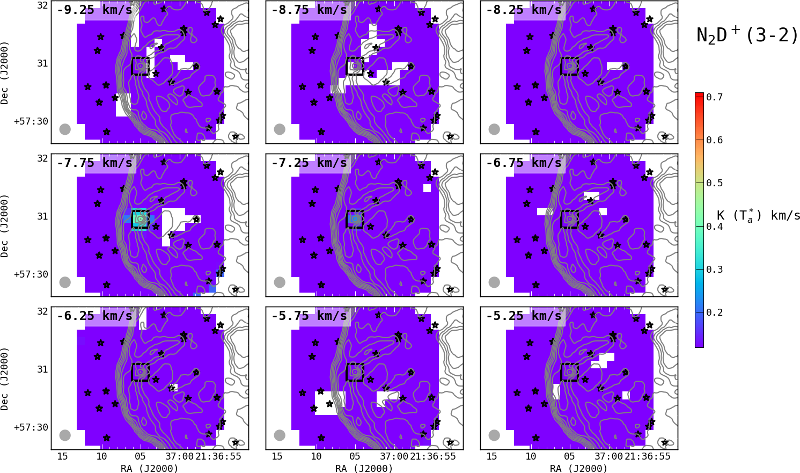} \\
\includegraphics[width=0.44\linewidth]{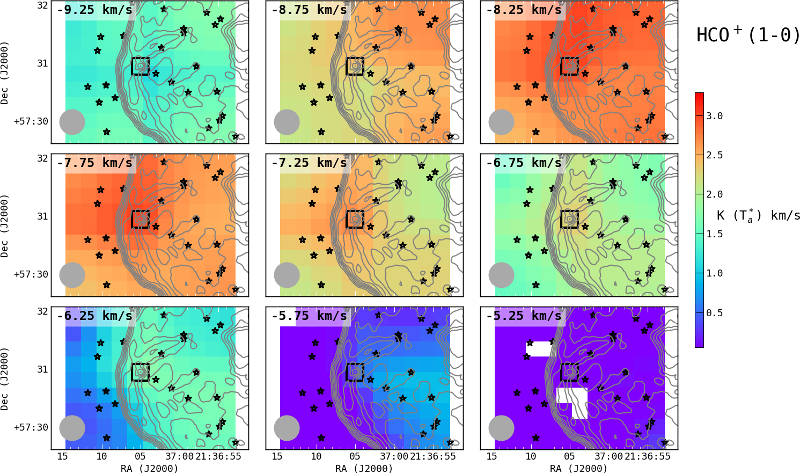} &
\includegraphics[width=0.44\linewidth]{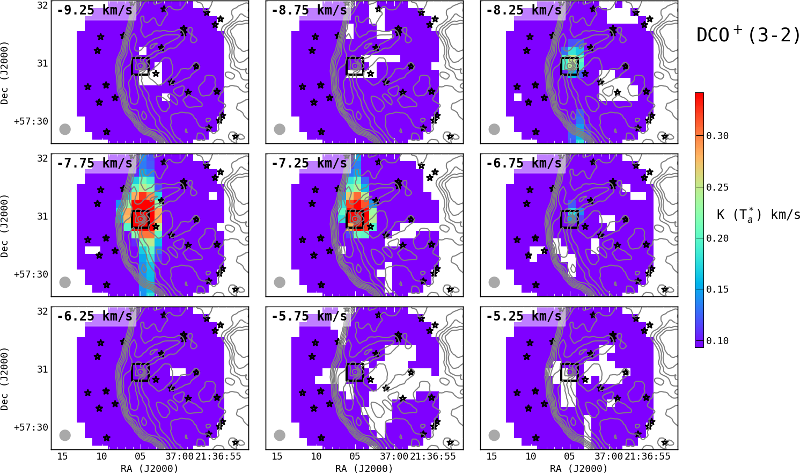} \\
\end{tabular}
\caption{Velocity integrated bit maps.  The colors show the line intensity integrated around 0.5 km/s per velocity bin between 3$\sigma$ and the maximum. White is used in case no line emission beyond 3$\sigma$ is detected in the bin (the bottom of the color scale would correspond usually to the noise level).  Note that the velocity bins displayed are much larger than the spectral resolution for the sake of space, as most of the lines are detected in between tens to over two hundred channels. The contours mark for reference the \emph{Herschel} 70~$\mu$m  emission on a log scale between 0.1 and 2 Jy/beam.
The beam size is shown for each line. Symbols: IC1396A-PACS-1 (large black square), other YSO candidates (small, black star symbols).
From left to right, top to bottom: N$_2$H$^+$, N$_2$D$^+$, HCO$^+$, DCO$^+$. \label{momentN2Hp-fig}}
\end{figure*}

\begin{figure*}
\centering
\begin{tabular}{cc}
\includegraphics[width=0.44\linewidth]{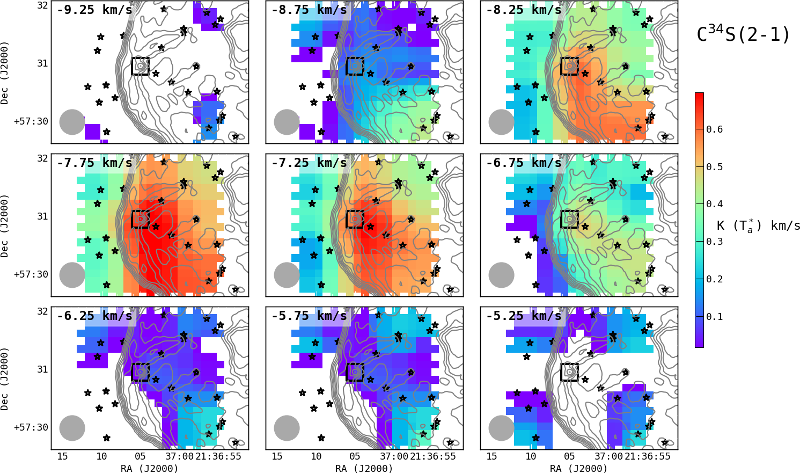} &
\includegraphics[width=0.44\linewidth]{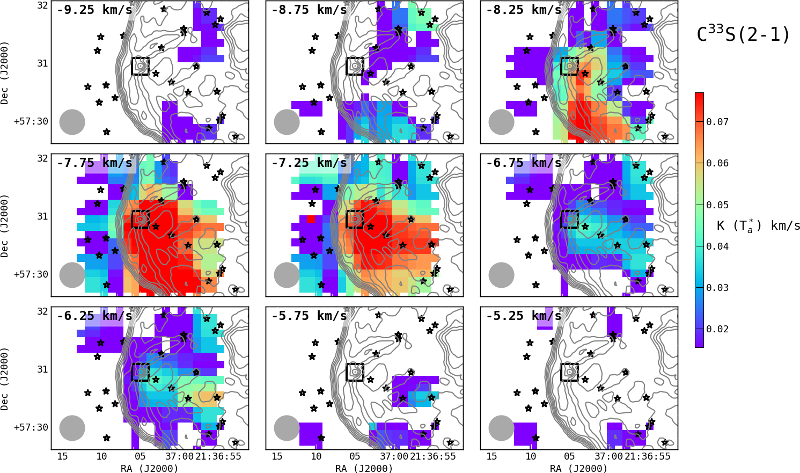} \\
\includegraphics[width=0.44\linewidth]{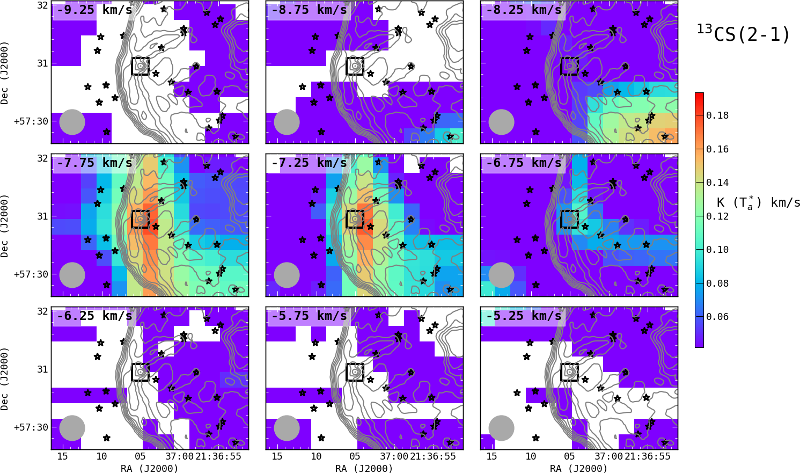} & 
\includegraphics[width=0.44\linewidth]{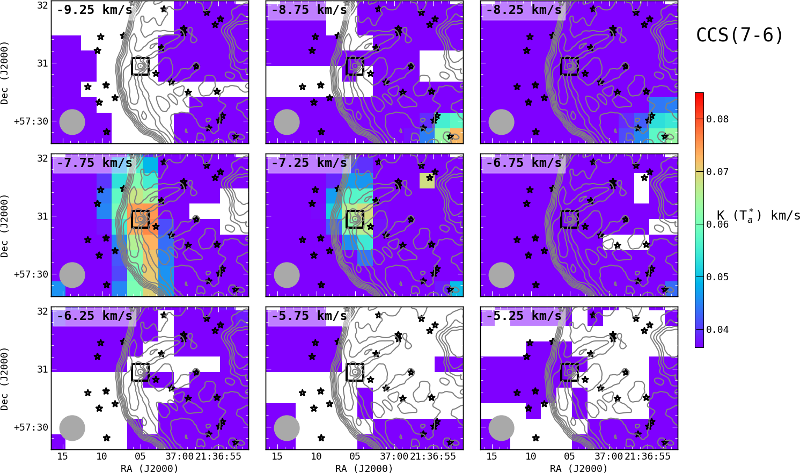} \\
\end{tabular}
\caption{Velocity integrated bit maps. The colors show the line intensity integrated around 0.5 km/s per velocity bin between 3$\sigma$ and the maximum. White is used in case no line emission beyond 3$\sigma$ is detected in the bin (the bottom of the color scale would correspond usually to the noise level).  Note that the velocity bins displayed are much larger than the spectral resolution for the sake of space, as most of the lines are detected in between tens to over two hundred channels. The contours mark for reference the \emph{Herschel} 70~$\mu$m  emission on a log scale between 0.1 and 2 Jy/beam.
The beam size is shown for each line. Symbols: IC1396A-PACS-1 (large black square), other YSO candidates (small, black star symbols).
From left to right, top to bottom: C$^{34}$S, C$^{33}$S, $^{13}$CS, CCS. \label{momentCS-fig}}
\end{figure*}

\begin{figure*}
\centering
\begin{tabular}{cc}
\includegraphics[width=0.44\linewidth]{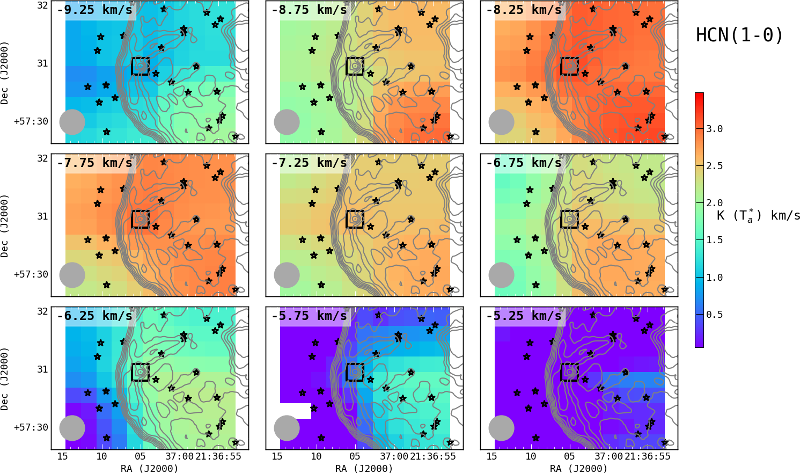} &
\includegraphics[width=0.44\linewidth]{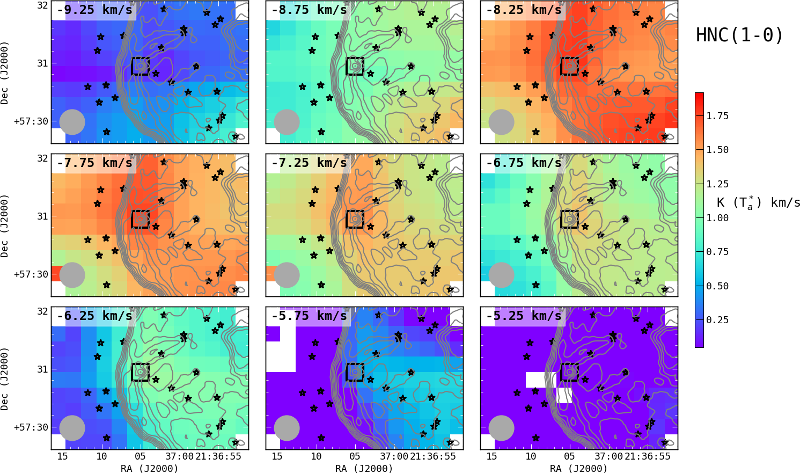} \\
\includegraphics[width=0.44\linewidth]{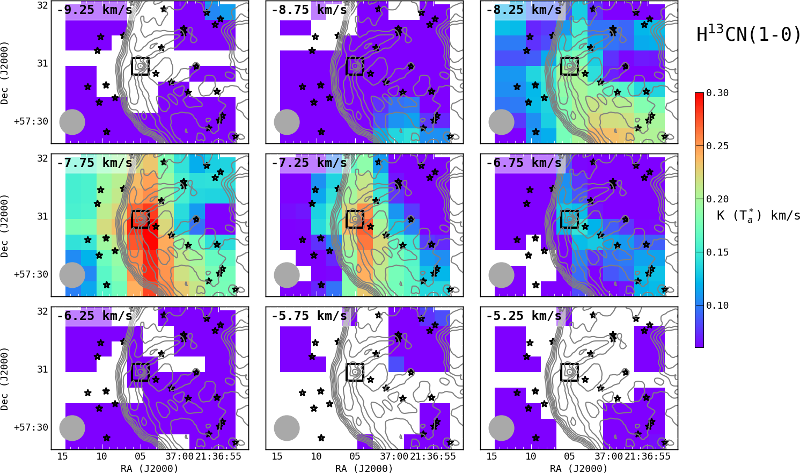} &
\includegraphics[width=0.44\linewidth]{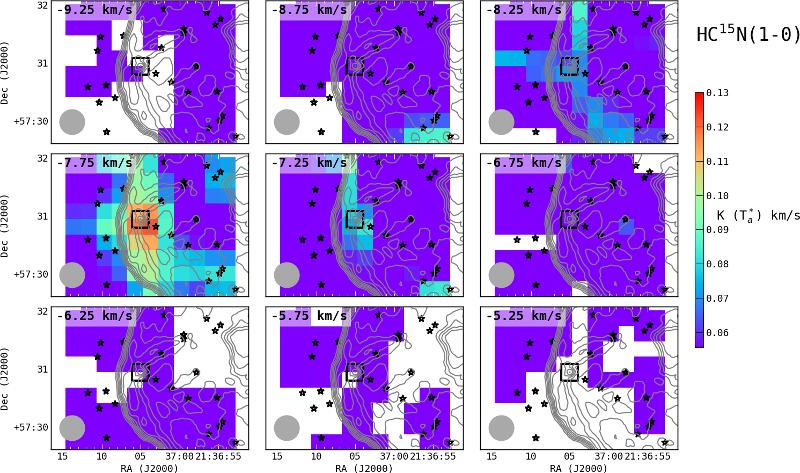} \\
\end{tabular}
\caption{Velocity integrated bit maps. The colors show the line intensity integrated around 0.5 km/s per velocity bin between 3$\sigma$ and the maximum. White is used in case no line emission beyond 3$\sigma$ is detected in the bin (the bottom of the color scale would correspond usually to the noise level).  Note that the velocity bins displayed are much larger than the spectral resolution for the sake of space, as most of the lines are detected in between tens to over two hundred channels. The contours mark for reference the \emph{Herschel} 70~$\mu$m  emission on a log scale between 0.1 and 2 Jy/beam.
The beam size is shown for each line. Symbols: IC1396A-PACS-1 (large black square), other YSO candidates (small, black star symbols).
From left to right, top to bottom: HCN, HNC, H$^{13}$CN, HC$^{15}$N. \label{momentHNC-fig}}
\end{figure*}

\begin{figure*}
\centering
\begin{tabular}{cc}
\includegraphics[width=0.44\linewidth]{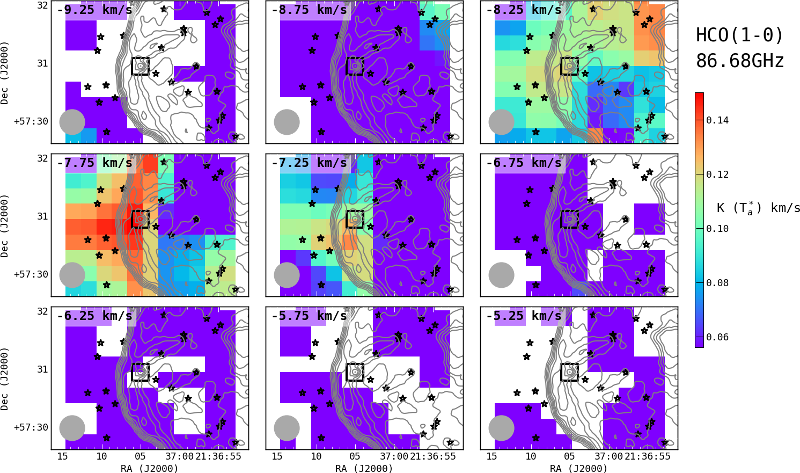} &
\includegraphics[width=0.44\linewidth]{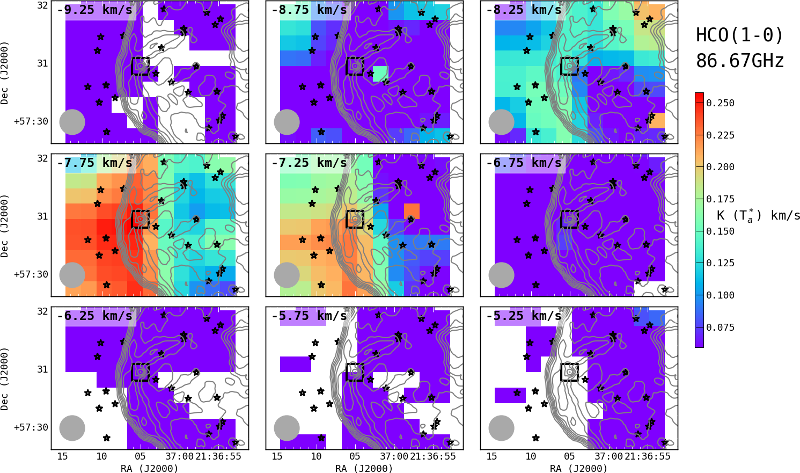} \\
\includegraphics[width=0.44\linewidth]{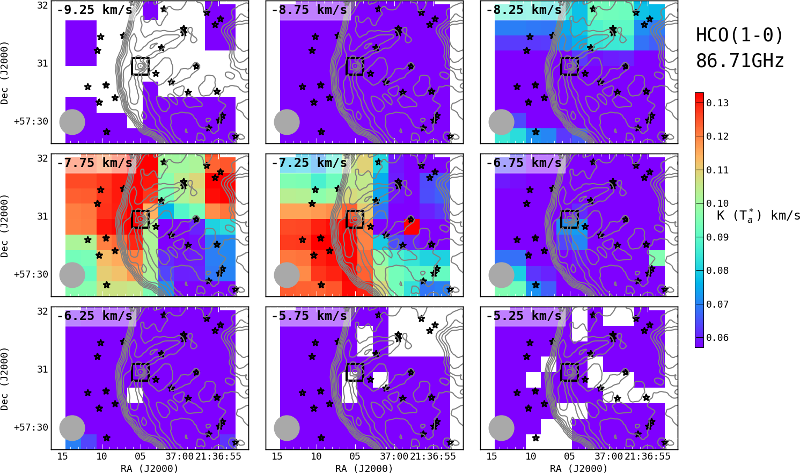} \\
\end{tabular}
\caption{Velocity integrated bit maps. The colors show the line intensity integrated around 0.5 km/s per velocity bin between 3$\sigma$ and the maximum. White is used in case no line emission beyond 3$\sigma$ is detected in the bin (the bottom of the color scale would correspond usually to the noise level).  Note that the velocity bins displayed are much larger than the spectral resolution for the sake of space, as most of the lines are detected in between tens to over two hundred channels. The contours mark for reference the \emph{Herschel} 70~$\mu$m  emission on a log scale between 0.1 and 2 Jy/beam.
The beam size is shown for each line. Symbols: IC1396A-PACS-1 (large black square), other YSO candidates (small, black star symbols).
From left to right, top to bottom: HCO, HCO 86.67 GHz, HCO 86.71 GHz.\label{momentHCO-fig}}
\end{figure*}

\begin{figure*}
\centering
\begin{tabular}{cc}
\includegraphics[width=0.44\linewidth]{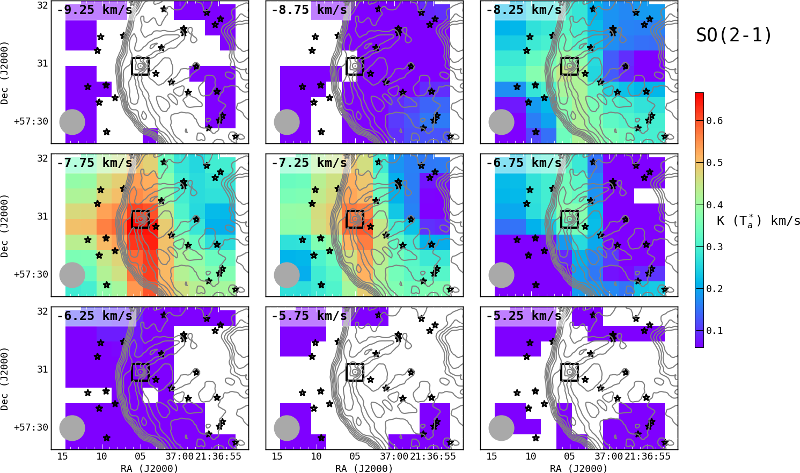} &
\includegraphics[width=0.44\linewidth]{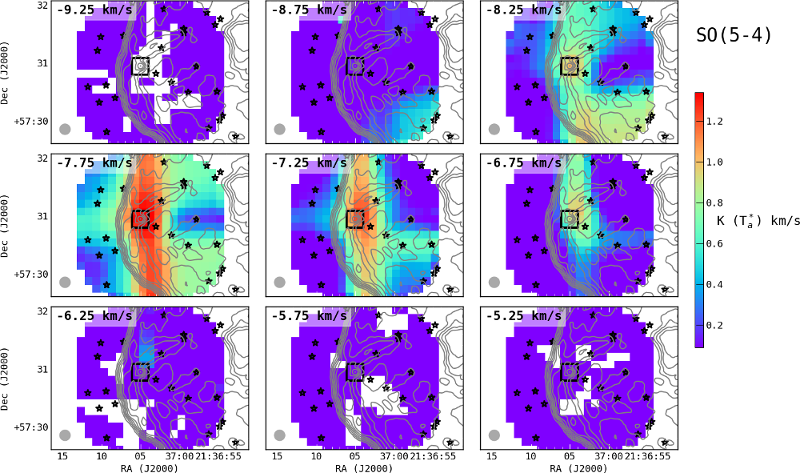} \\
\includegraphics[width=0.44\linewidth]{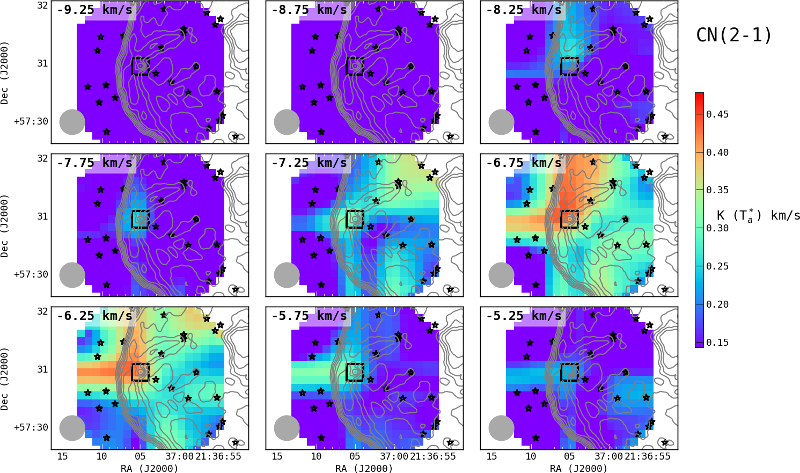} &
\includegraphics[width=0.44\linewidth]{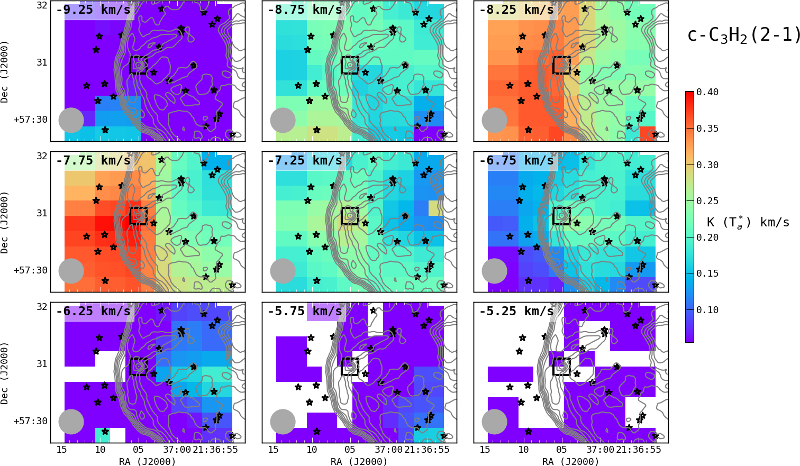} \\
\end{tabular}
\caption{Velocity integrated bit maps. The colors show the line intensity integrated around 0.5 km/s per velocity bin between 3$\sigma$ and the maximum. White is used in case no line emission beyond 3$\sigma$ is detected in the bin (the bottom of the color scale would correspond usually to the noise level). Note that the velocity bins displayed are much larger than the spectral resolution for the sake of space, as most of the lines are detected in between tens to over two hundred channels. The contours mark for reference the \emph{Herschel} 70~$\mu$m  emission on a log scale between 0.1 and 2 Jy/beam.
The beam size is shown for each line. Symbols: IC1396A-PACS-1 (large black square), other YSO candidates (small, black star symbols).
From left to right, top to bottom: SO(2-1), SO(5-4), CN, c-C$_3$H$_2$.\label{momentothers-fig}}
\end{figure*}

Figures \ref{momentCOs-fig} to \ref{momentothers-fig} show the velocity-integrated
bit maps for all the strong lines observed with the high-resolution configuration. 
In order to clarify the location of the emitting regions, all maps
are shown over the \emph{Herschel}/PACS 70~$\mu$m image. Note that for some strong lines,
a quadrant structure is seen in some of the momentum maps due to leakage from the 
strong source, stronger towards the edges where the S/N is lower. The analysis avoids
including these regions.

\section{Details on the multigaussian fits and derived line parameters \label{gaussfit-app}}

\begin{figure}
\centering
\includegraphics[width=0.43\linewidth]{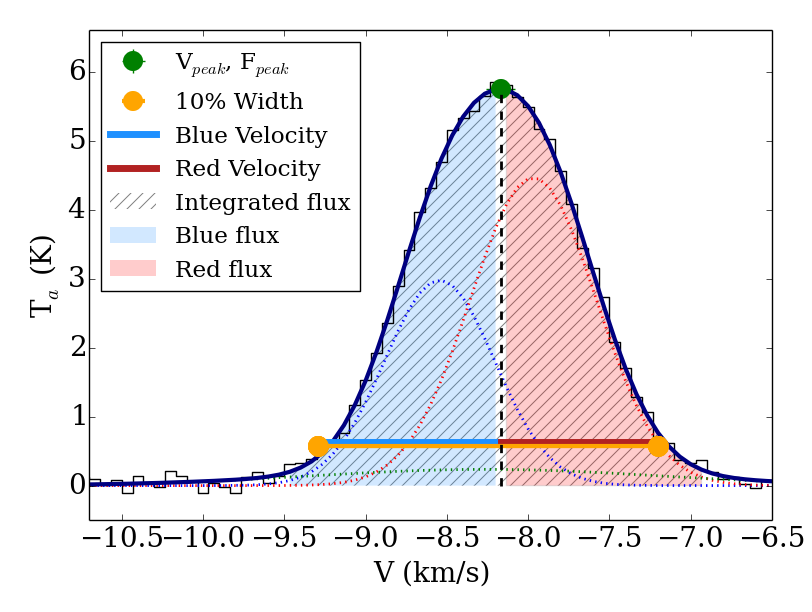}
\caption{Line parameters derived from the line fit (dark blue line) for a C$^{18}$O line,
with the data shown as the black step-plot. The peak velocity and flux (V$_{peak}$, F$_{peak}$)
are self-explanatory. The line width is measured at 10\% of the peak height. The blue and red side of the velocity 
width are used to determine the blue and red velocity, and the corresponding velocity asymmetry as the ratio of blue/red
velocity. The integrated flux is measured starting at 3$\sigma$. The blue and red parts of the flux are ratioed to
obtain the flux asymmetry. Although the underlying multi-Gaussian fit is highly degenerate (blue, green, and red dotted lines), if the final fit
adequately reproduces the line, the line parameters derived  (integrated flux, peak velocity and flux, line width, velocity and flux asymmetries) will be independent of the fit.
 \label{linepar-fig}}
\end{figure}

For the velocity-position analysis, the spectra are first extracted pixel-by-pixel and then fitted using custom Python
interactive routines to derive the line properties for each coordinate position. The fit consists on 1, 2, or 3 Gaussian
components, which can be chosen interactively by the user upon examination of the line and consideration of the $\chi^2$ value.
In particular, the lines are fitted with respect to their velocities using the function $f(v)$

\begin{equation}
f(v)=\sum_{i=1}^n A_i e^{-(v-v_i)^2/\sigma_i^2}, \label{gaussians-eq}
\end{equation}

where n=1, 2 or 3, A$_i$ is the amplitude,  v$_i$ the central velocity, and $\sigma_i$ the width of each Gaussian component.  Note that the fit cannot be directly compared to a physical model, but it is instead used to collect model-free information on how the line profile and intensity varies from pixel to pixel, which can be latter interpreted in terms of density, depletion, collapse, or expansion of the gas and to distinguish the presence of potential multiple components and multiple velocity structures.
Multi-Gaussian fits are usually highly degenerate, especially in the case of low S/N data, so we do not use the
Gaussian parameters directly, but derive several line properties from them (see Figure \ref{linepar-fig} for a graphical representation), 
including the velocity of the peak, line 
width measured at 10\% of the peak height, flux asymmetry (measured as the ratio of the blue vs red parts of the observed
velocity-integrated flux with respect to
the fitted line peak), and velocity asymmetry (measured as the ratio between the line width towards the blue vs the red parts of the line with
respect to the line peak). As long as the fits reproduce the line appropriately (with $\chi^2 \lesssim$1), 
these properties do not dependent on the individual Gaussians nor the number of Gaussians 
used for the fit, being consistent within the errors. The errors in the individual line parameters are derived from the comparison between the
fitted and the observed profiles, taking into account the noise in each particular spectrum. In addition, we also determine the
velocity-integrated flux for the whole line, by summing the flux in the various observed velocity bins, starting at 
the 3$\sigma$ level.

\end{appendix}
\end{document}